\documentclass[preprint]{aastex61}
\usepackage{natbib}
\usepackage{ mathrsfs }

\newcommand{\Msol}{M$_{\odot}$}
\newcommand{\nh}{NH$_{3}$}
\newcommand{\Rcl}{{R_{\rm cl}}}

\newcommand{\Sigmacl}{\bar \Sigma_{\rm cl}}
\newcommand{\Sigmas}{\Sigma_{s}}
\newcommand{\Sigmasmin}{\Sigma_{s,{\rm min}}}

\newcommand{\Sigmabar}{\bar \Sigma}
\newcommand{\Sigmatot}{\Sigma_{\perp}}

\begin{document}
\title{The Green Bank Ammonia Survey: Dense Cores Under Pressure in Orion A}
\author{Helen Kirk}
\affil{NRC Herzberg Astronomy and Astrophysics, 5071 West Saanich Rd, Victoria, BC, V9E 2E7, Canada}
\author{Rachel K. Friesen}
\affil{Dunlap Institute for Astronomy \& Astrophysics, University of Toronto, 50 St. George Street, Toronto, Ontario, Canada M5S 3H4}
\author{Jaime E. Pineda}
\affil{Max-Planck-Institut f\"{u}r extraterrestrische Physik, Giessenbachstrasse 1, D-85748, Garching, Germany}
\author{Erik Rosolowsky}
\affil{Department of Physics, University of Alberta, Edmonton, AB, Canada}
\author{Stella S. R. Offner}
\affil{Department of Astronomy, University of Massachusetts, Amherst, MA 01003, USA}
\author{Christopher D. Matzner}
\affil{Department of Astronomy \& Astrophysics, University of Toronto, 50 St. George Street, Toronto, Ontario, Canada M5S 3H4}
\author{Philip C. Myers}
\affil{Harvard-Smithsonian Center for Astrophysics, 60 Garden St., Cambridge, MA 02138, USA}
\author{James Di Francesco}
\affil{NRC Herzberg Astronomy and Astrophysics, 5071 West Saanich Rd, Victoria, BC, V9E 2E7, Canada}
\affil{Department of Physics and Astronomy, University of Victoria, 3800 Finnerty Road, Victoria, BC, Canada V8P 5C2}
\author{Paola Caselli}
\affil{Max-Planck-Institut f\"{u}r extraterrestrische Physik, Giessenbachstrasse 1, D-85748, Garching, Germany}
\author{Felipe O. Alves}
\affil{Max-Planck-Institut f\"{u}r extraterrestrische Physik, Giessenbachstrasse 1, D-85748, Garching, Germany}
\author{Ana Chac\'on-Tanarro}
\affil{Max-Planck-Institut f\"{u}r extraterrestrische Physik, Giessenbachstrasse 1, D-85748, Garching, Germany}
\author{How-Huan Chen}
\affil{Harvard-Smithsonian Center for Astrophysics, 60 Garden St., Cambridge, MA 02138, USA}
\author{Michael Chun-Yuan Chen}
\affil{Department of Physics and Astronomy, University of Victoria, 3800 Finnerty Road, Victoria, BC, Canada V8P 5C2}
\author{Jared Keown}
\affil{Department of Physics and Astronomy, University of Victoria, 3800 Finnerty Road, Victoria, BC, Canada V8P 5C2}
\author{Anna Punanova}
\affil{Max-Planck-Institut f\"{u}r extraterrestrische Physik, Giessenbachstrasse 1, D-85748, Garching, Germany}
\author{Young Min Seo}
\affil{Jet Propulsion Laboratory, NASA, 4800 Oak Grove Dr, Pasadena, CA 91109, USA}
\author{Yancy Shirley}
\affil{Steward Observatory, 933 North Cherry Avenue, Tucson, AZ 85721, USA}
\author{Adam Ginsburg}
\affil{National Radio Astronomy Observatory, Socorro, NM 87801, USA}
\author{Christine Hall}
\affil{Department of Physics, Engineering Physics \& Astronomy, Queen's University, Kingston, Ontario, Canada K7L 3N6}
\author{Ayushi Singh}
\affil{Department of Astronomy \& Astrophysics, University of Toronto, 50 St. George Street, Toronto, Ontario, Canada M5S 3H4}
\author{H\'{e}ctor G. Arce}
\affil{Department of Astronomy, Yale University, P.O. Box 208101, New Haven, CT 06520-8101, USA}
\author{Alyssa A. Goodman}
\affil{Harvard-Smithsonian Center for Astrophysics, 60 Garden St., Cambridge, MA 02138, USA}
\author{Peter Martin}
\affil{Canadian Institute for Theoretical Astrophysics, University of Toronto, 60 St. George St., Toronto, Ontario, Canada, M5S 3H8}
\author{Elena Redaelli}
\affil{Max-Planck-Institut f\"{u}r extraterrestrische Physik, Giessenbachstrasse 1, D-85748, Garching, Germany}

\begin{abstract}
We use gas temperature and velocity dispersion data from the Green Bank Ammonia Survey
and core masses and sizes from the James Clerk Maxwell Telescope
Gould Belt Survey
to estimate the virial states of dense cores within the Orion~A
molecular cloud.  Surprisingly, we find that almost none of the dense cores are sufficiently
massive to be bound when considering only the balance between self-gravity and the 
thermal and non-thermal motions present in the dense gas.  Including the additional
pressure binding imposed by the weight of the ambient molecular cloud material
and additional smaller pressure terms, however, 
suggests that most of the dense cores are pressure confined.
\end{abstract}

\section{Introduction}
Dense cores, $\sim$0.1~pc-sized condensations of gas and dust, are the birthplaces of 
stars \citep[e.g.,][]{Bergin07,DiFrancesco07,WardThomp07b}.  
One of the most important properties of a dense core is its 
degree of stability against gravitational
collapse.  Understanding which subset of cores in a molecular cloud population is likely to
form protostars in the near future, versus which subset of cores is likely to not form protostars
soon (or ever, unless local conditions change) has profound implications 
for star formation efficiency and the interpretation of the core mass function.
Detailed studies of individual cores \citep[including full 3-D modelling, e.g.,][]{Steinacker13} 
are essential for accurately determining all properties of an individual system.  
This approach, however, requires a prohibitive 
amount of data and modelling time for cloud-wide core population studies.
Instead, by adopting simple stability proxies, insight can be gained into the global
properties and variations of cores within a molecular cloud.  Even this approach requires a
non-negligible amount of information, including mass, 
temperature, non-thermal motions, and other
properties that are typically estimated from different types of observations.  
The Green Bank Ammonia
Survey \citep[GAS;][]{Friesen17} provides an important contribution to these studies.
Using the {\it Herschel} and JCMT Gould Belt Surveys \citep{Andre10,WardThomp07} to
identify high extinction
areas within nearby molecular clouds ($< 500$~pc) that likely have dense gas, 
GAS data provide gas temperature and velocity information to complement the (dust) column 
density information collected by the Gould Belt Surveys.  The large and uniform areal 
coverage of these three surveys provides an unprecedented 
opportunity to understand the broad 
stability properties of dense structures formed within nearby molecular clouds.

The Orion~A molecular cloud is one of the most distant clouds covered by GAS and is
located at a distance of $\sim 415$~pc from the Sun \citep{Menten07,Kim08}.
Orion~A is also the closest example of a molecular cloud forming high-mass stars. 
For example, there are several O stars forming within the Orion Nebula Cluster 
\citep[e.g.,][]{Hillenbrand97,Bally08}.
The Orion~A complex also hosts hundreds of mostly lower mass protostars 
\citep[e.g.,][]{Megeath12,Stutz13}
and many dense cores forming, or with the potential to form, additional protostars
\citep[e.g.,][]{Mezger90,Johnstone99,Ikeda07,Li07,Sadavoy10,Shimajiri11,Polychroni13,Salji15a,Mairs16,Lane16}.  The large census
of dense cores, and extensive observations available for the region, in combination with 
its status as the nearest higher-mass star-forming region, make Orion A an ideal cloud 
in which to
investigate core properties and boundedness.  

The dense cores within the OMC2 and OMC3 regions of Orion~A (i.e., the top portion
of the Integral Shaped Filament) have previously been
analyzed from a virial perspective by \citet{DiLi13}, using a combination of SCUBA data from 
\citet{Nutter07} to identify the dense cores and NH$_3$ data from the VLA and GBT
to estimate the velocity dispersion and kinetic temperature.  While the spatial 
resolution of the NH$_3$ data is quite good, 5\arcsec, the spectral resolution is
poor, with a FWHM of 0.6~km~s$^{-1}$, and \citet{DiLi13} find that many of the 
dense cores have intrinsic line widths below their resolution.  

Given the better velocity resolution and larger spatial coverage of our GAS observations,
it is worth re-visiting a virial analysis of Orion~A.
Here, we combine data from the JCMT Gould Belt Survey,
which identified dense cores and characterized their basic properties (size and mass) 
with new dynamical data from GAS which trace the motions of the dense gas (NH$_3$) 
associated with these dense cores.  

In Section~2, we present the GAS and JCMT GBS observations used in our analysis, along with
supplementary publicly available data (e.g., protostellar catalogues and total cloud column 
density maps).
In Section~3, we first perform a simple virial analysis comparing only self-gravity with
thermal and non-thermal support, and find that most of the dense cores appear unbound
under these assumptions.  We
then include confining pressure terms in our analysis and show that most of the cores 
are actually bound.
In Section~4, we discuss our results in the context of external pressure binding being
an important ingredient in all nearby star-forming regions, before concluding in Section~5.

\section{Observations}
\label{sec_obs}

\subsection{GBT NH$_3$ Observations}
\begin{figure}[htb]
\begin{tabular}{cc}
\includegraphics[width=3in]{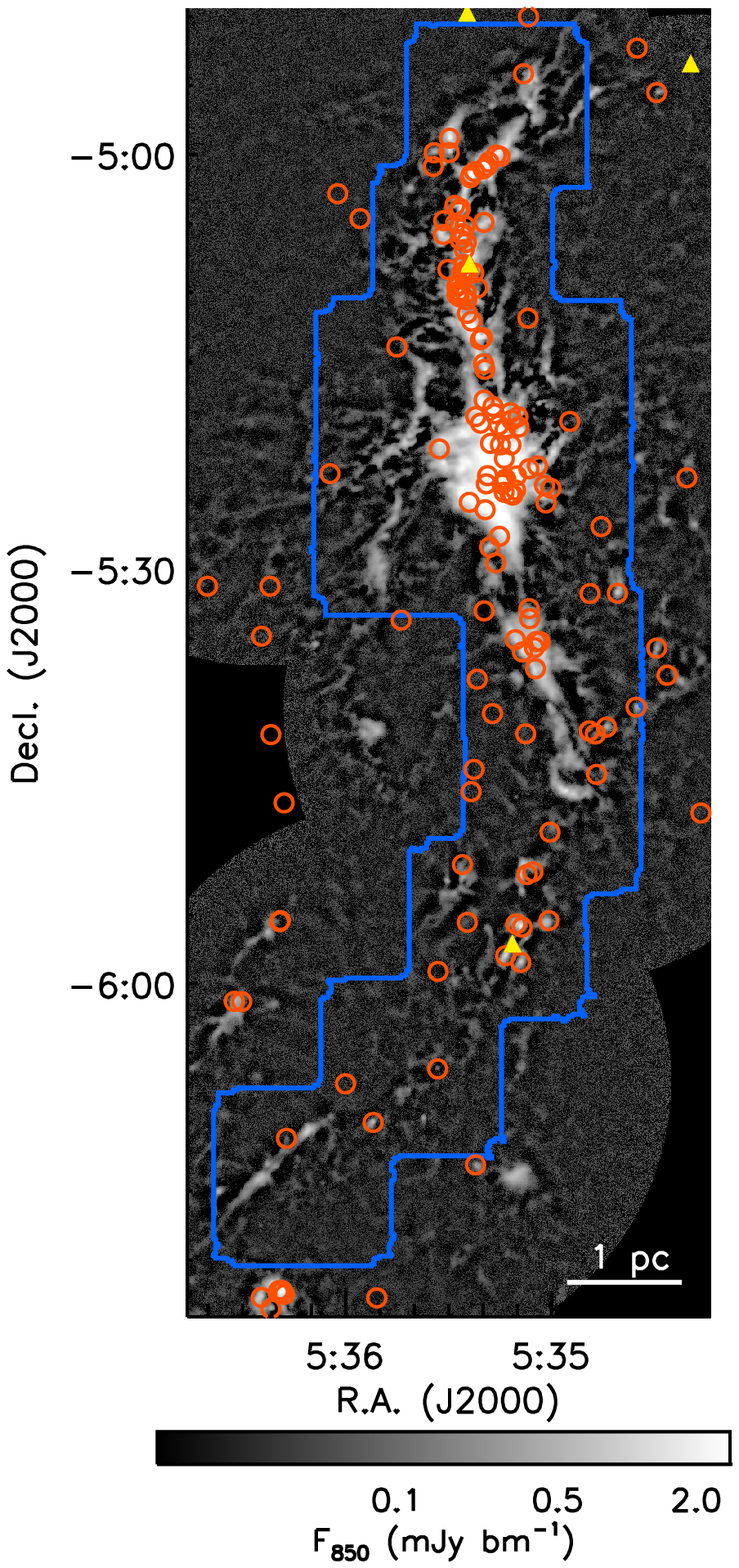} &
\includegraphics[width=3in]{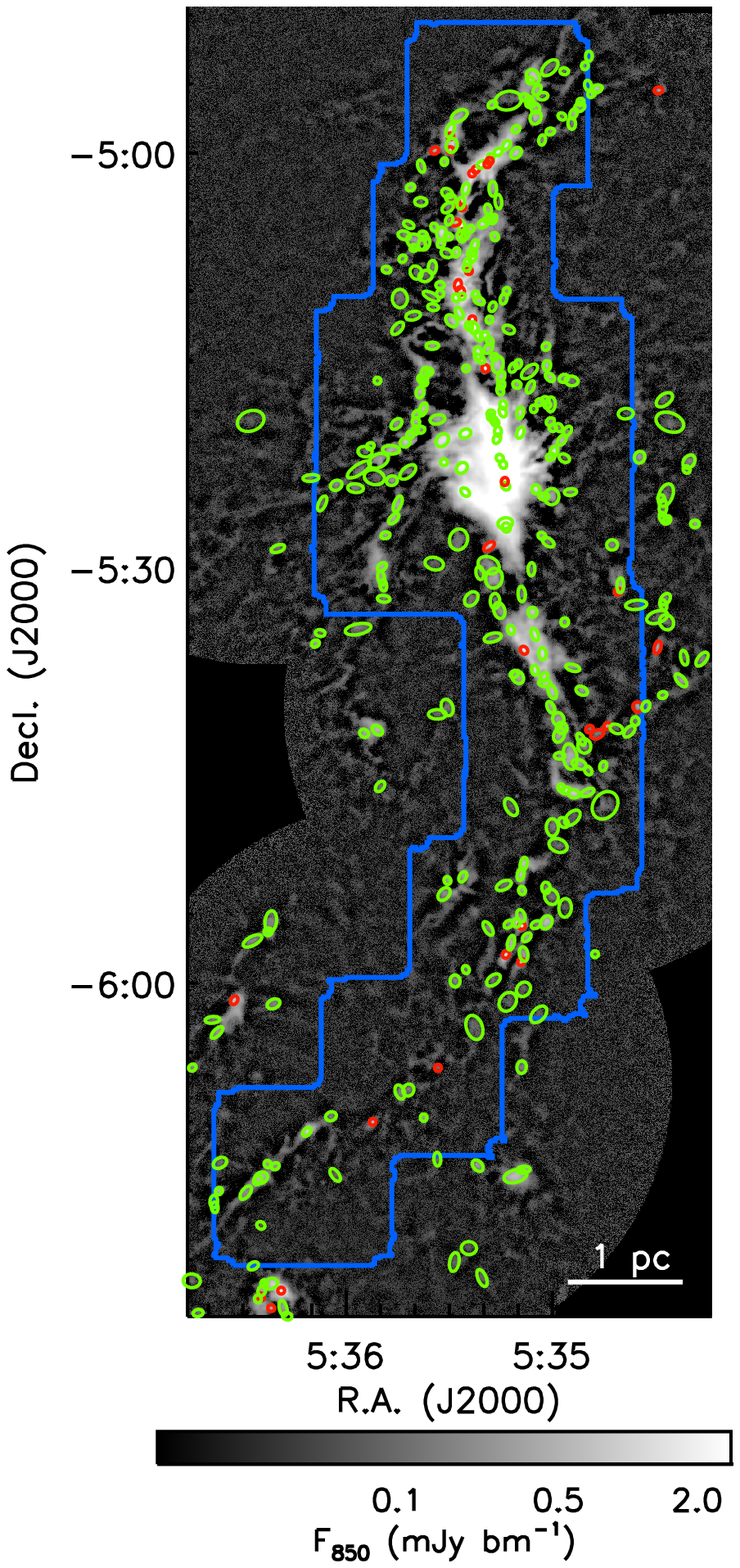} \\
\end{tabular}
\caption{An overview of the JCMT GBS 850~$\mu$m data over the portion of Orion~A
	observed by GAS.  The blue contour shows the areal coverage of GAS for
	the present DR1 analysis.  In the left panel, the dark orange circles 
	show YSOs identified using {\it Spitzer} data \citep{Megeath12}, while the 
	filled yellow triangles 
	show additional YSOs identified using {\it Herschel} data \citep{Stutz13}.
	In the right panel, the ellipses show the Gaussian 1-sigma contours fit to
	the culled {\it getsources} core catalogue of \citet{Lane16}.  Here, protostellar
	cores are shown in red while starless cores are shown in green.
	}
\label{fig_scuba2_gas}
\end{figure}

NH$_3$ observations were obtained through the Green Bank Ammonia Survey (GAS), 
a large project
to map the ammonia (1,1), (2,2), and (3,3) rotation-inversion transitions 
across the high-extinction regions of
nearby Gould Belt molecular clouds using the Green Bank Telescope's
K-band Focal Plane Array (KFPA).  
The survey strategy, data reduction procedure, and basic data properties
are described in detail in \citet{Friesen17}.  
The spatial resolution is 32\arcsec\ (0.064~pc), 
while the spectral resolution is 0.07~km~s$^{-1}$.
Observations of the northern portion of the
Orion~A cloud are one of the first four areas mapped and are presented in \citet{Friesen17} 
as `Data Release 1' (DR1)\footnote{DR1 maps and data products are all publicly
available at {\tt https://dataverse.harvard.edu/dataverse/GAS\_DR1}.}.  
The DR1 observations of Orion~A cover the
integral shaped filament (ISF) and areas slightly southward of it.  Additional
data further south were obtained at a later date and will be included in a second
data release.  
Figure~\ref{fig_scuba2_gas} shows an overview of the area mapped
overlaid on the JCMT SCUBA-2 850~$\mu$m image of the region,
with the dense cores (see Section~\ref{sec_cores}) and protostars (see Section~\ref{sec_proto})
overlaid in each panel. 

\subsection{JCMT SCUBA-2 data}
\label{sec_cores}
Although NH$_3$ is an excellent tracer of dense gas, defining dense cores 
from its emission alone can be
complicated by varying abundance levels.  At the distance of Orion
\citep[$\sim 415$~pc;][]{Menten07,Kim08},
the $\sim$32\arcsec\ resolution of our GAS data also presents a challenge in identifying
individual dense cores in such a complex and clustered environment.
Submillimetre continuum emission offers an alternative method to identifying dense cores
and estimating their sizes and masses.  There are, however several separate 
significant sources
of uncertainty in the conversion between flux density and mass,
as well as the potential for chance alignments of lower density structures
giving rise to apparent cores.
The James Clerk Maxwell Telescope Gould Belt Survey \citep[JCMT GBS;][]{WardThomp07}
observed 6.2 square degrees around the Orion A molecular cloud at 850~$\mu$m
and 450~$\mu$m with SCUBA-2, with resolutions of 14.6\arcsec\ and 9.8\arcsec\
\citep{Dempsey13}, respectively.  Observations of the northern and southern portions
of Orion~A were first published in \citet{Salji15a} and \citet{Mairs16}, respectively.
For our analysis, we use the dense core catalogue presented in \citet{Lane16}, which
covers the entire Orion~A complex.  

\citet{Lane16} analyzed the JCMT GBS 850~$\mu$m data for the entire Orion~A molecular
cloud, using a map which reaches the full survey depth but is slightly poorer at
recovering the largest-scale structures than the current best reduction method.
\citet{Lane16} identified dense cores using two independent methods,
{\it getsources} \citep{Menshchikov12} and {\it FellWalker} \citep{Berry15}, and
analyzed their clustering properties.  The {\it getsources} algorithm is a multi-scale,
multi-wavelength source extraction algorithm.  {\it Getsources} first decomposes emission
from each wavelength into a variety of scales, and then uses the combined information across
wavelengths to create a Gaussian-based model that characterizes the small-scale sources
and separates them from larger-scale emission features.
In contrast, the {\it FellWalker} algorithm separates peaks based on local gradients,
assigning each pixel to the peak that the local gradient points towards.  {\it FellWalker}
therefore does not assume any particular source geometry and in the \citet{Lane16} 
catalogue, no background subtraction was performed to remove large-scale structures.
Since Orion~A is home to quite 
complex emission structures, \citet{Lane16} found that in general {\it getsources}
performed better in isolating dense and compact structures.  
For our main analysis, we therefore adopt the {\it getsources}-based catalogue of 
\citet{Lane16}, although in Appendix~\ref{app_cores}, we present an analysis using
instead the {\it FellWalker}-based catalogue.
In Appendix~\ref{app_cores}, we demonstrate that our results are robust against
changes in the core catalogue used.

The \citet{Lane16} Orion~A catalogue provides the sizes, total fluxes, and 
peak positions of the
dense cores.  We approximate the
dense core radius as the geometric mean of the major and minor axis
FWHMs fit by {\it getsources}, noting that at a radius equal to the FWHM, a Gaussian
profile is at one eighth of its full height, and is therefore a reasonable
estimate of the full core extent.  We assume a distance of 415~pc to Orion~A,
and correct the \citet{Lane16} catalogue values from their assumed distance of
450~pc.
We also correct the core sizes for the telescope beam; for cores with sizes less
than half of the 14.6\arcsec\ SCUBA-2 850~$\mu$m beam FWHM, deconvolution is likely to be
unreliable, and so we report this value as the upper limit to their true size.
We convert the total flux of each 
core measured by {\it getsources} to a mass using 
\begin{equation}
M = 1.30 \Big( \frac{S_{850}}{1~{\rm Jy}} \Big) 
        \Big( \frac{\kappa_{850}}{0.012~{\rm cm}^2~{\rm g}^{-1}} \Big) 
        \Big(exp\Big(\frac{17~{\rm K}}{T_d}\Big) -1\Big)
        \Big( \frac{D}{450~{\rm pc}}\Big) ^2 M_{\odot}
\end{equation}
where $S_{850}$ is the total flux density at 850~\micron, $\kappa_{850}$ is the dust opacity at 
850~\micron, $T_d$ is the dust temperature, and $D$ is the distance.
We assume a constant dust grain opacity at 850~$\mu$m of 0.012~cm$^2$~g$^{-1}$,
consistent with previous JCMT and {\it Herschel} Gould Belt analyses, and 
which includes the standard dust to gas ratio of 0.01 \citep[e.g.,][]{JKirk13,Pattle15}. 
(Note that the popular \citet{Ossenkopf94} model~5 for icy grains would give a 
larger opacity of 0.018~cm$^2$~g$^{-1}$ at 850~$\mu$m.)
For each core, we adopt a dust temperature equal to the kinetic temperature derived
from our GAS NH$_3$ observations (see Section~2.4).  For cores where a kinetic temperature was not
able to be measured, we assume a value of 18~K, which is the 
mean temperature measured for the cores in NH$_3$.  Previous works, including
\citet{Lane16}, assumed a constant dust temperature of 15~K, which would give masses
roughly 30\% larger than those quoted here for a mean dust temperature of 18~K.  
Similarly, assuming instead a higher temperature
of 21~K would give masses that were 20\% smaller.
We note that while a 20\% to 30\% change in estimated mass can be significant for a 
detailed study of a single object, this level of difference has relatively little impact on our
population study due to the broad range in properties measured.  For example, most of the cores
would not change their status of being
gravitationally bound or unbound in Section~3.2 with only
a 20\% to 30\% change in their mass.
Much larger variations in the mass estimated for cores arise in regions of complex
emission structure such as Orion~A through the choice in algorithm used to identify cores.
In Appendix~\ref{app_cores}, the cores identified using {\it FellWalker} tend to be more massive 
with larger sizes than the {\it getsources} cores.  Highlighting the challenges of core
identification, there is not always a one-to-one correspondence between objects in the 
two catalogues, as {\it getsources} tends to split emission structures more finely than
{\it FellWalker} does.  
Despite these large uncertainties in defining individual cores and accurately
measuring their properties, our overall conclusions about the typical virial state of 
cores is similar using either algorithm, suggesting that our overall conclusions about
bulk virial properties are robust.

Finally, we note that some of the dense cores in \citet{Lane16}, although they are
clearly detected, have poor Gaussian fits.  While this was not important for the
original clustering analysis of \citet{Lane16}, 
poor estimates of the core radius could significantly
impact our virial analysis.  We therefore apply several cuts to the \citet{Lane16}
{\it getsources} catalogue, requiring dense cores in our present analysis to 
have parameters:\\
\texttt{
gs\_sig\_glob $\ge$1\\
gs\_sig\_mono8 $\ge$7\\
gs\_good $>$ 0\\
gs\_relp850 $>$ 3\\
gs\_relt850 $>$ 3\\
}
In short, these criteria imply `significant' fits (\texttt{gs\_sig\_glob, gs\_sig\_mono8})
with fitted peak fluxes of S/N $\ge 3$
(\texttt{gs\_relp850}) and fitted total fluxes of S/N $\ge 3$ 
(\texttt{gs\_relt850}),
and no {\it getsources} flags (\texttt{gs\_good}) 
associated with the fits.  
These fitting criteria reduce the original 919 dense cores in \citet{Lane16}
to 610, primarily excluding the faintest cores with the poorest fits.  In the culled
catalogue, the lowest mass core identified is 0.09~\Msol.  Of the 610 reliably fit
dense cores, roughly half (311) lie within the area mapped by GAS.
The median mass of the cores is 0.7~\Msol\ for those with reliable {\it getsources} 
sizes and reliable temperature and kinematic measures (see Section~2.4). 

Figure~\ref{fig_scuba2_gas} shows the GBS 850~$\mu$m emission over the area observed
with GAS, with the culled {\it getsources} catalogue overlaid.
The dearth of {\it getsources} cores in the centre of the ISF visible in
Figure~\ref{fig_scuba2_gas} is caused by two effects.  First, {\it getsources} 
excludes large-scale emission modes from its core-fitting routine, and hence a
significant amount of flux within the centre of the ISF is not attributed to any
{\it getsources} core.  
Second, the larger-scale emission within the centre of the
ISF increases the difficulty of fitting Gaussian models to the compact emission
features present, leading to a greater fraction of cores being culled in this particular
area.
We note, however, 
that our overall results are similar even when the full {\it getsources}
catalogue from \citet{Lane16} is used.

In Table~\ref{tab_core_props}, we summarize the properties of the cores identified
by {\it getsources} for which there are measurements of their GAS-derived 
properties (see following sections).

\subsection{Protostars}
\label{sec_proto}

\citet{Lane16} further classify their dense cores as being protostellar or starless
based on whether or not they are spatially coincident with a protostar identified
by \citet{Megeath12} using {\it Spitzer} observations or \citet{Stutz13} using
{\it Herschel} data.  Due to the highly clustered nature of Orion~A, \citet{Lane16}
required a separation of less than one beam radius (7.25\arcsec) from the dense
core's peak location for a core to be classified as protostellar.  
Note that this small separation requirement ensures that any given protostar
can be associated with a maximum of one dense core.
\citet{Meingast16} recently identified YSOs in Orion based on near infrared
observations using VISTA.  
We searched their catalogue for evidence of additional protostellar
cores, but found only three possible new associations with our core catalogue,
all of which were listed in \citet{Meingast16} as previously known class~II sources.  
For simplicity, we did not re-classify these three cores.
Protostellar
and starless dense cores are shown in Figure~\ref{fig_scuba2_gas} in the right-hand
panel, while the protostars themselves are shown in the left-hand panel.

\subsection{Deriving Dynamical Core Properties}
\begin{figure}[htb]
\includegraphics[height=5in]{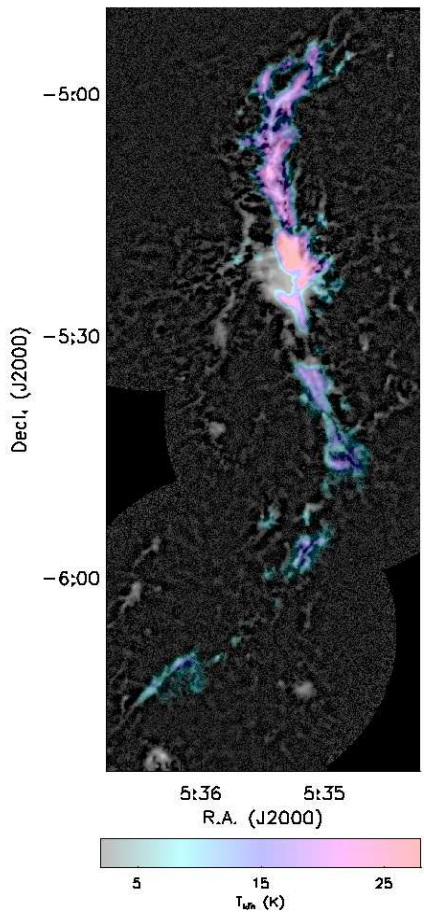}
\caption{A comparison of the GBS 850~$\mu$m data in Orion A (greyscale) with
	the kinetic temperature of NH$_3$ fitted to the GAS data (colourscale).
	The SCUBA-2 flux scale is the same as in Figure~\ref{fig_scuba2_gas}, while $T_{kin}$
	varies from 5~K (blue) to 30~K (red) in the colour scaling (see colour bar).
}
\label{fig_scuba2_Tkin}
\end{figure}

The GAS \nh\ (1,1) and (2,2) observations were used to estimate the dense gas properties,
as described in more detail in \citet{Friesen17}.  Our interest here is in the 
velocity dispersion, $\sigma$, and the kinetic temperature, $T_{kin}$.  
Figure~\ref{fig_scuba2_Tkin} shows a comparison of $T_{kin}$ overlaid 
with the SCUBA-2 850~$\mu$m
emission of Orion~A,
highlighting the well-known fact that the most active part of the ISF tends
to be noticeably warmer than its surroundings.  We note that for the GAS
DR1, all spectra are fit with a single velocity component.  A visual inspection
of the spectra suggest that a single velocity component is sufficient to describe
the majority of NH$_3$ spectra, even in Orion~A, however, several percent of the
spectra require a second velocity component to be well-described \citep{Friesen17}. 
Multiple velocity component fits are planned for a future GAS data release.

Using the GAS \nh-based property maps, we calculate the weighted mean
kinetic temperature and velocity dispersion
of each dense core.
For the {\it getsources}-based dense core catalogue used for the
bulk of our analysis, we consider all GAS pixels which lie within a radius of one
core FWHM from each core's peak (i.e., the same area as the core's full extent,
as discussed in Section~2.2), and calculate the weighted mean of each property
(weighting by the inverse square of the uncertainty in the fitted parameter).  
We calculate both the error in the
weighted mean as well as the weighted standard deviation, to measure the variation in
property values across each core.  These values are all listed in Table~\ref{tab_core_props}.
We also measured the kinetic temperature
and line width at the peak position for all of the cores, and found largely similar
results to those presented here.
A total of 29 protostars and 282 starless cores lie within the GAS observing
footprint.  Of these, 26 protostars and 211 starless cores have kinematic properties
measured, i.e., there was sufficient sensitivity in the ammonia data to estimate
line width and kinetic temperatures.  The velocity dispersions are similar for the 
starless and protostellar cores, with a mean and standard deviation of
0.38~km~s$^{-1}$ $\pm 0.20$~km~s$^{-1}$ and 0.37~km~s$^{-1}$ $\pm 0.17$~km~s$^{-1}$
for the starless and protostellar cores, respectively.  The kinetic temperatures
are also similar.
After removing less reliable measures where $T_{kin} > 30$~K, we find typical kinetic temperatures of
18~K $\pm$ 4~K for the starless cores, and 17~K $\pm$ 3~K for the protostellar cores.
Previous observations have tended to find smaller kinetic temperatures for 
starless cores relative to protostellar cores - for example, \citet{Jijina99} compiled
all available NH$_3$ observations in the literature and found median values of the
kinetic temperature of 12.4~K and 15~K for apparently starless and protostellar cores
respectively, while more recent work in the Perseus molecular cloud, reveals typical
kinetic temperatures of 10.6~K and 11.9~K for starless and protostellar cores respectively
\citep{Foster09}.  Both \citet{Jijina99} and \citet{Foster09} also note that environment
appears to play a stronger role than protostellar content on the kinetic temperature.
\citet{Jijina99} found median kinetic temperatures of 20.5~K and 12.0~K for cores
within and outside of clusters, while \citet{Foster09} measured typical kinetic temperatures
of 12.9~K and 10.8~K, for clustered and unclustered cores respectively, in Perseus.  
Since all of the dense cores we analyze
in Orion~A lie within a highly clustered environment, we might not expect to see
cores with lower kinetic temperatures.  
This general trend of higher kinetic
temperatures being present in more clustered environments is also obvious across the four
GAS DR1 regions \citep{Friesen17}, where the isolated B18 (Taurus) has the lowest
kinetic temperatures, and the moderately clustered NGC1333 (Perseus) and L1688 (Ophiuchus)
have intermediate kinetic temperatures. 
It is possible that our estimated T$_{kin}$ values in Orion~A are also are slightly 
elevated due to contamination from 
the warmer envelope material surrounding the dense cores.

\subsection{Total Column Density}
\begin{figure}
\plottwo{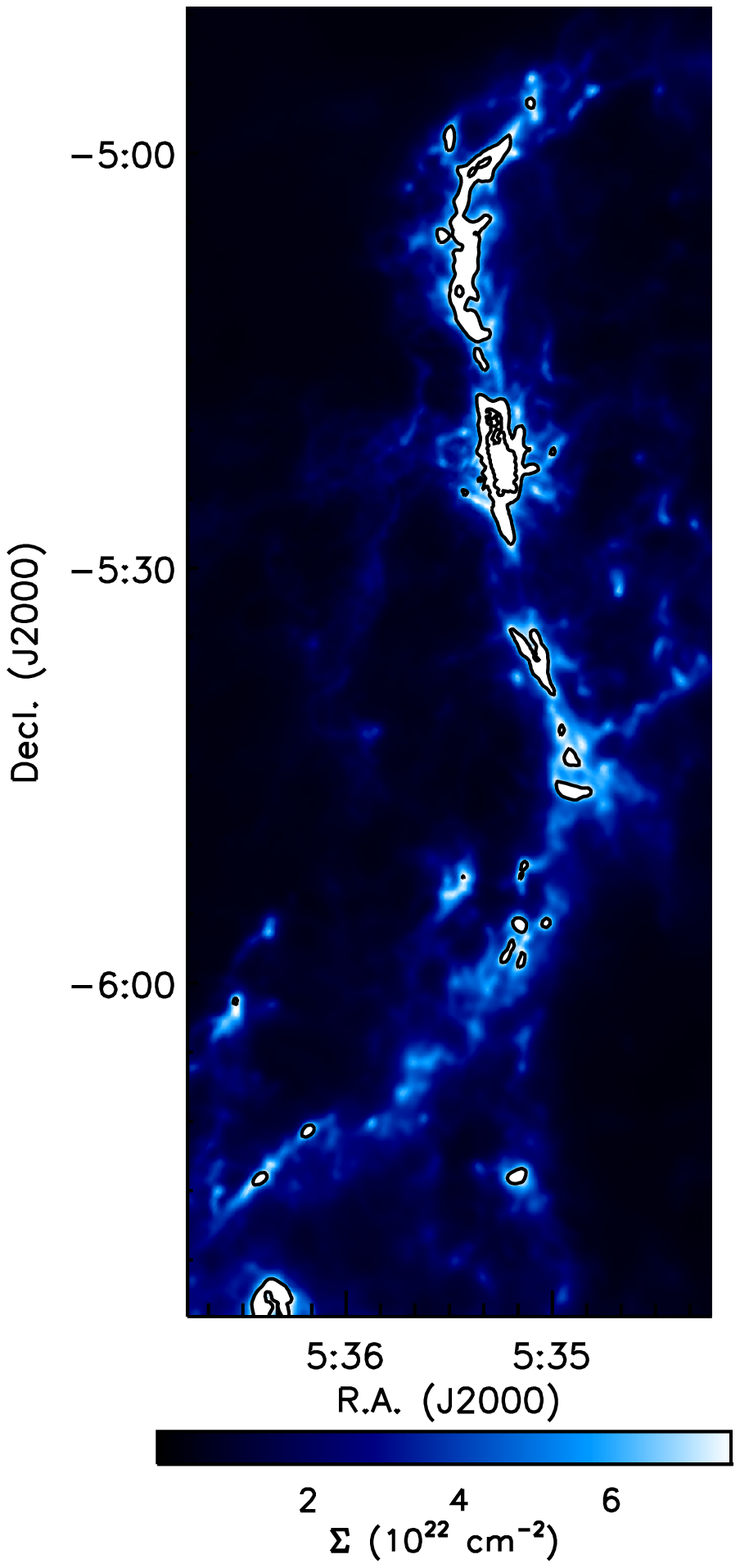}{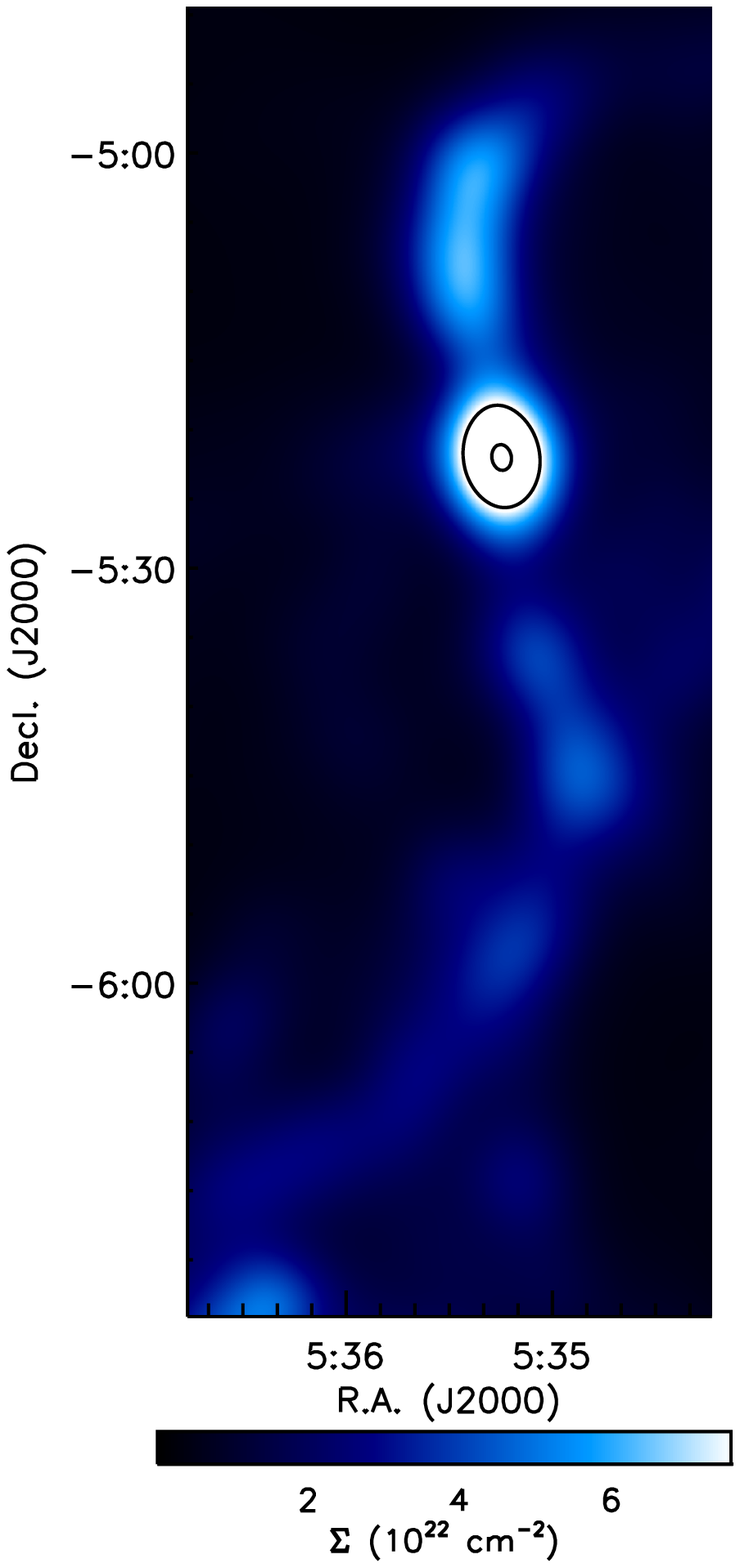}
\caption{Left: Original column density map from \citet{Lombardi14} shown over the
	same area as the previous figures.  Contours are shown for 
	$8 \times 10^{22}$~cm$^{-2}$ and $3 \times 10^{23}$~cm$^{-2}$.  
	Right: column density map containing only structures larger than 0.5~pc.
	Contours are shown for $8 \times 10^{22}$~cm$^{-2}$ and 
	$1.2 \times 10^{23}$~cm$^{-2}$.
}
\label{fig_lombardi}
\end{figure}

To estimate the ambient pressure due to the weight of surrounding molecular
cloud material on the dense cores, we require a map of the total column density
in Orion~A.  \citet{Lombardi14} derived such a column density map by performing
point-by-point modelling of the spectral energy distribution of flux measured
by both the {\it Planck} and {\it Herschel} Space Telescopes.  We use the
850~$\mu$m optical depth map derived in \citet{Lombardi14}, and convert into
column density using the equations and constants given in their 
Equations 12 and 16:
\begin{equation}
A_K = \gamma \tau_{850} + \delta
\end{equation}
and 
\begin{equation}
\frac{\Sigma}{A_K} = \mu \beta_K m_p
\end{equation}
where $A_K$ is the extinction in the K-band, $\gamma = 2640$~mag, $\delta = 0.012$~mag,
$\Sigma$ is the column density (in mass units), $\mu = 1.37$, 
$\beta_k = 1.67 \times 10^{22}$~cm$^{-2}$~mag$^{-1}$, and $m_p = 1.67\times 10^{-24}$~g\footnote{
We note that these equations assume that there is no grain growth, 
a process that might affect
the scaling at the highest column densities.}.
Note that the $\gamma$ and $\delta$ factors are fitted by \citet{Lombardi14}, and
they obtain slightly different values in Orion~B than in Orion~A.  We adopt the Orion~A 
values here.
The left panel of Figure~\ref{fig_lombardi} shows the column density map from \citet{Lombardi14} 
with the scalings above adopted.

Every position in the column density map may have contributions from structures on a 
variety of size scales.  To estimate the pressure on the cores from the weight
of the overlying molecular cloud, we need to consider only the portion of the column
density that is attributable to larger-scale structure.  In particular, the 
36\arcsec\ angular resolution of the column density map (matching the {\it Herschel} 
resolution at 500~$\mu$m) is comparable to the size of the dense cores, suggesting that
in the vicinity of the dense cores, a non-negligible portion of the column density
measured could arise from the cores themselves and not larger-scale structures.  
We therefore filter out the smallest scale structures from the column density map when 
estimating the cloud weight pressure.
To perform this filtering, we use a similar method to that in
\citet{Kainulainen14}. Namely, we use the {\it a Trous} wavelet transform to measure
the amount of structure on various scales\footnote{We use the IDL program {\it atrous.pro} 
developed by Erik Rosolowsky which is available at 
{\tt https://github.com/low-sky/idl-low-sky/tree/master/wavelet}.}.
Structures are measured on scales of $2^N$ pixels, by smoothing and
subtracting the smoothed image from the remainder, going from the largest
to the smallest scale.  The large-scale structure can then be represented as
the sum of all single-scale structure maps at sizes above the desired smoothing scale.

In the \citet{Lombardi14} maps, each pixel is 15\arcsec, corresponding to 0.03~pc
at a distance of 415~pc.  Dense cores tend to have sizes between 0.03~pc and 0.2~pc
\citep{Bergin07}, while larger-scale clumps and clouds span sizes from several tenths
of a parsec to more than ten parsecs.  
For a conservative estimate,
we assume that the structures of order 0.5~pc and larger
belong to the cloud rather than the dense core.  The closest smoothing scale to
0.5~pc is 16 pixels, which corresponds to 0.48~pc.  For our nominal best estimate
of the column density attributable to the cloud, we sum all structures of sizes
$\sim$0.5~pc and larger generated by the {\it a Trous} algorithm.  In Appendix~\ref{app_press},
we show that our qualitative conclusions remain similar even when we change
the limiting size scale for the cloud column density features.
Figure~\ref{fig_lombardi} shows the original \citet{Lombardi14} column density map,
as well as the 0.5~pc filtered version of the column density map, which we use
for the majority of our analysis.

Within the large-scale cloud structure, dense cores are also usually found to reside
within filaments.  Observations from {\it Herschel} suggest that filaments have
widths of order 0.1~pc \citep{Arzoumanian11, Andre14}, similar to the size of the dense cores,
which makes separating cores and filaments unfeasible using the simple filtering method
described above.  
In our analysis in Section~3.3, we approximate the bounding pressure of filaments using 
a different method than that discussed here.
Finally, we note that not all of the column density may belong to structures that 
the core is associated with, i.e., some of the column density may belong to additional
structures along the line of sight.  This is especially important to acknowledge in a 
region as complex as Orion~A.  While multiple structures along the line of sight are
likely more common at intermediate size scales than the largest cloud scale,
we note that our estimate of the cloud column density may be slightly over-estimated.

\section{Analysis}

\subsection{Linewidth-Size Relationship}
We first examine the relationship between the line widths and sizes of the dense cores.
The observed line width, $\sigma_{obs}$ has contributions from both  
turbulent and thermal energy, i.e.,
\begin{equation}
\sigma_{obs,NH3}^2 = \sigma_{turb}^2 + \sigma_{therm,NH3}^2
\end{equation}
We assume that the kinetic temperature for the mean gas is the same as we measure for  
NH$_3$, and calculate the total line width by subtracting the thermal contribution from
NH$_3$ and adding in the assumed total gas thermal contribution, i.e.,
\begin{equation}
\sigma_{tot} = \sqrt{\sigma_{obs} ^2 - \frac{k_B T_{kin}}{\mu_{\mathrm{NH3}} m_\mathrm{H}} + 
	\frac{k_B T_{kin}}{\mu_{\mathrm{mean}} m_\mathrm{H}} }
\end{equation}
where $k_B$ is the Boltzmann constant, $\mu_{\mathrm{NH3}}$ is the molecular weight
of NH$_3$ (17), $\mu_{\mathrm{mean}}$ the mean molecular weight, and $m_\mathrm{H}$ the mass 
of a hydrogen atom.  Following \citet{Kauffmann08}, we take $\mu_{\mathrm{mean}} = 2.37$.

Bulk motions within a core can also contribute to the total velocity dispersion,
and are usually quantified through measurement of the standard deviation in the centroid
velocity of the gas measured across the core.  This contribution would then be added
to the total velocity dispersion in quadrature.  We find that including this term
typically has a small effect on the resulting total velocity dispersions, adding
less than 0.1~km~s$^{-1}$ to the total velocity dispersion calculated without its inclusion
for more than 90\% of the cores.
The mean difference in the total velocity dispersion between including and excluding the
standard deviation in centroid velocity is 0.04~km~s$^{-1}$, while the median difference 
is 0.02~km~s$^{-1}$.
We therefore exclude the contribution from the change in line centroid to 
the total velocity dispersion that we use for our analysis.

\begin{figure}[htb]
\includegraphics[height=3in]{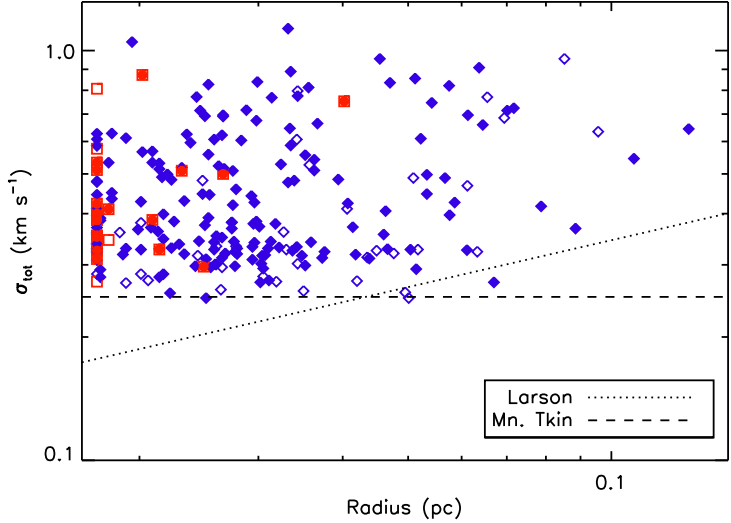}
\caption{ The total 1-D velocity dispersion of the dense cores compared with their effective 
	radii.  The dashed horizontal line shows the velocity dispersion of purely thermal gas
	at a temperature equal to the mean kinetic temperature of the cores,
	i.e., 18~K.  The
	dotted diagonal line shows the line width-size relationship reported by
	\citet{Larson81}, after scaling for the appropriate observed quantities,
	i.e., radius versus diameter and 1-D versus 3-D velocity dispersion.
	Red squares indicate protostellar cores while blue diamonds indicate starless
	cores.  Filled symbols indicate cores lying near the ISF,
	and open symbols show cores which lie further south.}
\label{fig_larson}
\end{figure}

Figure~\ref{fig_larson} shows the total line width and size measured for each core, using
Equation~5.  All of the cores have significantly supersonic velocity dispersions -- 
the velocity dispersion expected for a purely
thermal gas at the mean $T_{kin}$ value observed is shown by the horizontal dashed line,
and it lies about a factor of two lower than the smallest total velocity dispersions
measured.  These line widths are larger than is typical for the nearest molecular
clouds, such as those in Perseus, where cores tend to have roughly transsonic total line widths
\citep[e.g.,][]{Foster09}.  Orion, however, has long been recognized as an environment
where cores have larger non-thermal motions present in dense gas tracers such as
NH$_3$ \citep[e.g.,][]{Jijina99,Caselli95}.  These larger line widths do not appear to be an
artefact of the larger distance to Orion as compared to nearby clouds such as Perseus.
We compared the NH$_3$ line widths reported by \citet{DiLi13} using higher angular
resolution VLA observations ($\sim$5\arcsec\ versus our 32\arcsec) and see no evidence
that our measured line widths are systematically larger.  On the contrary, our measurements
tend to be slightly smaller, likely due to the poorer velocity resolution of the observations of
\citet{DiLi13}.

The dotted diagonal line in Figure~\ref{fig_larson} shows the standard \citet{Larson81} 
line width-size relationship measured for larger-scale structures using CO observations.
Clearly our population of cores does not follow Larson's scaling law, but instead shows
no variation as a function of core size, and only a large amount of scatter.
This lack of change in line width as a function of size is exactly what is expected for the 
behaviour of dense gas within
cores, as the material lies in a zone of coherence \citep[e.g.,][]{Goodman98b,Pineda10},
although the zone of coherence is typically also characterized by a thermally-dominated
velocity dispersion, which is not the case here.

\subsection{\nh\ motions versus self-gravity}
\label{sec_nh_grav}
We next compare the amounts of thermal and non-thermal support available to
each dense core to their self-gravity.
The thermal Jeans criterion is:
\begin{equation}
M_{J,T} = \frac{5 R}{2 G (\mu_{\mathrm{mean}} m_\mathrm{H} / k_B T_{kin})}
\end{equation}
\citep[e.g.,][]{Bertoldi92}
where $M_{J,T}$ is the (thermal) Jeans mass, $R$ is the radius, and $G$ is 
the gravitational constant.
If non-thermal support is included, we can re-write the equation as
\begin{equation}
M_{J,tot} = \frac{5 R \sigma_{tot}^2}{2 G}
\end{equation}
where $\sigma_{tot}$ is the total line width given by Equation~5.

\begin{figure}[htb]
\includegraphics[height=3in]{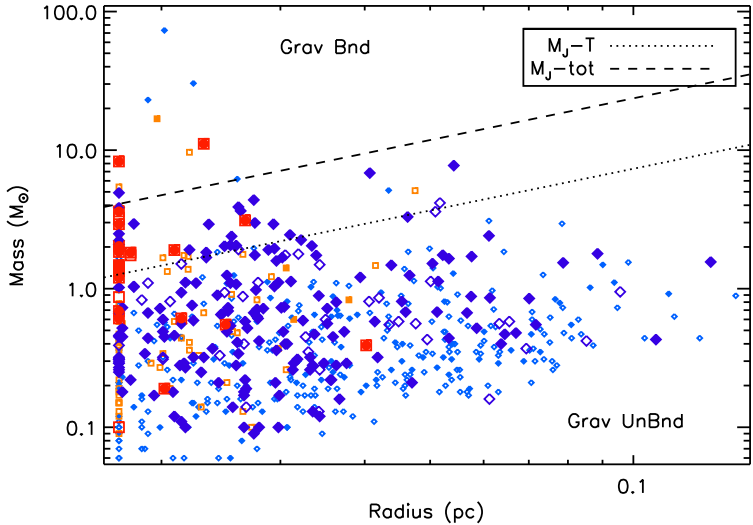}
\caption{A comparison of the masses and sizes of the dense cores.
	Starless cores are shown as blue diamonds while protostellar 
	cores are shown as red squares.  
	Filled symbols show cores associated with the ISF MST-based 
	cluster in \citet{Lane16}. 
	The dotted line
        indicates the thermal Jeans mass for the mean temperature measured for the cores.  
	The dashed line shows the Jeans mass when mean non-thermal motion is included as a 
	thermal pressure-like term.  
	Smaller, lighter symbols show cores that fell within the
	GAS footprint but did not have reliable kinematic properties measured.
}
\label{fig_nopress1}
\end{figure}

Figure~\ref{fig_nopress1} shows the mass and size measured for each dense 
core based on SCUBA-2 850~$\mu$m data compared to various Jeans stability criteria.
Here, we use the mean values of $T_{kin}$ and 
$\sigma_{nt}$ obtained for dense cores where a good fit to these parameters
was obtained.  Unsurprisingly, turbulent motions dominate over thermal motions in
all cores.  When non-thermal motions are considered in addition to the thermal pressure
supporting the dense cores, nearly all of the dense cores lie below the Jeans line
(dashed line in the figure).  This behaviour implies that the cores are gravitationally
unbound,
i.e., they have insufficient
self-gravity due to their mass to counteract the internal (turbulent) gas motions.  

\begin{figure}[htb]
\includegraphics[height=3in]{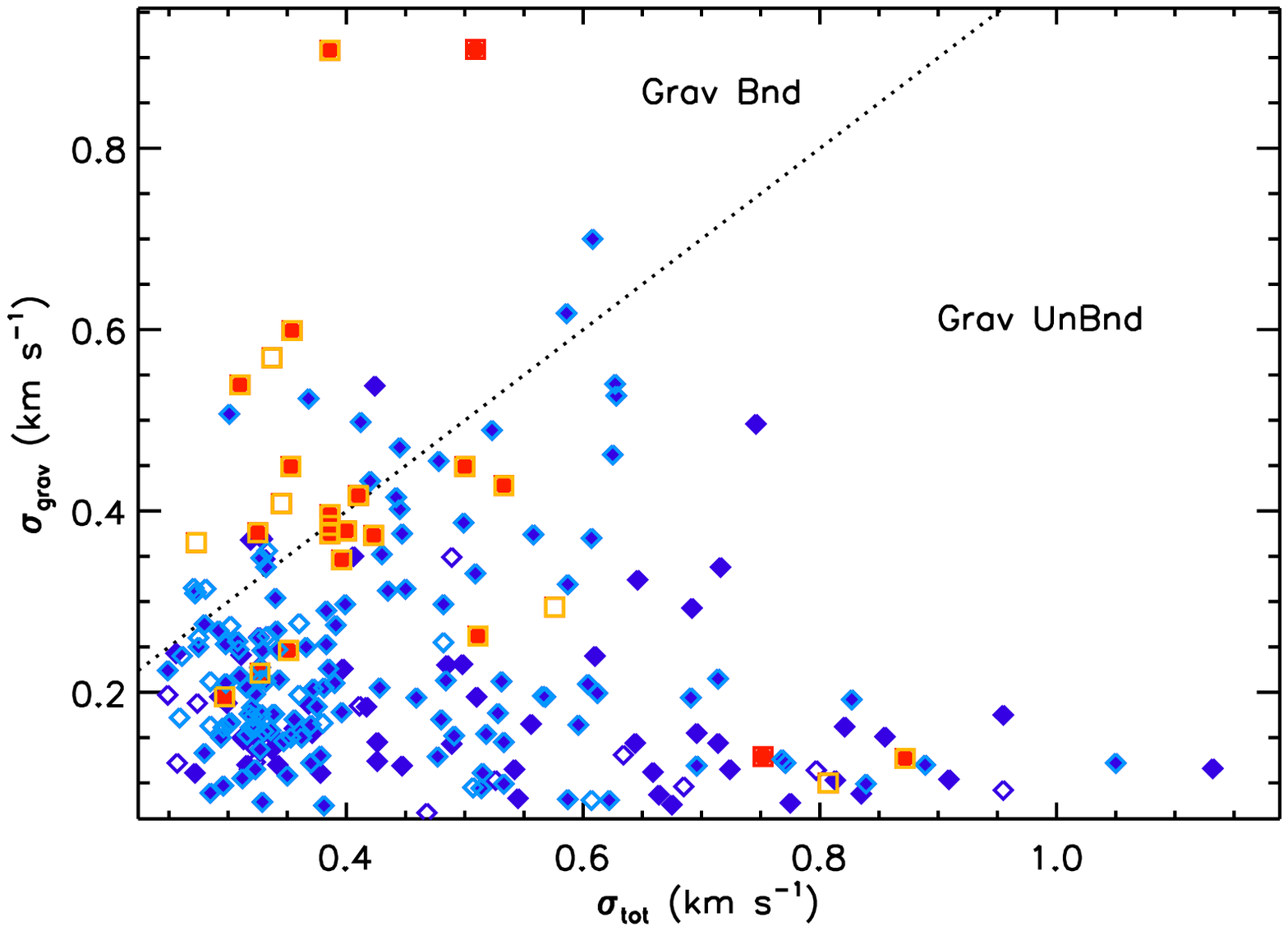}
\caption{A direct comparison of the velocity dispersion
        required to balance gravity (vertical axis; $\sigma_{grav}$)
        with the total velocity dispersion
        measured (horizontal axis).
	Blue diamonds denote starless cores while red squares denote
	protostellar cores.  Filled symbols show cores associated with the ISF, while
	open symbols show cores which are located further south in Orion A.  Cores
	with lighter outlines have upper limits to their true sizes reported, implying
	that the estimated values of $\sigma_{grav}$ are lower limits.
	Cores lying below the dotted line are gravitationally unbound.
	Both this figure and Figure~\ref{fig_nopress1} 
	clearly indicate that considering only the balance between gravity
	versus temperature and local non-thermal motions implies that most of the 
	dense cores are gravitationally unbound.  
}
\label{fig_nopress2}
\end{figure}

In Figure~\ref{fig_nopress2}, we show $\sigma_{tot}$ versus $\sigma_{grav}$, the dispersion
needed to balance gravity alone, to make a similar comparison on a core-by-core
basis and account for variations in the $T_{kin}$ and $\sigma_{tot}$ measured for each
core.  Here too, the final result is similar: most of the cores do not 
have sufficient mass to remain bound.

\begin{figure}[htb]
\includegraphics[height=3in]{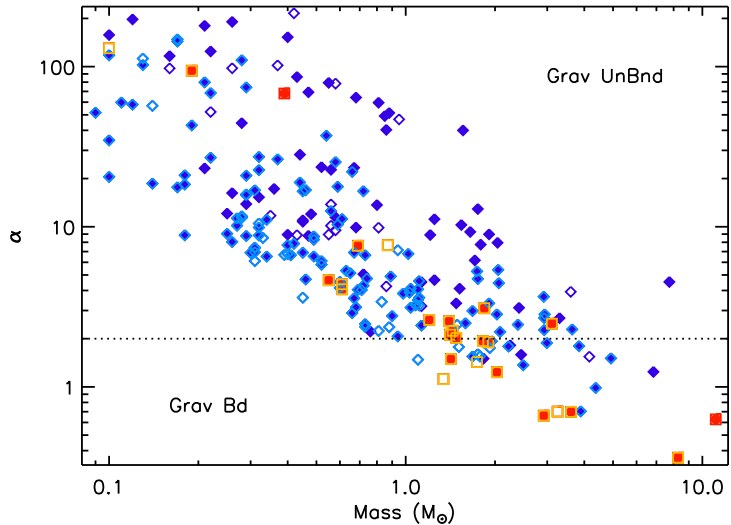}
\caption{A comparison of core masses and their estimated virial parameters.
	As in previous figures, blue diamonds denote starless cores while red squares denote
	protostellar cores.  Filled symbols show cores associated with the ISF, while
	open symbols show cores which are located further south in Orion A.  Cores
	with lighter outlines have upper limits to their true sizes reported, implying
	that the estimated values of $\alpha$ are also upper limits.
	Cores lying above the dotted line are gravitationally unbound.}
\label{fig_M_alpha}
\end{figure}

The cores that are gravitationally bound tend to have larger masses.
The virial ratio, defined as
\begin{equation}
\alpha = \frac{5 \sigma_{tot}^2 R}{G M}
\end{equation}
represents the degree to which a core is gravitationally bound, with bound cores
satisfying $\alpha < 2$ \citep[e.g.,][]{Bertoldi92}.  
Figure~\ref{fig_M_alpha} shows the virial parameter as a function of core mass, showing
that the most massive cores are the ones most likely to be gravitationally bound.
Similar behaviour has been seen in other studies of populations of dense cores,
such as the Pipe Nebula \citep{Lada08}.

Although the trend that we observe in virial ratios decreasing with increasing core
mass is consistent with previous studies, it is worth noting that most nearby dense core
populations such as Perseus tend to have overall lower virial ratios than what we measure 
\citep[e.g.,][]{Kirk07,Foster09}.  There are several reasons that dense cores in Orion could
appear to be less self-gravitating than in nearby clouds: 1) the larger distance to Orion
could lead to a greater confusion of multiple velocity components within the telescope's beam,
artificially increasing the measured core velocity dispersions; 2) the higher
mean density in Orion could imply that NH$_3$ traces kinematics beyond the core, implying that
the NH$_3$ velocity dispersion is larger than should be attributed to the core; or 
3) Orion does truly harbour more non-self-gravitating cores.

We can rule out the first possibility through
comparison with \citet{DiLi13}.  \citet{DiLi13} use a combination of VLA and GBT NH$_3$
observations to assess the virial state of dense cores in the OMC2 and OMC3 regions of
Orion~A.  
There, the angular resolution is 5\arcsec\ and the velocity resolution is 0.6~km~s$^{-1}$.
We compared the reported velocity dispersions for cores at nearly coincident positions in
\citet{DiLi13} and our own work, and found that our velocity dispersions tended to be 
slightly smaller.  The relatively poor velocity resolution in the \citet{DiLi13} observations
likely causes slight over-estimates in their velocity dispersion measurements.
The fact that we observe slightly lower core velocity dispersions, however, suggests that 
the spatial resolution of our GBT observations does not cause a significant increase to the 
velocity dispersions that we report.  Further comparisons with the \citet{DiLi13} results
are discussed in Section~\ref{sec_dili}.

The second possibility requires observations using
a higher effective density molecular tracer \citep[e.g.,][]{Shirley15}.
Ideally, such a tracer would be known to trace unambiguously much denser gas
than NH$_3$, such as a deuterated form of ammonia \citep[e.g.,][]{Crapsi07}.
The closest observations that we were able to find to satisfy this condition use N$_2$H$^+$.
N$_2$H$^+$ has a higher effective density than NH$_3$ \citep{Shirley15}, and 
NH$_3$ appears to be sensitive to gas at lower densities than N$_2$H$^+$.
For example, the lower-density filament B216 in Taurus is visible in NH$_3$ 
but not N$_2$H$^+$ \citep{Seo15}.  Due to a 
combination of chemical and optical depth effects, however, NH$_3$ {\it also} traces gas to
higher densities than N$_2$H$^+$ -- e.g., see the higher concentration of core emission
in NH$_3$ relative to N$_2$H$^+$ in \citet{Tafalla02} and the simulations of 
\citet{Gaches15}.  
While a full understanding of the chemistry behind the creation and destruction
for both N$_2$H$^+$ and NH$_3$ remain elusive \citep[see][for a recent test of models
of the latter]{Caselli17}, observations such as those described here are strong evidence
that NH$_3$ traces a wider range of densities than N$_2$H$^+$.
Therefore, while N$_2$H$^+$ is not strictly
a tracer of denser gas than NH$_3$, comparison between the two should allow us to test 
whether or not 
a significant amount of the NH$_3$ emission originates from the lowest densities 
of material to which it is sensitive.  We note that if the mean density in
Orion is sufficiently high that even N$_2$H$^+$ is sensitive to inter-core gas,
as has been found in some infrared dark clouds \citep[IRDCs; e.g.,][]{Henshaw13}, then
this comparison between NH$_3$ and N$_2$H$^+$ will not rule out the possibility
of contamination in the NH$_3$ spectra from non-core gas.

\citet{Tatematsu08} observed the N$_2$H$^+$ (1-0) emission line in OMC2 and OMC3
using the Nobeyama Radio Observatory, with a spatial resolution of 17\arcsec\ 
(with a spacing between observations of 41.1\arcsec) and a velocity resolution of 
0.12~km~s$^{-1}$.  We compared the N$_2$H$^+$ velocity dispersion reported by 
\citet{Tatematsu08} with the NH$_3$ velocity dispersion that we measure at the same
locations of a sample of their cores with a range of declinations and find very good
agreement, always within 0.02~km~s$^{-1}$.  This close correspondence in velocity
dispersions implies that the NH$_3$ and N$_2$H$^+$ are tracing similar zones of material.
Furthermore, a visual comparison of on-core and nearby off-core NH$_3$ spectra
suggest that the core emission is not being significantly biased to higher widths due
to non-core material.
While these tests suggest that the NH$_3$ emission is not dominated
by material at the lowest densities to which it is sensitive, observations 
of an unambiguously higher density tracer, such as emission from
deuterated NH$_3$, should be used to
verify this behaviour. 

With the data available, we argue that
it is reasonable to assume that the cores in Orion A 
represent a truly less self-gravitating population than is typically observed in nearby
molecular clouds.  The less self-gravitating cores certainly 
do not appear to be due to observational biases due to the greater distance to Orion~A.
Observations of a high density molecular tracer on small spatial scales are necessary, however,
to verify that the large line widths originate in gas associated with the dense cores.

\subsection{Pressure}

\subsubsection{Cloud Weight}
We next modify our analysis to include the fact that the ambient molecular cloud material
provides an additional confining pressure on the cores.
Under the assumption that the large-scale cloud can be roughly
approximated by a sphere, the column density at the position of each core then provides
an estimate for how deeply embedded the core is within the cloud.  Although the Orion~A
cloud is clearly not spherical, we argue that the approximation is acceptable for the
large-scale column density distribution, since, as we discussed in Section~2.5, we already
exclude small-scale column density features from consideration.
The pressure exerted on the ambient cloud is often expressed as
\begin{equation}
P_{cloud} = \pi G \bar{\Sigma} \Sigma
\end{equation}
where $P$ is the pressure, $\bar{\Sigma}$ is the mean cloud column density, and $\Sigma$ is the
column density at the location of the dense core \citep[see, for example,][]{Kirk06,McKee89}.
As we show in Appendix~C, this equation is strictly 
appropriate only for a spherical cloud with density varying as $\rho \propto r^{-k}$
with $k=1$. 
Other values of $k$ in the density scaling lead to a two-term expression for the pressure, 
given in Equation C2.
When the cloud's density distribution has an exponent of $1 < k < 3$, a reasonable range for
large-scale clouds, this second term is positive, so Equation 9 provides a lower limit 
to the full pressure.   Additional considerations, such as using $\Sigma / 2$ as a proxy
for how deeply embedded a dense core is within the cloud, may introduce some bias to the
cloud weight pressure estimate when $k \neq 1$.  See Appendix~C for further discussion of
the caveats associated with estimating the cloud weight pressure.

To calculate the mean cloud column density used in Equation~9,
we consider only the area within the large-scale column density map which was covered in
the GAS observations, yielding a value of $\bar{N} = 3.9\times10^{22}$~cm$^{-2}$
or $\bar{\Sigma} = 8.9\times10^{20}$~g~cm~$^{-2}$.

To compare the relative strengths of cloud pressure, core self-gravity, and thermal plus
non-thermal support, we use the formalism introduced in \citet{Pattle15}, expressing each
in terms of their energy density in the virial equation:
\begin{equation}
\Omega_P = -4 \pi P R^3 
\end{equation}
\begin{equation}
\Omega_G = \frac{-1}{2 \sqrt{\pi}} \frac{G M^2}{R} 
\end{equation}
\begin{equation}
\Omega_K = \frac{3}{2} M \sigma_{tot}^2
\end{equation}
where $\Omega_P$ is the pressure term, $\Omega_G$ is the gravitational term, and $\Omega_K$ 
is the
kinetic term.  The expression for $\Omega_G$ is derived in \citet{Pattle16a} for a core with
an infinitely extending Gaussian density distribution.  For a core with Plummer-like
density distribution \citep[e.g.,][]{Whitworth01}\footnote{A Plummer-like column density distribution has
$\Sigma \propto \big(\frac{R_{flat}}{\sqrt{r^2+R^2_{flat}}}\big)^{\eta}$ where $R_{flat}$ is the
radius of the central flat portion of the core and $\eta$ describes the outer power law
slope \citep[see][for more information]{Pattle16a}.} with an exponent of 4, the factor of 
$\sqrt{\pi}$ becomes $\pi$ instead,
i.e., implying slightly smaller $\Omega_G$ values than we estimate. 
For a constant density object, as is assumed when deriving the standard virial parameter
or Jeans mass discussed earlier, $\Omega_G$ is a factor of $\sim$ 3.5 larger.
A dense core is in virial equilibrium when 
\begin{math}
-(\Omega_G + \Omega_P) = 2 \Omega_K
\end{math}
.

\begin{figure}[htb]
\includegraphics[height=3in]{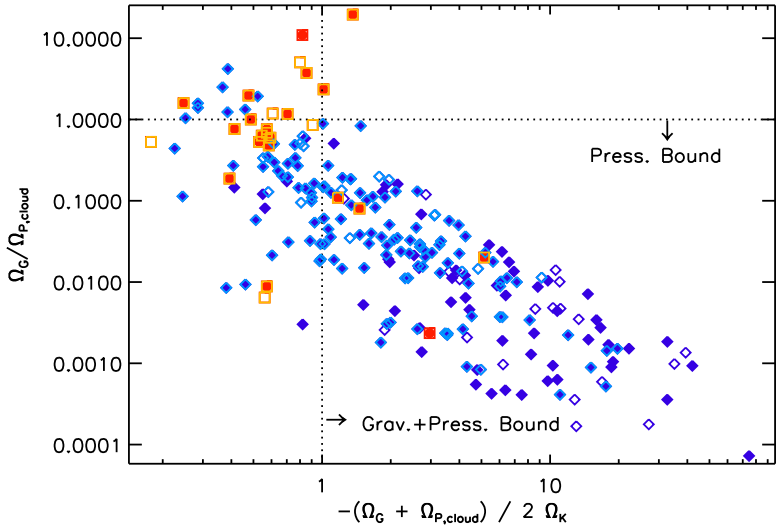}
\caption{A comparison of terms in the virial equation, following \citet{Pattle15}.  
	The vertical axis shows the ratio
	of self-gravity to pressure, i.e., the confinement ratio, showing 
	that in most cores, pressure plays 
	a more significant role in binding the cores than gravity does.  The horizontal axis
	shows the ratio of energy densities with cloud pressure included.  
	Inclusion of cloud pressure
	causes most of the dense cores to be bound, with the majority of that binding
	attributable to the weight of the overlying cloud material.
	See Figure~\ref{fig_nopress2} for the plotting conventions used.
	}
\label{fig_press}
\end{figure}

In Figure~\ref{fig_press}, we show the energy density ratio 
for the dense cores, compared to the
ratio of the gravitational and pressure energy densities.  The latter
ratio expresses whether gravity or pressure is the dominant element binding the core,
and is termed the `confinement ratio' in \citet{Pattle16a}.
As can be seen in Figure~\ref{fig_press},
the majority of dense cores are bound, with energy density ratios 
exceeding one.  As already
illustrated in Figure~\ref{fig_nopress2}, self-gravity alone contributes relatively little
to this binding; pressure dominates gravity for the vast majority of the cores.
A trend of decreasing confinement ratio with increasing energy density ratio 
is seen in Figure~\ref{fig_nopress2}.  We investigated the cause of this trend, and
found no correlation between $\Omega_{P,cloud}$ and either $\Omega_K$ or $\Omega_G$. 
$\Omega_G$, however, varies approximately as $\Omega_K^2$.  This latter trend appears to be driven
by the mass term in $\Omega_G$ and $\Omega_K$, which follows the same relative scaling.  
As Figure~\ref{fig_larson}
has already shown, there is no relationship between $R_{eff}$ and $\sigma_{tot}$, and both
have a relatively small range in values (approximately an order of magnitude).  The range of
masses is just over three orders of magnitude, and thus appears to be driving the 
correlation between
$\Omega_K$ and $\Omega_G$, which in turn is responsible for the trend seen in 
Figure~\ref{fig_nopress2}.

\subsubsection{Filament Pressure}
As discussed in Section~2.5, cores are typically found embedded within filaments in the
larger molecular cloud structure.  Like the larger molecular cloud, these filaments will also
help to confine the dense cores.  Since it is difficult to disentangle the column density
belonging to the filaments and the dense cores with simple techniques due to their similar 
size scales, we instead use previous observations to estimate a lower limit to the filament
confining pressure.  \citet{Tafalla15} observed filaments and cores in the Taurus L1495/B213
region using a combination of dust continuum and molecular emission line observations.
They used these observations to model the density profile of the filaments in locations where
cores were not present, using a cross sectional profile of
\begin{equation}
n(r) = \frac{n_0}{1 + (r/r_0)^\alpha}
\end{equation}
where $n$ is the density at radial separation $r$, $n_0$ is the central density, $r_0$ is the 
characteristic size, and $\alpha$ is the power law dependence \citep{Whitworth01}.  
Across eight different
filaments in Taurus, \citet{Tafalla15} find $n_0 \sim 6.5\times10^4$~cm$^{-3}$, 
$r_0 \sim 45$~\arcsec\ (at an assumed distance of 140~pc), and $\alpha \sim 3$, while
a characteristic temperature of 10~K was assumed.

\begin{figure}[htb]
\includegraphics[height=3in]{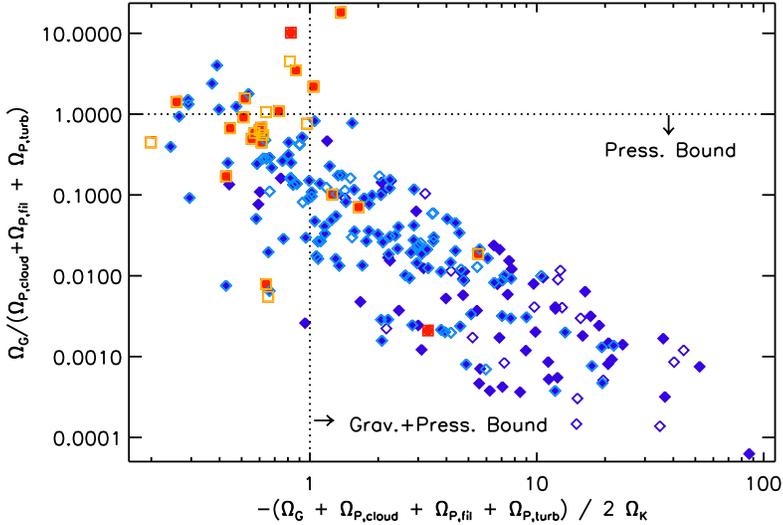}
\caption{A comparison of terms in the virial equation, similar to that shown in 
	Figure~\ref{fig_press}.  Here, $\Omega_P$ includes the 
	cloud weight pressure, 
	filament weight pressure, and turbulent pressure.  
	See Figure~\ref{fig_nopress2} for the plotting conventions used.
	}
\label{fig_incl_extra_press}
\end{figure}

The pressure induced by the weight of overlying material in an isothermal filament
can be expressed as 
\begin{equation}
P_{fil} = \frac{2}{\pi} G \Sigma_{\perp}^2
\end{equation}
(see Appendix~C)
where $\Sigma_{\perp}$ is the column density through the centre of the filament.  
The binding pressure
on cores from the filaments measured in Taurus can therefore be estimated using the 
\citet{Tafalla15} model estimates.  Filaments in Orion, however, 
are likely to be much more dense than
those found in Taurus, and therefore the Taurus estimate provides us with a lower limit to the
true filament pressure.  Generally, the pressure exerted by the cloud on cores is about 
a factor of 10 higher than the pressure exerted by these model filaments.
Figure~\ref{fig_incl_extra_press} shows the resulting energy density ratios with this additional
pressure included (see also the following section for a further pressure term 
added).

\subsubsection{Turbulent Pressure}
\label{sec_turb_press}
In addition to the weight of overlying material, turbulent pressure,
i.e., pressure induced from the higher velocity dispersion, lower column density material
surrounding the cores, may also be present.  We estimate the magnitude of this pressure
in an approximate way.  \citet{Shimajiri14} obtained $^{13}$CO~(1--0) and C$^{18}$O~(1--0)
observations of the northern portion of Orion A, extending slightly southward of the
main ISF.  According to their Figure~2, the C$^{18}$O~(1--0)
emission has a velocity dispersion $< 1.5$~km~s$^{-1}$ everywhere, with most of the
gas having velocity dispersions $< 0.75$~km~s$^{-1}$.  They assume a mean density
of material traced by C$^{18}$O~(1--0) emission of $5\times10^3$~cm$^{-3}$, based
on previous analysis by \citet{Ikeda09}.  Using these estimates, the turbulent
pressure on the dense cores can be written as
\begin{equation}
P_{turb} = \rho_{turb}~\sigma_{v,turb~gas}^2
\end{equation}
\citep[e.g.,][]{Pattle15}.
A more careful measurement of both the mean gas density and typical CO velocity dispersion
around each dense core (or even the velocity dispersion of the surrounding fainter
NH$_3$ emission) would be required to obtain a more precise estimate of the 
turbulent confining pressure.  
Nonetheless, using the present equation to approximate the influence of turbulent
pressure in combination with the filament weight pressure discussed above, 
Figure~\ref{fig_incl_extra_press} shows that all cores have slightly increased energy 
density ratios, with an additional six dense cores now appearing to be bound.
We emphasize that more precise, core-specific, estimates for both the turbulent pressure
and the filament weight pressure could yield additional binding, especially as the
latter pressure estimate is a lower limit.

\subsubsection{Bonnor-Ebert Sphere Pressure}
Cores are often approximated as
Bonnor-Ebert spheres \citep{Ebert55,Bonnor56}, a spherically symmetric, isothermal 
equilibrium model where self-gravity and an external binding pressure balance internal 
thermal motions.  In this model, the maximum external binding pressure for a stable
BE sphere model can be expressed as
\begin{equation}
P_{BE,crit} = 1.40 \frac{c_s^8}{G^3 M^2}
\end{equation}
\citep{Hartmann98}.
While we do not find any correlation between the critical binding pressure of a 
BE sphere and
the estimated
pressure on each core from the weight of the overlying cloud material, the values span
a similar range.  The mean and standard deviation of $log (P_{BE,crit}/k_B)$ is
$6.5 \pm 0.8$~log(K cm$^{-3}$).  This rough similarity in pressures between the
BE sphere model and those estimated from the weight of the cloud suggest that the BE
sphere model may provide a reasonable representation 
of the cores, although this conjecture should be tested
more carefully through measurements of the radial column density profile of the cores.

\subsection{Concentration}
\begin{figure}[htb]
\includegraphics[height=3in]{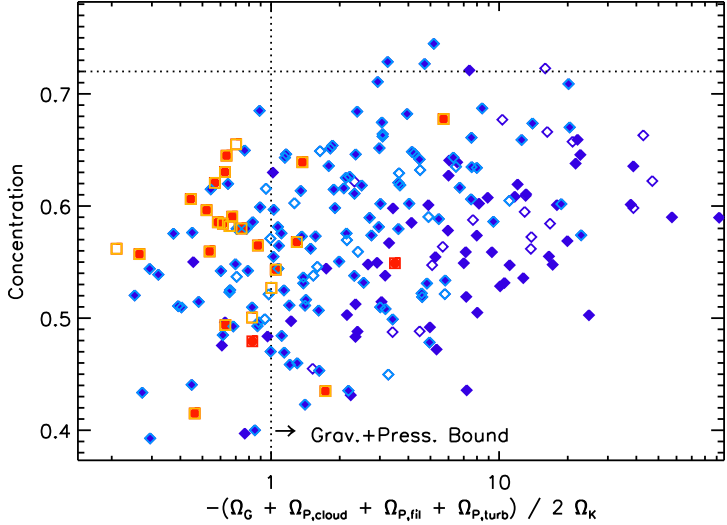}
\caption{A comparison of the central concentration of the dense cores with their
	energy density ratio as shown in Figure~\ref{fig_press}.  
	The horizontal dotted line
	marks the maximum stable concentration for the BE sphere model, while the
	vertical dotted line denotes the boundary between bound (right) and
	unbound (left) systems.  Note that very few of the dense cores have 
	concentrations exceeding that expected for a stable Bonnor Ebert sphere model.
	See Figure~\ref{fig_nopress1} for the plotting conventions used.
	}
\label{fig_concs}
\end{figure}

When only dust continuum information is available for cores, often their stability is
assessed in terms of their concentration, or peakiness.  
Following \citet{Johnstone01}, the concentration, $C$, can be written as
\begin{equation}
C = 1 - \frac{1.13 B^2 S_{tot}}{\pi R^2 F_{pk}}
\end{equation}
where $B$ is the telescope beamsize, $S_{tot}$ is the total flux
observed, and $F_{pk}$ is the peak flux density observed.
Typically, $B$ is expressed in arcsec, $S_{tot}$ in Jy, $R$ in arcsec, and $F_{pk}$ in
Jy~beam$^{-1}$.  
In the BE sphere model, when observed in 2D,
the minimum concentration is 0.33, representing 
a uniform density sphere, while the maximum concentration for a stable configuration is
0.72 \citep{Johnstone00b}.  In Figure~\ref{fig_concs}, we show the concentration 
measured for each dense core compared to its respective energy density ratio.  
Higher concentration
cores have more mass within a smaller size, and therefore would be expected to have 
stronger self-gravity, and hence a higher energy density ratio. 
As plotted in Figure~\ref{fig_concs},
there is an absence of low-concentration strongly bound cores and high-concentration
strongly unbound cores, both of which represent states that would be difficult to populate
or maintain.  There is no further correlation visible between the two quantities.

Some of the protostellar cores have low concentration values.  Previous studies,
such as \citet{vanKempen09}, have
found low concentration protostellar cores when the core evolution is more advanced, i.e.,
much of the core material has already been accreted onto the central protostar
or dispersed by protostellar outflows, making the
remaining envelope traced by the submillimetre continuum observations 
appear more diffuse.  Angular resolution and the
core identification algorithm used also play a role in the concentrations observed,
since both of these influence the peak and total flux associated with each core.

\subsection{Regional Variation}
We additionally checked whether or not
there are any bulk differences in the virial properties of dense cores associated with
the ISF and the remainder of the field.  \citet{Lane16} identified clusters of dense cores
across Orion~A using minimal spanning trees.  One of these clusters roughly encompasses
the ISF, and therefore provides a simple way to identify dense cores either 
associated or
not associated with the ISF.  In all of the previous figures showing dense core properties,
filled symbols denote cores associated with the ISF, while open symbols show cores not
associated with the ISF.  
Dense cores associated with the ISF tend to have slightly smaller energy density ratios 
than the
remaining cores, due to their typically smaller sizes and therefore much 
smaller $\Omega_P$ values but the two populations largely overlap.  The starless cores in the ISF
have typical energy density ratios 
of $log (-\Omega_G - \Omega_{P,tot})/2\Omega_K = 0.39 \pm 0.42$, 
while the starless cores outside of the ISF
have typical values of $0.62 \pm 0.42$ (median and median absolute deviation quoted 
for both).
The non-ISF population that we analyze is presently small (39 starless cores), but the
full GAS map of Orion~A will extend much further south than the DR1 map analyzed here,
and will allow for a more stringent test of whether or not virial properties vary between
the ISF and the rest of Orion~A.

\subsection{Comparison to \citet{DiLi13} }
\label{sec_dili}
As noted in Section~\ref{sec_nh_grav}, \citet{DiLi13} observed NH$_3$ in cores in
the Orion~A OMC2 and OMC3 regions and assessed their virial nature.  While \citet{DiLi13}
also find that many of their dense cores are not self-gravitating ($\alpha > 2$),
they do not report virial ratios as high as those 
found in our analysis.  The relative lack of high virial
ratios in \citet{DiLi13} can be attributed to our greater sensitivity to lower mass
cores.  For example, 
almost none of the \citet{DiLi13} cores have masses below 1~\Msol, while
more than half of our cores (65\%) have masses below 1~\Msol.  As discussed
earlier, higher mass cores tend to be more self-gravitating, so it is reasonable that the
lower mass cores in our sample 
have larger virial ratios than those reported in \citet{DiLi13}.

In their full virial analysis, \citet{DiLi13} include a turbulent pressure term estimated
assuming a mean gas density of $10^4$~cm$^{-3}$ and a mean velocity dispersion of 
1~km~s$^{-1}$.  This implies a slightly larger turbulent pressure than the one we estimate
in Section~\ref{sec_turb_press}, which we noted is typically smaller than the 
pressure binding provided by the overlying cloud material that we estimated for each core 
individually.
\citet{DiLi13} also include a magnetic support term in their analysis, assuming 
a typical field strength of 0.1~mG based on nearby observations of the Zeeman effect
in CN.  Including both the turbulent pressure binding and magnetic support reduces the
number of super-virial cores in their analysis, although nearly half (47\%) of their
cores still appear to be unbound, with a `critical mass ratio' above one.  
The overall
conclusions of \citet{DiLi13} therefore seem consistent with our own results, given
the different core samples and different assumptions used for the virial analysis.

New results from the BISTRO survey \citep{WardThomp17} suggest that the total magnetic 
field strength may be substantially larger than assumed in the \citet{DiLi13} analysis.
Using the Chandrasekhar-Fermi method \citep{ChandFermi53}, \citet{Pattle17b} estimate a 
magnetic field strength in the OMC-1 region, i.e., the central densest portion of the ISF) 
to be 6.6$\pm$4.4~mG in the plane of the sky.  At this level, the magnetic field would
contribute significantly to the energy density of each core.
Inclusion of the magnetic pressure would serve to increase the amount of internal
support against the gravitational collapse of the cores, thus in 
Figure~\ref{fig_incl_extra_press}, points would move up and to the left.
The magnetic pressure may not be well-approximated by a 
constant field strength value over the large area of Orion~A included in our analysis, as
the field strength derived only for the densest part of the cloud may not be representative
of the entire region.  Therefore, while we exclude the magnetic support term from our
virial analysis, we note that this should be re-visited when magnetic field strength
estimates are available for a larger extent of Orion A.

\section{Discussion}
\label{sec_disc}
The potential for dense cores to be pressure-bound due to the weight of the overlying
molecular cloud material has been considered for several decades 
\citep[e.g.,][]{Elmegreen89,Bertoldi92}, and some early observations supported this
hypothesis \citep[e.g.,][]{Keto86}.
Other types of external pressure have also been considered, including inter-core
ram pressure \citep{Miettinen10}, ram pressure generated by core accretion
\citep{NaranjoRomero15,Seo15}, radiation pressure \citep{Seo16}, 
and turbulent shocks \citep{Gong15}.
A related issue is the role of
tidal forces within molecular clouds.  As discussed in \citet{BP09a}, tidal disruption
should be considered for a full virial analysis of molecular clouds.  On the 
scale of cores, however, tidal compression is more likely than tidal disruption, and is
generally expected to contribute less than self-gravity for smaller cores.  We therefore
expect tidal effects to increase slightly the total number of bound cores
in our sample.
Beyond the role of external pressure, whether dense cores are in stable equilibrium,
unstable equilibrium \citep[e.g.,][]{Field11}, or not in equilbrium at all, as
suggested by some numerical simulations \citep[e.g.,][]{BP03,NaranjoRomero15},
is an additional issue to consider.  Our analysis makes the simplest and usual
assumption that the dense cores evolve sufficiently slowly that the virial
equation parameters measured provide a reasonable approximation of the full virial
state of the cores.  

On the kinematic side, we made the simple assumption that the entire velocity
dispersion of the core acts to provide internal thermal plus non-thermal pressure
support.  While this assumption is reasonable for the thermal pressure, the non-thermal
velocity dispersion almost certainly does not represent only isotropic small-scale motions
that provide additional pressure support.  For example, some of the non-thermal velocity 
dispersion could be caused by infall motions within the core, which can lead to broader
emission lines in dense gas tracers \citep[e.g.,][]{Myers05,Bailey15}.  For the protostellar
cores, where we know that infall motions must be present, in addition to some of the
starless cores where infall has begun but has not yet produced a detectable central
protostar, the observed line widths will be larger due to infall motion
\citep[e.g.,][]{Caselli02a}.  Our present analysis therefore over-estimates the 
amount of internal pressure support,
and hence under-estimates the level of binding of these cores.  Mapping or pointed
observations of a reliable infall tracer such as HCN \citep[e.g.,][]{Sohn07} could
be used to help remove this bias in our analysis.

A final subtle consideration is the degree to which the
column density and velocity measurements we use correspond to single discrete
cores in three-dimensional space.  \citet{Beaumont13} ran a careful comparison
of numerical simulations of a molecular cloud as viewed in true 3-D (position-position-position,
or PPP) space versus the view inferred from synthetic $^{13}$CO (1-0) observations of the
position-position-velocity (or PPV) structures.  They found that the lack of perfect 
correlation between PPP and PPV structures led to scatter in the virial parameters
by a factor of about two.  Since NH$_3$ traces denser gas than $^{13}$CO, the
correspondence between structures is likely moderately 
better, as line-of-sight confusion is
expected to play a smaller role in dense gas, which has a lower volume filling fraction.
Nevertheless, the high average density of Orion~A allows for the
possibility that some of the apparent cores in our SCUBA-2 based catalogue 
could be chance alignments of
somewhat dense gas along the line of sight rather than representing a true core
or include some contamination from the material surrounding a true core.
Since our analysis requires not only NH$_3$~(1,1) but also NH$_3$~(2,2) emission for cores to
be reasonably well detected (at S/N $> 3$), 
we expect that there should be a minimal number of unreal cores in
our analysis, but it is possible that a few of our cores with the faintest and broadest
NH$_3$ spectra fall into this category.

\subsection{Previous Studies With Pressure}

Our conclusion that Orion~A dense cores are significantly bounded by the pressure of the
surrounding cloud material is a result that has been found in other nearby molecular
clouds as well, spanning a range of environments.  
One example is the nearby Pipe Nebula, a star-forming environment signficantly different
from Orion~A.  Dense 
cores in the Pipe have typical volume densities of less than $10^4$~cm$^{-3}$,
sizes of $\sim$0.2~pc, non-thermal velocity dispersions of $\sim$0.2~km~s$^{-1}$, and 
inhabit a cloud with a mean extinction of A$_V \sim$ 4~mag and a total mass of 
$\sim 10^4$~\Msol\ that is
located a mere 130~pc away \citep[][and references therein]{Alves07,Lada08}.  
\citet{Lada08} used pointed NH$_3$ or C$^{18}$O observations, in combination with
dense core sizes and masses estimated using extinction mapping, to demonstrate that
most Pipe dense cores have insufficient mass to be gravitationally bound.
Furthermore, the vast majority of the internal pressure support 
in Pipe cores was found to be from thermal
pressure rather than non-thermal motions.  They also found a surprising similarity
in the total internal pressures of Pipe cores across the cloud and argued 
that the most
likely source of pressure on Pipe core boundaries is that caused by the weight of the
overlying cloud material.  Indeed, typical internal core pressures tend to be slightly larger
in the `bowl' of the Pipe, compared to the `stem'; the former area is also associated
with higher mean cloud column densities.
Magnetic pressure may also play an important role in confining cores in the
Pipe, as argued by \citet{FAlves08}.
In contrast to the Pipe, Orion~A has a mean cloud column density contributing to the
cloud external pressure that is much higher (a factor of $> 5$),
and non-thermal motions and perhaps additional sub-core bulk motions 
provide significant contributions to the internal pressures
of its dense cores.

Several other studies examining the role of cloud pressure have been performed in 
star-forming regions which have properties intermediate to the Pipe and Orion~A
(i.e., cloud mass, dense core volume density, size, and non-thermal velocity dispersion).
In the Ophiuchus molecular cloud, three independent analyses using different datasets all came
to the conclusion that external cloud pressure is important in confining many 
cores in the cloud.  Specifically,
\citet{Johnstone00b} argued this point 
based on Bonnor-Ebert sphere modelling of dust continuum
observations at 850~$\mu$m and the level of expected pressure from ambient cloud material, 
\citet{Maruta10} used Nobeyama H$^{13}$CO$^+$ observations, while \citet{Pattle15} used a 
combination of SCUBA-2 850~$\mu$m dust continuum data and IRAM 30~m N$_2$H$^+$ and C$^{18}$O
observations.  In the Perseus molecular cloud, \citet{Kirk07} analyzed the dense cores,
while \citet{Sadavoy15} analyzed the larger B1-E structure, both also finding evidence
that pressure confinment is important.  Note, however, that \citet{Foster09}
also analyzed Perseus dense cores and found instead that self-gravity alone was sufficient
to confine the dense cores.  This difference is primarily
due to the larger masses derived for the dense cores by \citet{Foster09}
using CSO Bolocam data, rather than higher resolution JCMT SCUBA data of \citet{Kirk07}.
\citet{Foster09} furthermore did not estimate the magnitude of cloud pressure to see how it 
compares with core self-gravity, and it is likely that such an estimate would
provide a larger amount of binding than they estimated from self-gravity alone.
In the Taurus L1495 / B218 region, \citet{Seo15} analyzed the virial properties of
filaments and smaller structures using NH$_3$ observations, and found that most
of the small NH$_3$ structures are gravitationally unbound when considering just
the balance between self-gravity, and thermal plus non-thermal support.  They
argued that many of these structures, especially the youngest ones,
appear to be pressure confined, and that cloud pressure is a reasonable candidate
for the source of that pressure with ram pressure from material inflowing to the system
being their other main candidate.
In the Cepheus clouds, \citet{Pattle16b} performed a virial analysis using a 
combination of JCMT SCUBA-2 850~$\mu$m dust continuum and archival $^{13}$CO
observations to argue that cloud pressure is required to bind most of the dense cores.
Similarly, \citet{Kirk16a} find suggestive evidence that cloud pressure is important
for dense cores in Orion~B, although there, the analysis lacks kinematic information
to provide precise estimates of the dense cores' temperatures and non-thermal 
motions\footnote{GAS will provide NH$_3$ observations in a future data release,
however.}.

For more distant (non-Gould Belt) star-forming regions, \citet{Kauffmann13} performed
a self-consistent virial analysis combining observations from a variety of previous surveys.
While external pressure was not included in their analysis, a comparison of self-gravity
and non-thermal motions showed that the majority of structures were unbound.  Structures
that were bound were either much larger entities (masses above 100~\Msol), or were in
regions of high-mass star formation, or both. \citet{Kauffmann13} speculate that
the structures, which do appear to be gravitationally unbound, may in fact be pressure bound.
Additionally, \citet{Barnes16} analyze a large sample of massive molecular clumps
observed with the CHaMP survey in HCO$^+$ and $^{12}$CO emission to demonstrate that the
less dense gas traced by $^{12}$CO appears to provide significant pressure binding to
the more compact clumps traced by HCO$^+$, at a sufficient level that most of the clumps
are near virial equilibrium.
Thus, while it may be the case that high mass stars form under a different mode,
dense cores which form low- to intermediate- mass stars appear to be well-represented
by a large virial parameter, requiring the presence of some additional factor
such as external pressure to remain bound.

Some numerical simulations have also illustrated the importance of cloud pressure
on the formation and evolution of dense cores.  \citet{Lucas17} show that for
otherwise identical simulations of star-forming regions, the absence of a cloud
envelope either at the beginning of formation, or the removal of such an envelope
during evolution, leads to less dense substructure forming and to lower markedly 
star formation efficiencies.

Despite the differences in the analyses described above, 
it is striking that all of these studies come to a similar 
conclusion: pressure from the ambient cloud material on dense cores is important
and often is required to keep the cores bound.
The fact that this conclusion 
holds across such different star-forming environments from analyses 
using different techniques suggests that this phenomenon is a 
universal feature of star formation.

The one major component missing from most virial analyses of core populations
is the support provided by magnetic fields.  The magnetic field strength remains challenging
to observe directly, although estimates can be made based on the field alignment, as
measured by polarization data.  
Polarization observations are beginning to be available over larger areas of molecular
clouds at lower resolution through telescopes such as Planck \citep[e.g.,][]{Planck15}
and Blast-Pol \citep[e.g.,][]{Fissel16}, and at higher resolution with the JCMT's POL-2 
\citep[through the BISTRO survey;][]{WardThomp17,Pattle17b}.

\section{Conclusion}
\label{sec_conc}
We perform a simple virial analysis of the dense cores in Orion~A and find that most  
of them are bound.  We likely underestimate the true number that are bound, 
since our estimate of the binding pressure from the weight of filaments surrounding the
dense cores is a lower limit to the true value.  
The majority of the dense core binding comes in the form of pressure, mostly 
from the weight of the ambient cloud material rather than self-gravity or
other forms of binding pressure.
Qualitatively, the picture that
the pressure from surrounding
molecular cloud material is important, or essential, to keeping dense
cores bound appears common to a variety of local molecular cloud environments.
Thanks to the efforts of a variety of recent surveys of nearby molecular clouds, 
there are high-quality, large-area maps available to allow
for a uniform assessment of dense core sizes and masses 
\citep[e.g., the JCMT and {\it Herschel} Gould Belt Surveys;][]{WardThomp07,Andre10}, 
dense core protostellar content 
\citep[e.g., the {\it Spitzer} c2d and Gould Belt Surveys and the {\it Herschel} Gould Belt 
Survey][]{Evans09,Harvey08,Andre10}, and ambient cloud material
\citep[e.g., using variants on the Near Infrared Colour Excess, or NICE technique;][]{Lombardi01}.
The Green Bank Ammonia Survey \citep[GAS;][]{Friesen17} 
provides the necessary components to perform
the type of virial analysis presented here, measuring both the temperatures and 
internal velocities
of dense gas associated with the cores.  The results of a similar virial analysis performed
across all of the molecular clouds mapped by GAS should provide an answer to the question
of whether or not
the pressure of molecular cloud material is indeed an essential component to
(local) star formation.

\appendix

\section{Core Catalogues}
\label{app_cores}
Defining dense cores is generally challenging \citep[e.g.,][]{Pineda09}
and is especially difficult in Orion~A due to the high level of larger-scale
structure present along the ISF.  To test whether or not 
our conclusions are robust against uncertainties
in the dense core definitions, 
we re-ran our analysis using two independently-derived core catalogues:
the {\it FellWalker}-based catalogue from \citet{Lane16} and a Hessian-based core identification
algorithm presented and adopted in \citet{Salji15a}.  Both of these catalogues are based 
on the JCMT GBS SCUBA-2 850~$\mu$m data observations of Orion~A, similar to the 
\citet{Lane16} {\it getsources}-based catalogue used for our main analysis.
Below, we describe each of these catalogues in detail and present a virial analysis
based on each.  We note that we do not include a similar analysis for the 
structure catalogues presented in \citet{Mairs16}, as that analysis does not
extend far enough north in Orion~A to match the GAS coverage. 

\subsection{\citeauthor{Lane16} {\it FellWalker} Catalogue}
As discussed in Section~2.2, \citet{Lane16} presented two dense core catalogues for
Orion~A, one using {\it getsources}, that we adopted for our main analysis,
and also one using {\it FellWalker}, that we examine here.
{\it FellWalker} identifies peaks based on local gradients, effectively associating
any pixel lying along an upward path to a peak to that peak.  There is no
assumed shape for each core, and each pixel can be assigned to a maximum of
one peak.  When calculating the \nh-based properties of each core, analagously
to Section~2.4, we use the full core footprint.  Otherwise, we follow an
identical procedure to that presented for the {\it getsources}-based dense
core catalogue.

In Figure~\ref{fig_press_othercats}, we show the results from the virial analysis using
instead the {\it FellWalker} dense core catalogue.  As in our main analysis, we see that 
nearly all of the dense cores are bound (energy density ratios greater than one).  
Only seven of the cores 
have sufficient self-gravity to be entirely bound without
the pressure of the cloud (i.e., $-\Omega_G / \Omega_K \ge 2$), so our conclusion that
pressure binding is required for the majority of dense cores holds.

\subsection{\citeauthor{Salji15a} Catalogue}
\citet{Salji15a} analyzed the region around the ISF in Orion~A
using an early JCMT GBS data release that did not reach the final survey depth and
an early reduction method which is less sensitive to larger-scale structure in the
cloud (`Internal Release 1').  
To identify dense cores, \citet{Salji15a} developed a new structure detection
algorithm, tuned to identify only compact and roughly round sources.  Sources identified
with protostars were also removed from their final published catalogue.  The remaining
cores were classified as either starless or prestellar based on a thermal stability
analysis using only SCUBA-2 data (and no kinematic information).  To derive
the \nh -based properties for each core, we follow a similar procedure to that
in the main analysis: we consider all pixels within the 1-$\sigma$ contour of the
peak core position.  Since \citet{Salji15a} did not publish dense core axial 
ratios or a rotation angle, we assume the cores are well-described by a single
radius.

\begin{figure}[htb]
\plottwo{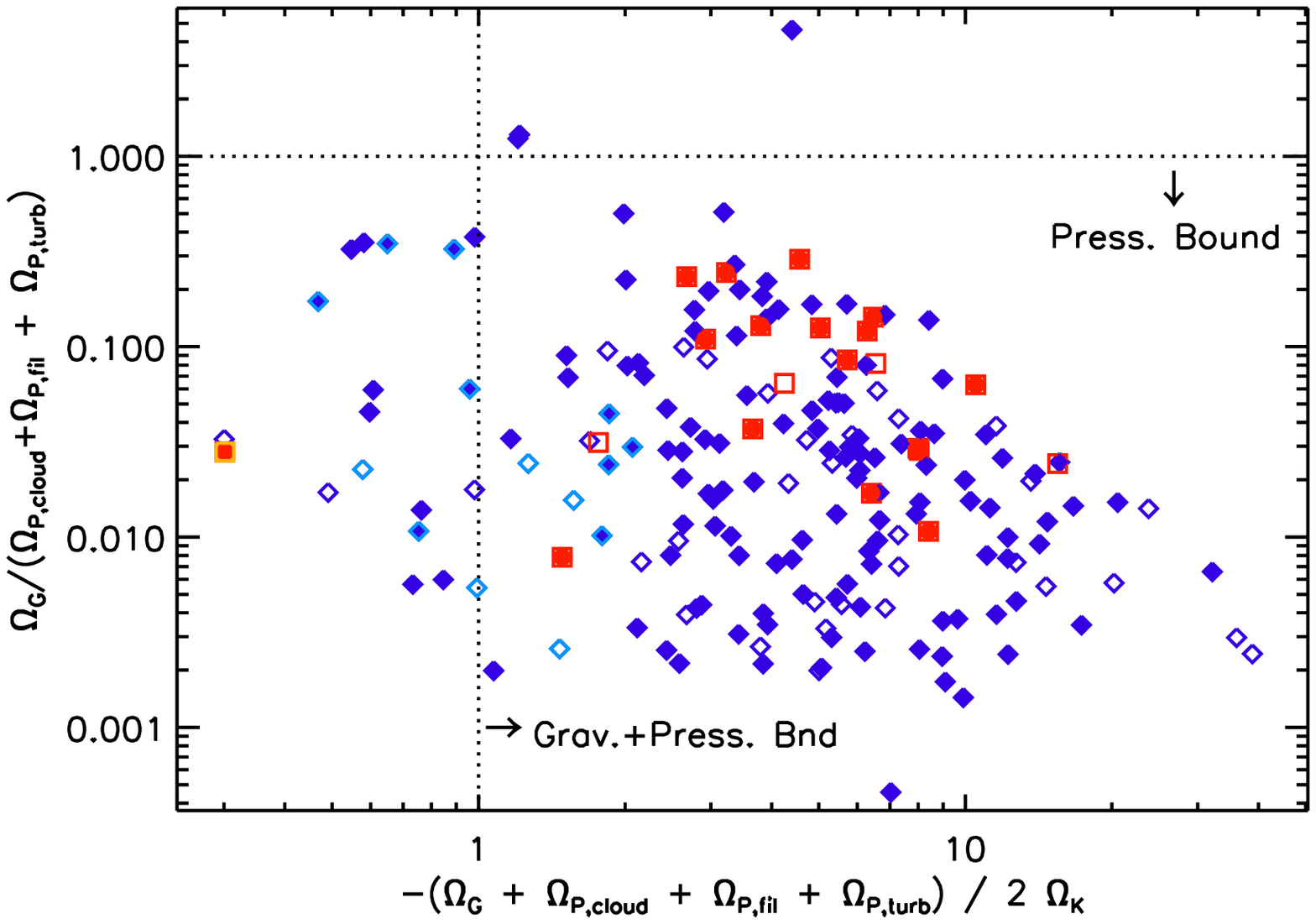}{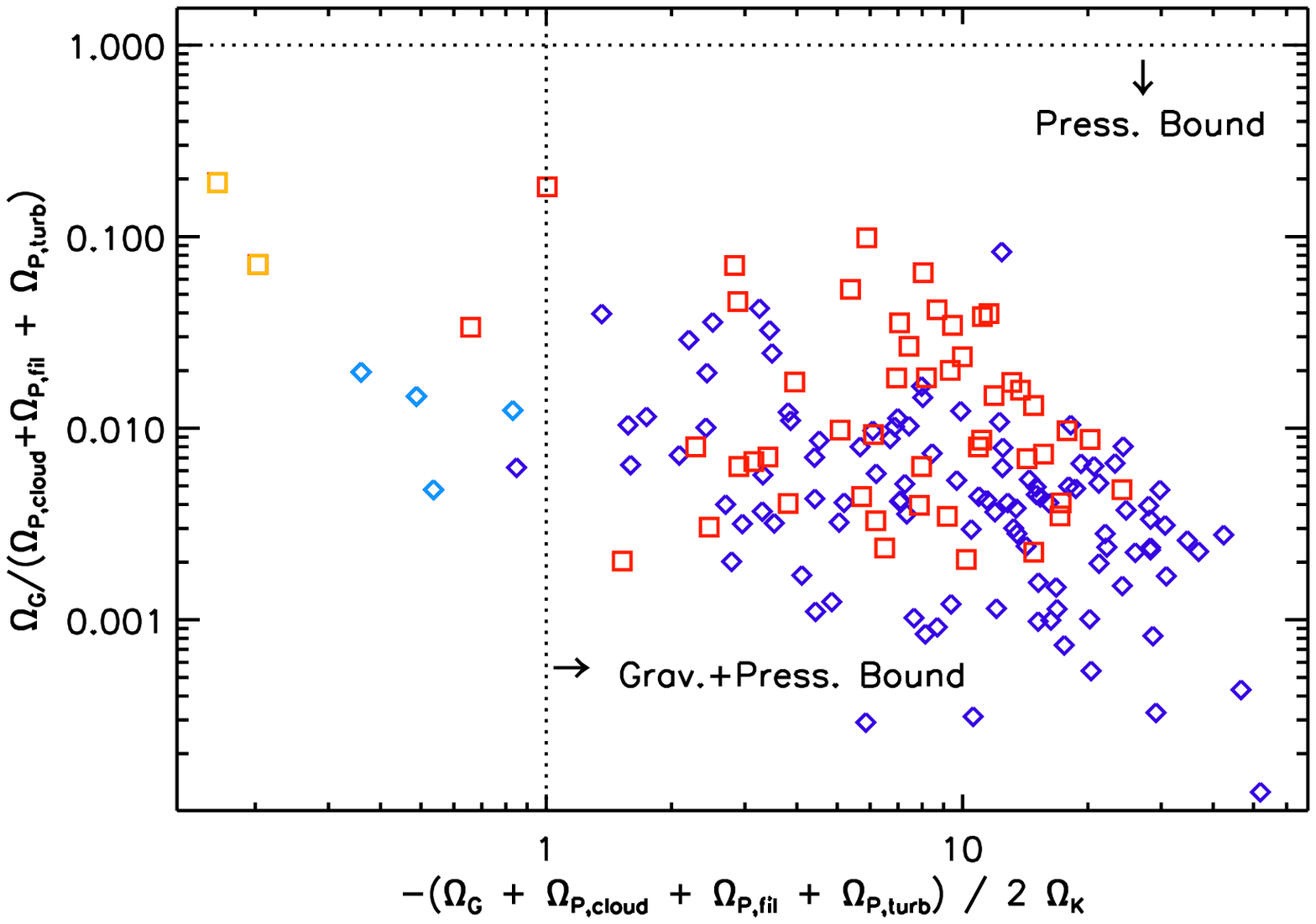}
\caption{A comparison of the energy density ratio and confinement ratio for dense
	cores identified using {\it FellWalker} by \citet{Lane16} (left panel) and
	by \citet{Salji15a} using a self-developed method (right panel).  
	See Figure~\ref{fig_nopress2} for the plotting conventions used.
	Note that the right panel follows a different colour convention than
	previous figures: blue points show starless cores, while red points show
	prestellar cores.  {\it Proto}stellar cores were eliminated from the
	\citet{Salji15a} catalogue and not published.
	}
\label{fig_press_othercats}
\end{figure}

The right panel of Figure~\ref{fig_press_othercats} shows the virial analysis for the
\citet{Salji15a} dense cores.  Here, the cores appear to follow a
much more linear trend, suggesting that higher self-gravity and higher cloud pressure
values tend to go together.  The distinction that \citet{Salji15a} made between
starless and prestellar cores does
appear to be largely borne out in our virial analysis, with the prestellar cores
tending to lie higher (more bound by both self-gravity and cloud pressure) than
the starless cores.  Although the qualitative appearance of the plot differs 
dramatically from our main analysis, we find that our main conclusion still
holds: the majority of dense cores are bound (94\%), and none
would appear to be bound if the pressure from the cloud were ignored.

\section{Cloud Pressure Estimate}
\label{app_press}

As discussed in Section~2.5, one complication in estimating the external cloud pressure is 
determining whether gas on different scales is associated with the core or the cloud.
Attributing {\it all} of the material
observed to cloud-wide material would lead us to over-estimate the level of pressure
exerted on the cores.  In our main analysis, we filtered the \citet{Lombardi14}
column density map to trace only structures on $\sim$0.5~pc scales and larger.
Here, we examine the impact of that choice of scale on our final results.  
We note that our test of the choice of scale also indirectly tests the choice
in method for measuring the large-scale column density.  The variations between
methods at a single spatial scale is likely to be much smaller than the variation
induced by significant changes to the scale adopted.

\begin{figure}[htb]
\plottwo{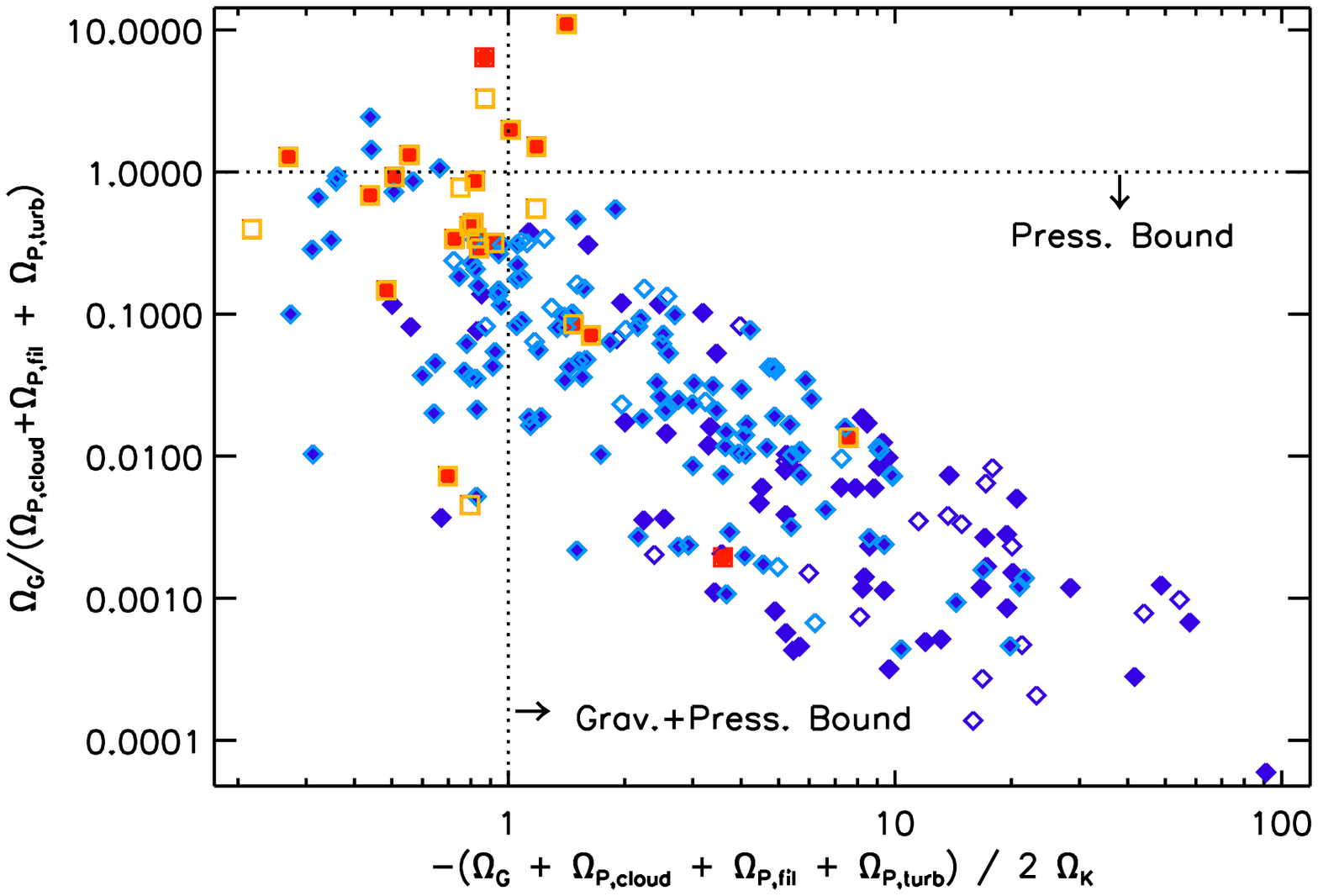}{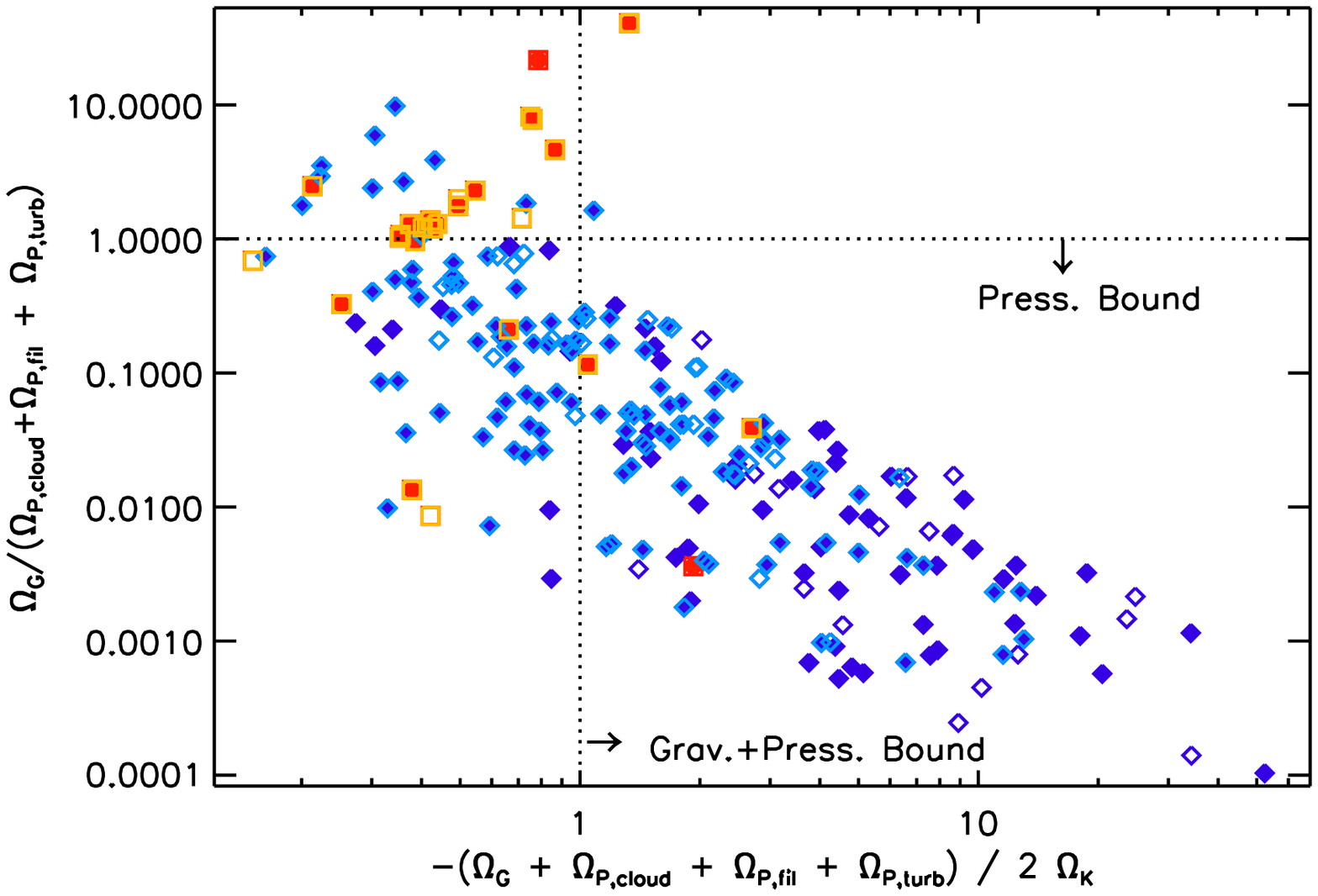}
\caption{Examining the effect of different smoothing scales on the cloud pressure derived for
	each core.  The left panel shows the cloud pressure derived including the cloud 
	column density
	up to scales a factor of two smaller than the nominal value (i.e., roughly 0.25~pc).  
	The right panel shows the cloud pressure derived including the cloud column density up to 
	scales a factor of two larger than the nominal value (i.e., 1~pc).  There
	is only a minimal shift in the points compared to Figure~\ref{fig_press}; in both
	cases, nearly all cores remain sub-virial and pressure-bound.
	See Figure~\ref{fig_nopress2} for the plotting conventions used.
}
\label{fig_press_smotest}
\end{figure}

The {\it a Trous} algorithm we apply for filtering can be applied on length scales of
$2^N$ pixels.  For context, 0.48~pc corresponds to 16 pixels
in the original \citet{Lombardi14} map.
In Figure~\ref{fig_press_smotest}, we show a comparable analysis to that in
Figure~\ref{fig_press} but with the cloud pressure calculated instead from a column
density map filtered to 0.24~pc (8 pixels) and 0.97~pc (32 pixels).  Although
points do shift for each of the different column density
choices, minimal changes to the energy density ratio distribution above or below one occur.

\begin{figure}[htb]
\plottwo{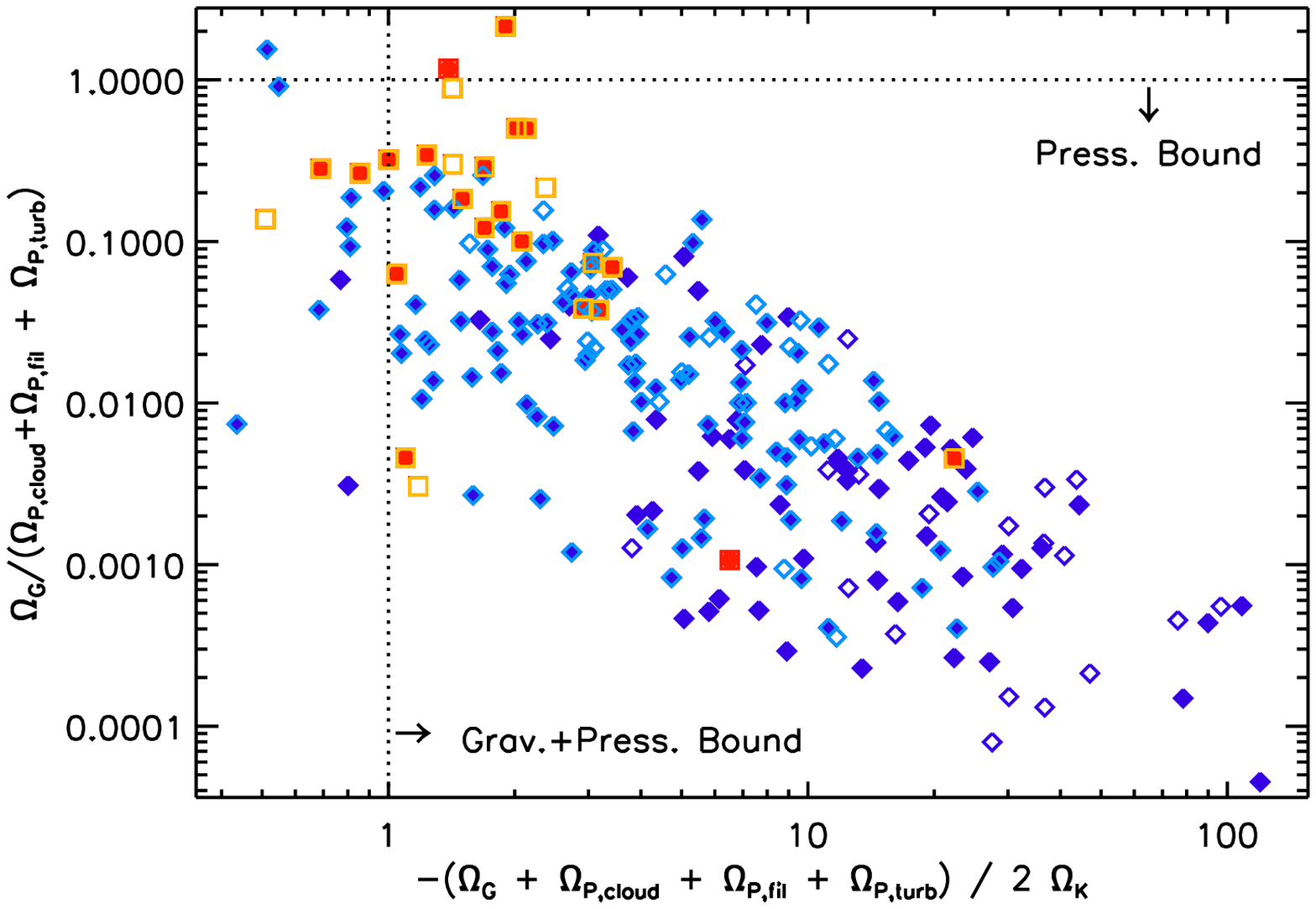}{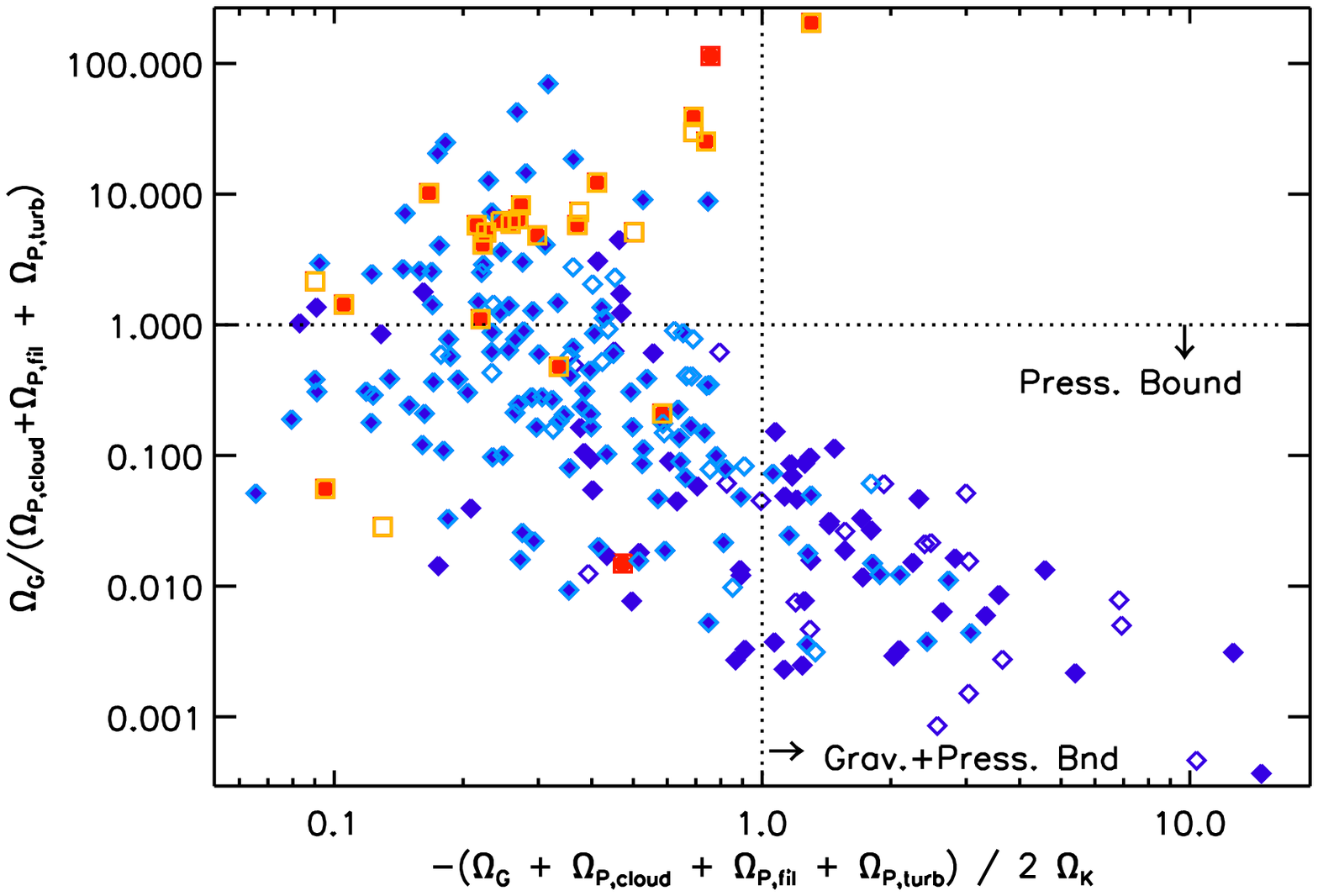}
\caption{Examining the effect of different smoothing scales on the cloud pressure derived for
	each core.  The left panel shows the cloud 
	pressure derived using the total cloud column density
	measured, i.e., with no filtering applied.  
	The right panel shows the cloud pressure derived including only the maximum smoothing
	scale for the cloud column density, i.e., 8~pc.
	In the former case, nearly all dense cores are sub-virial and pressure-bound,
	while in the latter case, many of the cores are super-virial, while pressure
	continues to dominate over gravity regardless of the level of confinement 
	on the core.
	See Figure~\ref{fig_nopress2} for the plotting conventions used.
	}
\label{fig_press_smotest2}
\end{figure}

In Figure~\ref{fig_press_smotest2}, we examine the most extreme cases, comparing
the results if instead the full, un-filtered column density map or
an extremely filtered map (7.7~pc or 256 pixels) were used.  In the case of no filtering,
we again see that most of the cores have energy density ratios above one.  
Only in
the extremely filtered map, do we find that the majority of cores 
are unbound.
In other words,  
a significant amount of cloud material must be excluded from our cloud pressure
estimate for most of the dense cores to be not pressure-confined.

These tests show that barring an extreme allocation of core-like and cloud-like material in the
\citet{Lombardi14} extinction map, we come to the same conclusion: pressure from the
ambient cloud material strongly dominates over self-gravity in keeping the dense cores
bound.  Given the fact that the amount of cloud pressure estimated for each core varies
by a significant amount (often a factor of 10 between the non-filtered and 
maximally-filtered cases), we may also surmise that changing
the overarching assumption in our method, i.e., assuming that the cloud is spherically 
symmetric, is also unlikely to change this conclusion, although the precise amount of
pressure-binding attributable to each core would vary under different models.

\section{Cloud Pressure Derivation}

Here we derive a useful expression for the hydrostatic pressure on a protostellar core embedded 
within a larger cloud.  We begin with a spherical cloud  of radius $\Rcl$, with a density 
profile $\rho(r)$, enclosed mass $m(r) = \int_0^r 4\pi \rho(r') r'^2 \,dr'$, and mean column 
(or more precisely, mass surface density) $\Sigmabar(r) = m(r)/(\pi r^2)$  within $r$. The 
bounding pressure is $P(\Rcl) = P_s$.  From any point, the column to the surface (along a radial 
path) is $\Sigmas(r) = \int_r^\Rcl \rho(r')\, dr'$.  The wisdom of these definitions is that the 
hydrostatic relation can be rewritten entirely in terms of column density:
\begin{equation} \label{eq:HydrostaticDifferential}
dP = - {G m \rho \over r^2}\, dr = \pi G \Sigmabar\, d\Sigmas, 
\end{equation} 
which can be integrated if the relation between $\Sigmabar$ and $\Sigmas$ is known.  Note also that
 \[ d\Sigmas = - \rho\, dr = - {dm\over 4\pi r^2} = -d{m\over 4\pi r^2} - {m\over 2\pi r^3}dr 
= -{d\Sigmabar\over 4} - {\Sigmabar\over2} {dr\over r}. \]  
Suppose we adopt a power-law density profile, $\rho \propto r^{-k}$, for which $\Sigmabar\propto 
r^{1-k}$; then $4(k-1) d\Sigmas = (3-k) d\Sigmabar$.  If $\Sigmacl = \Sigmabar(\Rcl)$ is the 
mean column of the entire cloud,   
\[ \Sigmabar = \Sigmacl + {4(k-1)\over 3-k}\Sigmas. \]  
Equation (\ref{eq:HydrostaticDifferential}) can then be integrated: 
\begin{equation} \label{eq:HydrostaticPressure}
P(r) - P_s =  \pi  G \Sigmacl \Sigmas + {2(k-1)\over 3-k} \pi G \Sigmas^2. 
\end{equation} 
The pressure is related to the mean cloud column and the column to the surface; for the 
characteristic GMC profile $k=1$, equation (\ref{eq:HydrostaticPressure}) reduces to the 
expression given by \citet{McKee89}.

Within our power-law density profile, one could also use the expression 
$P(r) = \left(\Rcl/r \right)^{k_P} P_s$ where $k_P = 2(k-1)$ \citep{McKee03}.    
The advantage of expression (\ref{eq:HydrostaticPressure}) is that it avoids making 
reference to the radius $r$, and is therefore more closely related to observed quantities 
and more robust to deviations from spherical symmetry.  

It is also useful to consider non-spherical geometries like sheets and cylinders, for which the 
maximum column $\Sigmatot$ perpendicular to the axis (for cylinders) or along the surface normal 
(for sheets) is the relevant scale; this 
quantity is twice the maximum value of $\Sigmas$. For a sheet 
confined by a surface pressure $P_s$, the overpressure is 
$P(\Sigmas) - P_s = 2\pi G \Sigmas (\Sigmatot- \Sigmas)$, for which the maximum (midplane) 
value is $ (\pi/2)G\Sigmatot^2$.  Within an unmagnetized isothermal filament 
\citep{Stodolkiewicz63,Ostriker64}, the central pressure is 
{at least}\footnote{In our expression $\Sigmatot$ neglects the column of the bounding medium, 
and $P$ reaches this lower limit when the bounding pressure is effectively zero.}  
$ (2/\pi) G \Sigmatot^2$. 

When estimating the pressure on a given core, one generally does not know the values of $\Sigmas$  
and $\Sigmacl$  directly from observations, but  rather the projected column  $\Sigma(x,y)$ 
without any indication how much column is in the background.  For this reason, we now consider 
how $P$ should be inferred from the column map. 

\subsection{Evaluation for an observed core}

What is the most reasonable range of pressures to associate with a specific core, given a map of 
the projected column $\Sigma$?  A natural choice is to associate $\Sigmacl$ with the mean of 
$\Sigma$ within the cloud boundaries, and $\Sigmas$ with one-half of $\Sigma$ at the location 
in question, but there are several ambiguities.  First, the external pressure $P_s$ must be 
guessed at.  Second, our $\Sigma$ map may be contaminated by emission unrelated to the 
structure of interest.  Third, if the density distribution is not isotropic, $\Sigma$ is 
affected by inclination.  Fourth, one usually does not know how much of $\Sigma$ is in front 
of the core and hence what value to associate with $\Sigmas/\Sigma$.  Finally, observations 
may suffer from intrinsic uncertainties involving finite resolution, spatial filtering, 
emission-to-mass conversions, and the like.   

We concentrate on the fourth ambiguity (foreground and background). The natural assumption 
that half of $\Sigma$ is in the foreground tends to maximize $P$, but it is more rigorous 
to adopt a model and some prior expectations regarding the occurrence of cores within clouds, 
and derive a probability distribution from $P$.  Let us assume: 
($i$) No cores exist with $\Sigmas<\Sigmasmin$, a situation corresponding to extinction thresholds 
observed in various regions \citep{Pak98,Johnstone04} perhaps because of FUV 
photo-ionization \citep{McKee89}. 
 ($ii$) The probability of finding a core scales as $\rho^\alpha$ per unit volume, 
a scenario 
accommodating a constant star formation rate per unit mass ($\alpha=1$) and a dynamically-limited 
rate $(\alpha=2)$.  
  
The differential cumulative probability is then $d\mathscr{P} \propto \rho^\alpha dz$, where $z$ 
is distance along the line of sight, except $d\mathscr{P}=0$ where $P<P(\Sigmasmin)$.  

For the special case $\alpha=1$, $d\mathscr{P}\propto d\Sigma$ so the column toward the core is 
uniformly distributed within its allowed range.   In the case of a planar sheet, $\Sigmas$ is 
uniformly distributed from $\Sigmasmin$ to $\Sigmatot/2$.  The median pressure within this 
distribution is lower than the midplane value by $(\pi/2)G(\Sigmatot/2-\Sigmasmin)^2$. 

To go further, we require relations among $P$, $\rho$, and $z$.  Hence, we adopt here the 
spherical model: then  $P\propto \rho^{\gamma_p}$ where $\gamma_p = k_P/k = 2-2/k$ is the 
polytropic exponent.  Also, $P\propto r^{-k_P} \propto (z^2 + \varpi^2)^{-k_P/2}$ if $\varpi$ is 
the projected offset from the cloud center and $z=0$ at the cloud center.  Working out $dz$ in 
terms of $dP$, 
 \begin{equation} \label{eq:ProbabilityOfPressure}
  d\mathscr{P} \propto {P^{(\alpha/2-1)k/(k-1)} \over |z(P)|} \, dP \propto {P^{(\alpha/2-1)k/(k-1)} \over \left[(P/\max P)^{-2/k_P}-1\right]^{1/2}} \, dP.   
 \end{equation}
 Note that for the dynamical star formation rate ($\alpha=2$) the numerator is constant and 
$d\mathscr{P}\propto |z(P)|^{-1}\,dP$ so long as $k\neq1$.    
 
The next step is to determine the maximum possible pressure by determining the maximum value of 
$\Sigmas$ along the line of sight, given the observed value of $\Sigma$, i.e., 
$\max P = P(\max \Sigmas$).  Often we can ignore the cloud boundary and extend a line-of-sight 
integration to infinity.  This assumption is especially reasonable if $P(\Sigmasmin) \gg P_s$, which is 
usually the case for giant molecular clouds.  We then obtain
\begin{equation} \label{eq:SigmasMax}
{\Sigma \over \max \Sigmas} = \pi^{1/2}(k-1)  { \Gamma[(k-1)/2] \over \Gamma(k/2)} \simeq k+1 + (\pi-3)(k-1)(3-k). 
\end{equation}
The middle expression comes from integrating through an infinite sphere, and the quadratic 
approximation is exact for $k=1,2,3$ and errs by only $0.6\%$ over the entire range.  It shows 
that the usual assumption ($\max \Sigmas = \Sigma/2$) is correct when $k=1$, but amounts to an 
over-prediction in steeper density profiles. 

\begin{figure}
\includegraphics[width=4in]{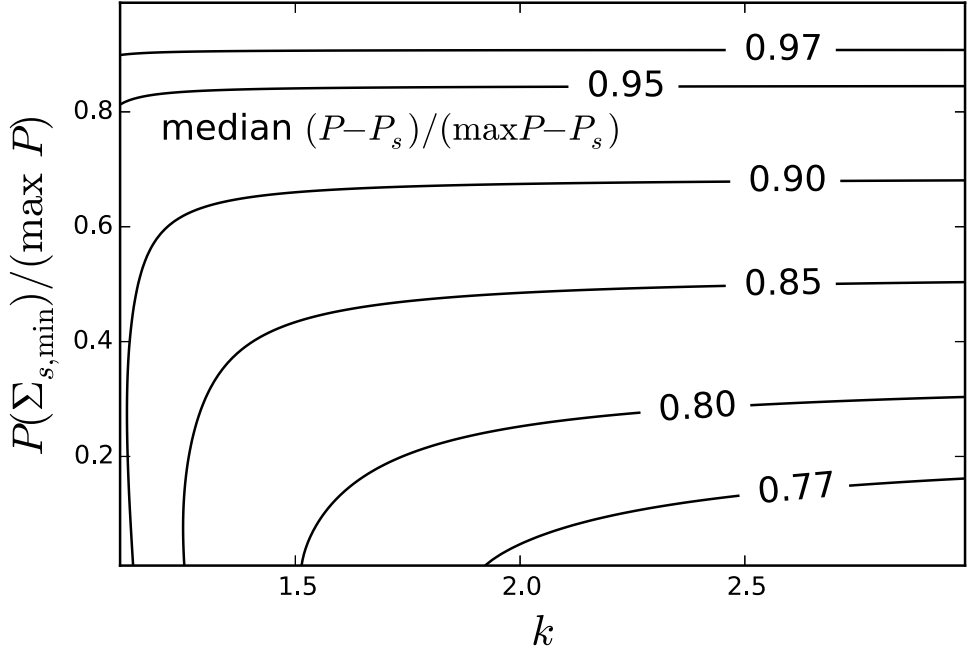} 
\caption{Median over-pressure, relative to the theoretical maximum, for cores detected within 
spherical clouds, assuming the probability per unit volume scales as $\rho^2$ for columns 
$\Sigmas>\Sigmasmin$ to the surface, i.e. pressures $P > P(\Sigmasmin)$.  Theoretical maximum and 
minimum pressures are determined from $\Sigmasmin$ and $\max \Sigmas$ (eq.\ \ref{eq:SigmasMax}) 
via equation (\ref{eq:HydrostaticPressure}), and the probability density follows equation 
(\ref{eq:ProbabilityOfPressure}).  This plot assumes $P(\Sigmasmin)\ll P(\Sigma/2)$. }
\label{fig:MedianPressure}
\end{figure}

Once $\max \Sigmas$ and $\max P$ have been determined, equation (\ref{eq:ProbabilityOfPressure}) 
provides the probability density.  In Figure \ref{fig:MedianPressure}, we plot the factor relating 
the median over-pressure to the maximum value. 
 
\acknowledgements{
RKF is a Dunlap Fellow at the Dunlap Institute for Astronomy \& Astrophysics.  The Dunlap 
Institute is funded through an endowment established by the David Dunlap family and the 
University of Toronto.  JEP, PC, ACT, and AP acknowledge the financial support of the European 
Research Council (ERC; project PALs 320620).  EWR and CDM are supported by Discovery 
Grants from NSERC of Canada.  The National Radio Astronomy Observatory is a facility 
of the National Science Foundation operated under cooperative agreement by Associated 
Universities, Inc.
SSRO is supported by NSF grant AST-1510021.
The JCMT has historically been operated by the Joint Astronomy Centre on behalf of the 
Science and Technology Facilities Council of the United Kingdom, the National Research 
Council of Canada and the Netherlands Organisation for Scientific Research. Additional 
funds for the construction of SCUBA-2 were provided by the Canada Foundation for 
Innovation. The identification number for the programme under which the SCUBA-2 data 
used in this paper is MJLSG41.
This research made use of Astropy, a community-developed core Python package for 
Astronomy \citep{Robitaille13}.
Figures in this paper were creating using the NASA IDL astronomy library
\citep{idlastro} and the Coyote IDL library ({\tt http://www.idlcoyote.com/index.html})
}

\facility{Green Bank Telescope, JCMT (SCUBA-2)}
\software{Astropy \citep{Robitaille13}, Starlink\citep{Currie14}, atrous.pro 
({\tt https://github.com/low-sky/idl-low-sky/tree/master/wavelet})  }

\bibliographystyle{apj}
\bibliography{oriona}{}

\clearpage
\startlongtable
\movetabledown=1.25in
\begin{longrotatetable}
\begin{deluxetable}{ccccccccccccccccc}
\tablecolumns{17}
\tablewidth{0pt}
\tabletypesize{\footnotesize}
\tablecaption{Dense Core Properties\label{tab_core_props}}
\tablehead{
\colhead{ID\tablenotemark{a}} &
\colhead{R.A.\tablenotemark{a}} &
\colhead{Decl.\tablenotemark{a}} &
\colhead{M\tablenotemark{a}} &
\colhead{R$_{eff}$\tablenotemark{a}} &
\colhead{C\tablenotemark{a}} &
\colhead{Pr?\tablenotemark{a}} &
\multicolumn{3}{c}{$\sigma_{obs}$ (km/s)\tablenotemark{b}} &
\multicolumn{3}{c}{$T_{kin}$(K)\tablenotemark{b}} &
\colhead{log$(P_c/k_B)$\tablenotemark{c}} &
\colhead{Vir.\tablenotemark{d}}&
\colhead{$-\Omega_G / \Omega_K$\tablenotemark{d}} &
\colhead{$\Omega_G / \Omega_{P,c}$\tablenotemark{d}}
\\
\colhead{} &
\colhead{(J2000)} &
\colhead{(J2000)} &
\colhead{(M$_{\odot}$)} &
\colhead{(pc)} &
\colhead{} &
\colhead{} &
\colhead{mn} &
\colhead{err} &
\colhead{std} &
\colhead{mn} &
\colhead{err} &
\colhead{std} &
\colhead{log(K/cm$^3$)} &
\colhead{Rat.}&
\colhead{ }&
\colhead{ }
}
\startdata
   5 &  83.84726 &  -5.02459 &   8.28 &  0.017 & 0.64 & Y &  0.311 &  0.002 &  0.041 &  17.5 &   0.1 &   0.8 & 7.3 &  1.4E+00  &  2.6E+00  &  2.0E+01 \\
   9 &  83.86135 &  -5.16620 &  11.10 &  0.023 & 0.48 & Y &  0.435 &  0.001 &  0.067 &  23.3 &   0.1 &   1.5 & 7.3 &  8.2E-01  &  1.5E+00  &  1.1E+01 \\
  10 &  83.82113 &  -5.32181 &   4.92 &  0.017 & 0.51 & N &  0.542 &  0.001 &  0.075 &  25.2 &   0.0 &   0.8 & 7.5 &  3.9E-01  &  6.2E-01  &  4.2E+00 \\
  12 &  83.81898 &  -5.32483 &   3.84 &  0.017 & 0.58 & N &  0.517 &  0.001 &  0.084 &  25.5 &   0.0 &   1.0 & 7.5 &  3.7E-01  &  5.2E-01  &  2.5E+00 \\
  16 &  83.81604 &  -5.33532 &   2.06 &  0.025 & 0.69 & N &  0.477 &  0.001 &  0.084 &  28.1 &   0.0 &   1.5 & 7.5 &  8.5E-01  &  2.1E-01  &  1.4E-01 \\
  17 &  83.82469 &  -5.31807 &   2.93 &  0.017 & 0.54 & N &  0.563 &  0.001 &  0.064 &  25.2 &   0.0 &   0.9 & 7.5 &  2.9E-01  &  3.5E-01  &  1.6E+00 \\
  18 &  83.81026 &  -5.31189 &   3.64 &  0.026 & 0.62 & N &  0.450 &  0.001 &  0.109 &  23.7 &   0.0 &   0.8 & 7.4 &  6.1E-01  &  4.1E-01  &  5.0E-01 \\
  19 &  83.86445 &  -5.15861 &   3.11 &  0.027 & 0.63 & Y &  0.423 &  0.001 &  0.074 &  23.7 &   0.1 &   1.7 & 7.3 &  5.8E-01  &  3.8E-01  &  4.8E-01 \\
  21 &  83.84284 &  -5.01978 &   3.61 &  0.017 & 0.56 & Y &  0.272 &  0.001 &  0.052 &  17.0 &   0.1 &   0.7 & 7.3 &  8.5E-01  &  1.3E+00  &  3.7E+00 \\
  23 &  83.82478 &  -5.00483 &   2.49 &  0.017 & 0.62 & N &  0.341 &  0.001 &  0.033 &  17.8 &   0.0 &   0.4 & 7.2 &  5.2E-01  &  6.9E-01  &  1.9E+00 \\
  30 &  83.83122 &  -5.25931 &   1.82 &  0.018 & 0.60 & Y &  0.329 &  0.001 &  0.032 &  19.9 &   0.1 &   0.7 & 7.2 &  4.9E-01  &  4.9E-01  &  1.0E+00 \\
  32 &  83.82470 &  -5.31361 &   2.94 &  0.018 & 0.39 & N &  0.565 &  0.001 &  0.071 &  25.1 &   0.0 &   1.0 & 7.4 &  2.9E-01  &  3.3E-01  &  1.4E+00 \\
  33 &  83.85801 &  -5.09514 &   2.92 &  0.017 & 0.54 & Y &  0.212 &  0.000 &  0.031 &  17.0 &   0.0 &   0.9 & 7.3 &  1.0E+00  &  1.4E+00  &  2.3E+00 \\
  34 &  83.83522 &  -5.01365 &   4.37 &  0.027 & 0.58 & N &  0.290 &  0.001 &  0.060 &  17.2 &   0.0 &   0.6 & 7.3 &  1.0E+00  &  9.5E-01  &  8.9E-01 \\
  35 &  83.85934 &  -5.06551 &   2.03 &  0.017 & 0.58 & Y &  0.275 &  0.002 &  0.096 &  16.4 &   0.1 &   0.9 & 7.3 &  7.1E-01  &  7.6E-01  &  1.2E+00 \\
  36 &  83.87365 &  -4.97983 &   1.84 &  0.017 & 0.56 & Y &  0.478 &  0.003 &  0.131 &  18.6 &   0.1 &   1.5 & 7.1 &  2.5E-01  &  3.0E-01  &  1.6E+00 \\
  39 &  83.85962 &  -5.17263 &   2.08 &  0.017 & 0.51 & N &  0.401 &  0.001 &  0.083 &  22.6 &   0.1 &   1.5 & 7.3 &  3.8E-01  &  4.3E-01  &  1.2E+00 \\
  40 &  83.86029 &  -5.08760 &   3.89 &  0.026 & 0.57 & N &  0.202 &  0.000 &  0.013 &  16.5 &   0.0 &   0.5 & 7.3 &  1.5E+00  &  1.3E+00  &  8.3E-01 \\
  41 &  83.78702 &  -5.93102 &   3.25 &  0.017 & 0.50 & Y &  0.264 &  0.001 &  0.092 &  14.6 &   0.1 &   0.8 & 7.0 &  8.0E-01  &  1.3E+00  &  5.1E+00 \\
  42 &  83.84709 &  -5.20066 &   1.48 &  0.017 & 0.58 & Y &  0.301 &  0.000 &  0.073 &  19.4 &   0.0 &   0.4 & 7.2 &  5.7E-01  &  4.7E-01  &  6.9E-01 \\
  43 &  83.78889 &  -5.33678 &   2.93 &  0.029 & 0.51 & N &  0.357 &  0.001 &  0.086 &  22.6 &   0.1 &   0.9 & 7.5 &  1.3E+00  &  4.1E-01  &  1.8E-01 \\
  44 &  83.82886 &  -5.01357 &   1.41 &  0.017 & 0.64 & Y &  0.311 &  0.001 &  0.064 &  17.5 &   0.0 &   0.7 & 7.2 &  5.9E-01  &  4.4E-01  &  6.0E-01 \\
  47 &  83.85118 &  -5.14224 &   1.40 &  0.017 & 0.62 & Y &  0.346 &  0.001 &  0.080 &  19.7 &   0.1 &   0.5 & 7.3 &  5.3E-01  &  3.7E-01  &  5.3E-01 \\
  49 &  83.83959 &  -5.21998 &   2.99 &  0.027 & 0.59 & N &  0.342 &  0.001 &  0.055 &  19.8 &   0.0 &   0.9 & 7.2 &  7.6E-01  &  5.0E-01  &  4.9E-01 \\
  50 &  83.85229 &  -5.13106 &   2.22 &  0.017 & 0.51 & N &  0.374 &  0.001 &  0.045 &  19.5 &   0.0 &   0.7 & 7.3 &  4.6E-01  &  5.3E-01  &  1.3E+00 \\
  54 &  83.80960 &  -5.30330 &   1.91 &  0.022 & 0.52 & N &  0.423 &  0.001 &  0.077 &  23.3 &   0.0 &   0.6 & 7.4 &  6.2E-01  &  2.8E-01  &  3.0E-01 \\
  55 &  83.83234 &  -5.21963 &   1.10 &  0.023 & 0.64 & N &  0.298 &  0.001 &  0.062 &  19.3 &   0.0 &   1.1 & 7.2 &  1.0E+00  &  2.7E-01  &  1.5E-01 \\
  56 &  83.82595 &  -5.00943 &   1.44 &  0.017 & 0.59 & Y &  0.327 &  0.001 &  0.051 &  17.7 &   0.0 &   0.6 & 7.2 &  5.4E-01  &  4.2E-01  &  6.3E-01 \\
  57 &  83.87378 &  -4.99644 &   0.69 &  0.017 & 0.61 & Y &  0.453 &  0.004 &  0.193 &  18.5 &   0.2 &   1.5 & 7.1 &  3.9E-01  &  1.2E-01  &  1.9E-01 \\
  58 &  83.84570 &  -5.21045 &   6.84 &  0.041 & 0.50 & N &  0.347 &  0.000 &  0.049 &  19.7 &   0.0 &   0.5 & 7.2 &  1.1E+00  &  7.6E-01  &  5.1E-01 \\
  59 &  83.80631 &  -5.96555 &   1.74 &  0.018 & 0.58 & Y &  0.271 &  0.002 &  0.054 &  15.1 &   0.1 &   1.0 & 7.1 &  6.1E-01  &  6.6E-01  &  1.2E+00 \\
  63 &  83.74833 &  -5.35940 &   2.05 &  0.026 & 0.44 & N &  0.552 &  0.004 &  0.087 &  21.4 &   0.2 &   1.2 & 7.2 &  4.1E-01  &  1.7E-01  &  2.7E-01 \\
  65 &  83.80722 &  -5.44595 &   7.74 &  0.054 & 0.40 & N &  0.689 &  0.003 &  0.112 &  27.3 &   0.1 &   2.9 & 7.3 &  7.0E-01  &  2.1E-01  &  1.7E-01 \\
  68 &  83.86681 &  -5.17436 &   1.04 &  0.018 & 0.54 & N &  0.369 &  0.001 &  0.091 &  22.1 &   0.1 &   1.5 & 7.3 &  5.5E-01  &  2.3E-01  &  2.6E-01 \\
  69 &  83.85507 &  -5.04317 &   1.04 &  0.029 & 0.73 & N &  0.215 &  0.001 &  0.040 &  16.4 &   0.1 &   0.8 & 7.3 &  4.3E+00  &  3.0E-01  &  3.7E-02 \\
  70 &  83.73424 &  -5.76767 &   2.92 &  0.023 & 0.43 & N &  0.586 &  0.002 &  0.088 &  15.9 &   0.1 &   0.8 & 7.1 &  2.5E-01  &  2.6E-01  &  1.0E+00 \\
  72 &  83.81564 &  -4.99858 &   1.11 &  0.022 & 0.60 & N &  0.324 &  0.001 &  0.049 &  18.0 &   0.1 &   0.8 & 7.2 &  9.2E-01  &  2.6E-01  &  1.6E-01 \\
  73 &  83.78455 &  -5.59903 &   1.90 &  0.021 & 0.49 & Y &  0.318 &  0.001 &  0.037 &  15.9 &   0.1 &   0.7 & 7.1 &  5.7E-01  &  4.9E-01  &  7.6E-01 \\
  77 &  83.77707 &  -4.91264 &   0.68 &  0.017 & 0.58 & N &  0.224 &  0.001 &  0.052 &  18.7 &   0.1 &   1.1 & 6.8 &  5.8E-01  &  3.0E-01  &  3.5E-01 \\
  81 &  83.67071 &  -5.52846 &   1.42 &  0.017 & 0.56 & Y &  0.250 &  0.005 &  0.039 &  14.3 &   0.6 &   1.1 & 6.7 &  4.8E-01  &  6.3E-01  &  2.0E+00 \\
  82 &  83.84701 &  -5.12541 &   2.39 &  0.025 & 0.51 & N &  0.374 &  0.001 &  0.043 &  19.3 &   0.0 &   0.8 & 7.3 &  7.6E-01  &  3.9E-01  &  3.4E-01 \\
  83 &  83.77700 &  -4.94065 &   1.83 &  0.023 & 0.48 & N &  0.221 &  0.001 &  0.039 &  17.7 &   0.1 &   0.9 & 7.0 &  8.4E-01  &  6.2E-01  &  5.9E-01 \\
  84 &  83.83829 &  -5.24769 &   1.59 &  0.030 & 0.58 & N &  0.239 &  0.001 &  0.036 &  19.6 &   0.1 &   0.5 & 7.2 &  1.9E+00  &  3.7E-01  &  1.1E-01 \\
  91 &  83.77017 &  -5.40508 &   1.25 &  0.025 & 0.48 & N &  0.639 &  0.010 &  0.077 &  23.6 &   0.4 &   2.2 & 7.4 &  5.6E-01  &  8.4E-02  &  8.1E-02 \\
  93 &  83.78789 &  -5.97440 &   1.34 &  0.017 & 0.53 & Y &  0.181 &  0.001 &  0.026 &  14.0 &   0.1 &   0.7 & 7.0 &  9.1E-01  &  8.4E-01  &  8.6E-01 \\
  95 &  83.80558 &  -4.99034 &   0.49 &  0.025 & 0.73 & N &  0.295 &  0.002 &  0.066 &  17.8 &   0.1 &   1.7 & 7.2 &  2.8E+00  &  1.1E-01  &  2.0E-02 \\
  97 &  83.77125 &  -5.62290 &   1.08 &  0.026 & 0.65 & N &  0.267 &  0.001 &  0.047 &  15.0 &   0.1 &   0.6 & 7.1 &  1.6E+00  &  2.9E-01  &  1.0E-01 \\
  99 &  84.04762 &  -6.17912 &   1.92 &  0.026 & 0.62 & N &  0.266 &  0.001 &  0.041 &  13.3 &   0.1 &   0.6 & 6.9 &  8.0E-01  &  5.4E-01  &  5.1E-01 \\
 100 &  83.85114 &  -5.20867 &   1.68 &  0.023 & 0.58 & N &  0.355 &  0.000 &  0.046 &  19.6 &   0.0 &   0.5 & 7.2 &  7.1E-01  &  3.2E-01  &  2.9E-01 \\
 101 &  83.86029 &  -5.10059 &   2.45 &  0.031 & 0.51 & N &  0.235 &  0.001 &  0.044 &  17.6 &   0.0 &   1.0 & 7.3 &  2.1E+00  &  5.9E-01  &  1.6E-01 \\
 105 &  83.84746 &  -5.11968 &   2.03 &  0.025 & 0.40 & N &  0.377 &  0.001 &  0.034 &  19.1 &   0.1 &   0.7 & 7.3 &  7.8E-01  &  3.3E-01  &  2.7E-01 \\
 107 &  83.76559 &  -5.30012 &   0.73 &  0.018 & 0.54 & N &  0.187 &  0.004 &  0.091 &  16.7 &   0.4 &   1.9 & 7.1 &  1.2E+00  &  4.0E-01  &  1.9E-01 \\
 108 &  83.84002 &  -5.24101 &   2.25 &  0.032 & 0.50 & N &  0.224 &  0.001 &  0.022 &  19.7 &   0.1 &   0.7 & 7.2 &  1.9E+00  &  5.2E-01  &  1.6E-01 \\
 109 &  83.84306 &  -5.17037 &   0.66 &  0.017 & 0.51 & N &  0.194 &  0.001 &  0.036 &  19.0 &   0.1 &   1.1 & 7.3 &  1.5E+00  &  3.3E-01  &  1.3E-01 \\
 110 &  83.88081 &  -5.01031 &   0.52 &  0.018 & 0.55 & N &  0.298 &  0.006 &  0.220 &  19.9 &   0.6 &   2.1 & 7.1 &  9.0E-01  &  1.6E-01  &  9.9E-02 \\
 114 &  83.80598 &  -5.45109 &   1.91 &  0.029 & 0.55 & N &  0.660 &  0.004 &  0.098 &  25.6 &   0.2 &   3.2 & 7.2 &  4.1E-01  &  1.0E-01  &  1.5E-01 \\
 116 &  83.85353 &  -5.13791 &   1.03 &  0.018 & 0.55 & N &  0.360 &  0.001 &  0.067 &  19.8 &   0.0 &   0.6 & 7.3 &  6.4E-01  &  2.4E-01  &  2.3E-01 \\
 117 &  83.70287 &  -4.92851 &   0.60 &  0.020 & 0.60 & N &  0.214 &  0.009 &  0.053 &  20.4 &   0.9 &   2.9 & 6.7 &  6.7E-01  &  2.3E-01  &  2.0E-01 \\
 121 &  83.96774 &  -6.16719 &   0.87 &  0.017 & 0.56 & Y &  0.536 &  0.007 &  0.101 &  14.7 &   0.3 &   1.9 & 6.9 &  1.8E-01  &  1.2E-01  &  5.3E-01 \\
 122 &  83.77141 &  -5.57990 &   1.13 &  0.031 & 0.65 & N &  0.157 &  0.001 &  0.043 &  17.0 &   0.1 &   1.4 & 7.1 &  3.6E+00  &  3.9E-01  &  5.7E-02 \\
 125 &  83.86634 &  -5.11958 &   0.65 &  0.024 & 0.58 & N &  0.245 &  0.002 &  0.069 &  19.2 &   0.1 &   1.4 & 7.3 &  3.3E+00  &  1.8E-01  &  2.9E-02 \\
 127 &  83.78911 &  -5.31285 &   1.11 &  0.030 & 0.62 & N &  0.288 &  0.002 &  0.065 &  21.2 &   0.1 &   1.7 & 7.4 &  3.3E+00  &  2.1E-01  &  3.2E-02 \\
 128 &  83.75322 &  -5.27003 &   0.52 &  0.027 & 0.66 & N &  0.220 &  0.003 &  0.056 &  17.9 &   0.2 &   1.7 & 6.9 &  2.4E+00  &  1.5E-01  &  3.3E-02 \\
 129 &  83.88409 &  -5.09695 &   3.30 &  0.046 & 0.51 & N &  0.331 &  0.002 &  0.044 &  18.5 &   0.1 &   1.1 & 7.2 &  2.7E+00  &  3.5E-01  &  6.8E-02 \\
 130 &  83.87380 &  -4.97238 &   1.74 &  0.034 & 0.54 & N &  0.417 &  0.002 &  0.126 &  19.4 &   0.1 &   2.5 & 7.0 &  9.0E-01  &  1.8E-01  &  1.1E-01 \\
 133 &  84.15429 &  -6.25039 &   0.83 &  0.019 & 0.53 & N &  0.288 &  0.017 &  0.123 &  15.5 &   1.9 &   3.0 & 6.9 &  5.5E-01  &  2.8E-01  &  3.4E-01 \\
 134 &  83.80037 &  -5.57433 &   0.90 &  0.024 & 0.55 & N &  0.187 &  0.001 &  0.046 &  18.0 &   0.1 &   0.9 & 7.0 &  1.7E+00  &  3.4E-01  &  1.1E-01 \\
 137 &  83.86283 &  -5.19270 &   0.74 &  0.030 & 0.64 & N &  0.202 &  0.001 &  0.045 &  19.7 &   0.0 &   0.5 & 7.2 &  5.6E+00  &  2.0E-01  &  1.8E-02 \\
 139 &  83.84413 &  -5.11082 &   0.61 &  0.028 & 0.67 & N &  0.391 &  0.002 &  0.042 &  19.3 &   0.1 &   1.2 & 7.3 &  2.8E+00  &  8.4E-02  &  1.5E-02 \\
 142 &  83.88501 &  -5.10955 &   0.45 &  0.018 & 0.61 & N &  0.313 &  0.002 &  0.056 &  18.1 &   0.1 &   0.9 & 7.2 &  1.2E+00  &  1.4E-01  &  6.0E-02 \\
 144 &  83.83382 &  -5.08503 &   1.75 &  0.027 & 0.49 & N &  0.448 &  0.003 &  0.109 &  19.4 &   0.1 &   1.0 & 7.2 &  7.9E-01  &  2.0E-01  &  1.4E-01 \\
 147 &  83.82629 &  -5.20081 &   0.31 &  0.017 & 0.65 & N &  0.232 &  0.001 &  0.041 &  18.0 &   0.1 &   1.3 & 7.2 &  2.1E+00  &  1.3E-01  &  3.4E-02 \\
 153 &  83.78853 &  -5.86895 &   1.75 &  0.031 & 0.57 & N &  0.194 &  0.001 &  0.061 &  13.7 &   0.1 &   0.8 & 6.9 &  1.8E+00  &  5.9E-01  &  2.0E-01 \\
 154 &  83.73949 &  -5.76451 &   1.02 &  0.017 & 0.52 & N &  0.545 &  0.002 &  0.076 &  15.8 &   0.1 &   0.8 & 7.1 &  2.3E-01  &  1.4E-01  &  4.4E-01 \\
 161 &  83.87602 &  -5.10059 &   0.72 &  0.020 & 0.56 & N &  0.280 &  0.002 &  0.057 &  18.5 &   0.1 &   0.9 & 7.2 &  1.3E+00  &  2.2E-01  &  8.8E-02 \\
 164 &  83.75679 &  -5.25595 &   1.10 &  0.025 & 0.50 & N &  0.317 &  0.005 &  0.139 &  17.4 &   0.3 &   2.0 & 6.9 &  7.1E-01  &  2.3E-01  &  2.0E-01 \\
 165 &  83.68338 &  -5.69030 &   1.20 &  0.017 & 0.41 & Y &  0.341 &  0.007 &  0.127 &  13.5 &   0.5 &   1.5 & 7.0 &  4.1E-01  &  3.6E-01  &  7.6E-01 \\
 166 &  83.86629 &  -5.06040 &   0.61 &  0.017 & 0.57 & Y &  0.271 &  0.002 &  0.088 &  16.7 &   0.1 &   1.0 & 7.3 &  1.2E+00  &  2.3E-01  &  1.1E-01 \\
 167 &  83.86085 &  -5.06032 &   1.95 &  0.029 & 0.54 & N &  0.247 &  0.001 &  0.083 &  16.4 &   0.1 &   0.8 & 7.3 &  2.1E+00  &  4.9E-01  &  1.3E-01 \\
 169 &  83.76035 &  -5.41373 &   0.66 &  0.025 & 0.54 & N &  0.663 &  0.012 &  0.121 &  23.4 &   0.5 &   1.8 & 7.3 &  7.1E-01  &  4.3E-02  &  3.1E-02 \\
 170 &  83.82585 &  -5.22697 &   0.31 &  0.017 & 0.62 & N &  0.237 &  0.001 &  0.044 &  19.6 &   0.1 &   1.4 & 7.2 &  1.8E+00  &  1.3E-01  &  3.6E-02 \\
 172 &  83.78475 &  -5.85627 &   1.78 &  0.032 & 0.56 & N &  0.181 &  0.001 &  0.015 &  14.0 &   0.2 &   0.9 & 6.9 &  2.0E+00  &  6.0E-01  &  1.8E-01 \\
 174 &  83.76215 &  -5.60311 &   1.68 &  0.030 & 0.51 & N &  0.160 &  0.001 &  0.061 &  16.2 &   0.1 &   0.9 & 7.1 &  2.6E+00  &  6.1E-01  &  1.3E-01 \\
 177 &  83.85584 &  -5.09951 &   0.67 &  0.019 & 0.61 & N &  0.236 &  0.001 &  0.046 &  17.6 &   0.0 &   1.0 & 7.3 &  1.7E+00  &  2.6E-01  &  8.3E-02 \\
 179 &  83.85463 &  -5.25948 &   0.27 &  0.017 & 0.65 & N &  0.270 &  0.010 &  0.122 &  21.3 &   0.8 &   2.3 & 7.1 &  1.6E+00  &  9.3E-02  &  3.0E-02 \\
 180 &  83.82183 &  -5.07955 &   0.54 &  0.025 & 0.49 & N &  0.785 &  0.011 &  0.127 &  22.5 &   0.4 &   2.6 & 7.2 &  6.0E-01  &  2.5E-02  &  2.1E-02 \\
 181 &  83.89253 &  -4.99704 &   0.61 &  0.021 & 0.43 & Y &  0.220 &  0.004 &  0.055 &  19.4 &   0.3 &   2.3 & 7.0 &  1.5E+00  &  2.2E-01  &  8.0E-02 \\
 182 &  83.77374 &  -4.90811 &   2.41 &  0.061 & 0.55 & N &  0.223 &  0.001 &  0.051 &  18.8 &   0.1 &   1.4 & 6.8 &  5.4E+00  &  3.0E-01  &  2.9E-02 \\
 189 &  83.72439 &  -5.77180 &   2.04 &  0.033 & 0.50 & N &  0.608 &  0.002 &  0.170 &  16.0 &   0.1 &   0.8 & 7.1 &  5.5E-01  &  1.2E-01  &  1.2E-01 \\
 191 &  83.75644 &  -5.30353 &   0.31 &  0.017 & 0.60 & N &  0.256 &  0.008 &  0.119 &  16.3 &   0.5 &   1.5 & 7.1 &  1.6E+00  &  1.3E-01  &  4.0E-02 \\
 192 &  83.86733 &  -5.08365 &   0.55 &  0.025 & 0.68 & Y &  0.197 &  0.000 &  0.010 &  16.4 &   0.0 &   0.5 & 7.3 &  5.1E+00  &  2.0E-01  &  2.0E-02 \\
 193 &  83.82760 &  -5.23540 &   0.63 &  0.028 & 0.52 & N &  0.203 &  0.001 &  0.030 &  21.1 &   0.1 &   0.9 & 7.2 &  4.0E+00  &  1.8E-01  &  2.2E-02 \\
 194 &  83.79358 &  -5.92738 &   1.51 &  0.021 & 0.50 & N &  0.254 &  0.001 &  0.041 &  14.3 &   0.1 &   0.6 & 7.0 &  8.3E-01  &  5.3E-01  &  4.7E-01 \\
 197 &  83.87226 &  -5.20200 &   1.25 &  0.047 & 0.55 & N &  0.220 &  0.001 &  0.069 &  19.8 &   0.1 &   1.1 & 7.2 &  9.8E+00  &  2.0E-01  &  1.0E-02 \\
 198 &  83.77046 &  -4.95716 &   1.52 &  0.050 & 0.50 & N &  0.228 &  0.002 &  0.057 &  18.3 &   0.2 &   2.1 & 7.0 &  6.6E+00  &  2.3E-01  &  1.8E-02 \\
 199 &  83.80018 &  -5.95131 &   1.49 &  0.034 & 0.63 & N &  0.221 &  0.001 &  0.070 &  14.1 &   0.1 &   1.2 & 7.1 &  3.1E+00  &  3.9E-01  &  6.6E-02 \\
 203 &  83.82688 &  -5.08047 &   0.58 &  0.027 & 0.58 & N &  0.644 &  0.007 &  0.168 &  20.7 &   0.2 &   1.9 & 7.2 &  9.9E-01  &  3.7E-02  &  1.9E-02 \\
 207 &  83.76871 &  -4.92399 &   0.42 &  0.028 & 0.64 & N &  0.216 &  0.001 &  0.052 &  18.3 &   0.1 &   1.3 & 6.9 &  3.6E+00  &  1.2E-01  &  1.7E-02 \\
 208 &  83.78403 &  -4.93252 &   0.32 &  0.022 & 0.65 & N &  0.233 &  0.002 &  0.055 &  18.3 &   0.1 &   1.6 & 6.9 &  2.4E+00  &  1.1E-01  &  2.3E-02 \\
 211 &  83.72458 &  -4.95249 &   0.98 &  0.028 & 0.52 & N &  0.259 &  0.004 &  0.060 &  16.7 &   0.2 &   1.1 & 6.8 &  1.1E+00  &  2.4E-01  &  1.3E-01 \\
 213 &  83.92177 &  -5.04128 &   0.94 &  0.021 & 0.46 & N &  0.184 &  0.005 &  0.041 &  14.8 &   0.4 &   1.5 & 6.9 &  1.1E+00  &  4.5E-01  &  2.7E-01 \\
 220 &  83.75208 &  -5.66773 &   0.73 &  0.030 & 0.51 & N &  0.293 &  0.005 &  0.212 &  17.6 &   0.3 &   2.0 & 7.1 &  2.7E+00  &  1.4E-01  &  2.7E-02 \\
 221 &  83.73876 &  -5.34577 &   0.72 &  0.028 & 0.52 & N &  0.551 &  0.008 &  0.135 &  20.5 &   0.4 &   1.4 & 7.1 &  9.8E-01  &  5.6E-02  &  2.9E-02 \\
 223 &  83.75273 &  -4.97396 &   0.28 &  0.021 & 0.68 & N &  0.284 &  0.004 &  0.078 &  16.8 &   0.3 &   1.7 & 6.9 &  1.9E+00  &  8.1E-02  &  2.2E-02 \\
 225 &  83.78123 &  -5.94482 &   0.88 &  0.026 & 0.62 & N &  0.171 &  0.001 &  0.064 &  13.0 &   0.1 &   1.2 & 7.0 &  3.1E+00  &  4.0E-01  &  6.7E-02 \\
 226 &  83.76019 &  -5.63125 &   0.69 &  0.025 & 0.60 & N &  0.227 &  0.001 &  0.048 &  14.9 &   0.1 &   0.7 & 7.1 &  2.6E+00  &  2.3E-01  &  4.7E-02 \\
 228 &  83.75243 &  -5.36597 &   0.56 &  0.021 & 0.51 & N &  0.464 &  0.005 &  0.200 &  22.1 &   0.2 &   1.6 & 7.3 &  1.1E+00  &  7.5E-02  &  3.6E-02 \\
 231 &  83.79408 &  -4.99982 &   1.21 &  0.039 & 0.55 & N &  0.422 &  0.003 &  0.117 &  19.1 &   0.2 &   1.7 & 7.1 &  2.5E+00  &  1.1E-01  &  2.1E-02 \\
 232 &  83.82275 &  -4.94168 &   0.32 &  0.023 & 0.53 & N &  0.446 &  0.007 &  0.086 &  23.3 &   0.4 &   3.7 & 7.0 &  1.1E+00  &  4.2E-02  &  1.9E-02 \\
 234 &  83.78691 &  -5.88847 &   0.81 &  0.021 & 0.55 & N &  0.184 &  0.003 &  0.061 &  13.9 &   0.2 &   1.1 & 7.0 &  1.3E+00  &  4.2E-01  &  1.9E-01 \\
 235 &  83.74957 &  -5.33127 &   0.29 &  0.017 & 0.65 & N &  0.426 &  0.014 &  0.145 &  16.4 &   0.9 &   4.4 & 7.2 &  1.0E+00  &  5.9E-02  &  2.9E-02 \\
 239 &  83.75886 &  -5.29740 &   0.34 &  0.019 & 0.62 & N &  0.228 &  0.005 &  0.099 &  16.3 &   0.4 &   1.5 & 7.1 &  2.1E+00  &  1.5E-01  &  3.5E-02 \\
 240 &  83.71620 &  -4.93830 &   0.26 &  0.017 & 0.57 & N &  0.223 &  0.007 &  0.083 &  18.1 &   0.6 &   2.2 & 6.8 &  1.0E+00  &  1.2E-01  &  6.1E-02 \\
 241 &  83.77792 &  -4.93596 &   0.46 &  0.017 & 0.56 & N &  0.232 &  0.001 &  0.045 &  17.9 &   0.1 &   1.0 & 7.0 &  8.9E-01  &  2.0E-01  &  1.3E-01 \\
 245 &  83.82808 &  -5.06771 &   0.22 &  0.027 & 0.71 & N &  0.647 &  0.006 &  0.118 &  21.8 &   0.3 &   1.6 & 7.2 &  2.6E+00  &  1.4E-02  &  2.6E-03 \\
 246 &  83.87723 &  -5.87634 &   1.10 &  0.019 & 0.52 & N &  0.194 &  0.002 &  0.065 &  12.0 &   0.3 &   1.0 & 6.8 &  8.2E-01  &  6.3E-01  &  6.2E-01 \\
 247 &  83.81210 &  -5.27932 &   0.49 &  0.020 & 0.45 & N &  0.328 &  0.008 &  0.123 &  25.2 &   0.6 &   1.4 & 7.3 &  1.5E+00  &  1.1E-01  &  3.8E-02 \\
 249 &  83.84900 &  -5.09092 &   0.30 &  0.017 & 0.66 & N &  0.229 &  0.001 &  0.052 &  16.9 &   0.1 &   1.0 & 7.3 &  2.8E+00  &  1.4E-01  &  2.5E-02 \\
 250 &  83.87190 &  -4.99081 &   1.65 &  0.053 & 0.60 & N &  0.439 &  0.003 &  0.188 &  18.5 &   0.1 &   1.4 & 7.1 &  4.2E+00  &  1.0E-01  &  1.2E-02 \\
 254 &  83.90602 &  -6.16198 &   0.45 &  0.017 & 0.60 & N &  0.195 &  0.006 &  0.095 &  14.4 &   0.8 &   4.0 & 6.9 &  1.0E+00  &  2.6E-01  &  1.5E-01 \\
 256 &  83.79314 &  -5.27891 &   0.46 &  0.021 & 0.47 & N &  0.507 &  0.015 &  0.164 &  21.8 &   0.8 &   4.3 & 7.2 &  8.9E-01  &  5.5E-02  &  3.2E-02 \\
 257 &  83.75793 &  -5.23305 &   0.44 &  0.019 & 0.54 & N &  0.572 &  0.014 &  0.134 &  15.7 &   1.0 &   2.8 & 6.8 &  2.4E-01  &  5.0E-02  &  1.1E-01 \\
 259 &  83.74219 &  -5.33073 &   0.29 &  0.022 & 0.64 & N &  0.439 &  0.015 &  0.134 &  16.0 &   0.9 &   3.0 & 7.1 &  1.5E+00  &  4.5E-02  &  1.5E-02 \\
 260 &  83.81226 &  -5.26960 &   0.67 &  0.027 & 0.50 & N &  0.285 &  0.004 &  0.059 &  21.2 &   0.2 &   2.9 & 7.2 &  3.0E+00  &  1.4E-01  &  2.3E-02 \\
 264 &  83.75742 &  -5.92680 &   4.16 &  0.052 & 0.49 & N &  0.258 &  0.001 &  0.061 &  13.5 &   0.1 &   0.6 & 7.0 &  2.9E+00  &  6.1E-01  &  1.2E-01 \\
 265 &  84.10731 &  -6.23729 &   0.94 &  0.025 & 0.54 & N &  0.440 &  0.002 &  0.046 &  12.8 &   0.1 &   0.7 & 6.9 &  5.8E-01  &  1.3E-01  &  1.3E-01 \\
 266 &  83.91805 &  -5.11422 &   0.32 &  0.017 & 0.48 & N &  0.315 &  0.008 &  0.132 &  19.2 &   0.5 &   2.4 & 7.0 &  9.3E-01  &  9.5E-02  &  5.4E-02 \\
 269 &  83.83881 &  -5.17787 &   0.71 &  0.028 & 0.53 & N &  0.180 &  0.001 &  0.030 &  18.7 &   0.1 &   0.9 & 7.2 &  5.2E+00  &  2.3E-01  &  2.3E-02 \\
 274 &  83.70746 &  -5.77024 &   0.59 &  0.032 & 0.57 & N &  0.479 &  0.002 &  0.180 &  16.3 &   0.1 &   0.9 & 7.1 &  2.4E+00  &  5.3E-02  &  1.1E-02 \\
 282 &  84.09522 &  -6.21789 &   0.39 &  0.017 & 0.57 & N &  0.297 &  0.002 &  0.127 &  13.9 &   0.1 &   0.6 & 6.9 &  8.1E-01  &  1.4E-01  &  9.5E-02 \\
 283 &  83.85865 &  -5.87194 &   1.11 &  0.028 & 0.54 & N &  0.274 &  0.002 &  0.080 &  12.0 &   0.2 &   0.9 & 6.9 &  1.2E+00  &  2.9E-01  &  1.4E-01 \\
 287 &  83.66783 &  -5.52075 &   0.89 &  0.034 & 0.59 & N &  0.260 &  0.005 &  0.081 &  14.2 &   0.6 &   1.5 & 6.7 &  2.0E+00  &  1.9E-01  &  5.1E-02 \\
 291 &  83.77259 &  -5.30811 &   0.18 &  0.018 & 0.62 & N &  0.157 &  0.004 &  0.056 &  17.9 &   0.5 &   2.0 & 7.3 &  6.0E+00  &  1.1E-01  &  8.9E-03 \\
 292 &  83.88603 &  -5.15899 &   0.27 &  0.022 & 0.48 & N &  0.230 &  0.004 &  0.062 &  22.5 &   0.3 &   2.3 & 7.2 &  4.4E+00  &  8.4E-02  &  9.6E-03 \\
 293 &  83.90896 &  -5.01580 &   0.40 &  0.031 & 0.66 & N &  0.200 &  0.006 &  0.035 &  15.5 &   0.6 &   1.6 & 6.9 &  6.1E+00  &  1.2E-01  &  1.0E-02 \\
 294 &  84.10313 &  -6.23431 &   3.60 &  0.051 & 0.45 & N &  0.448 &  0.002 &  0.046 &  12.9 &   0.1 &   0.8 & 6.9 &  1.2E+00  &  2.4E-01  &  1.1E-01 \\
 300 &  83.88596 &  -5.16940 &   0.29 &  0.017 & 0.44 & N &  0.246 &  0.003 &  0.058 &  22.0 &   0.2 &   2.0 & 7.2 &  1.9E+00  &  1.1E-01  &  2.8E-02 \\
 302 &  83.82228 &  -5.04308 &   0.56 &  0.035 & 0.59 & N &  0.491 &  0.006 &  0.114 &  22.7 &   0.3 &   3.1 & 7.2 &  3.7E+00  &  4.2E-02  &  5.7E-03 \\
 306 &  83.77675 &  -5.28182 &   0.31 &  0.032 & 0.59 & N &  0.279 &  0.011 &  0.117 &  21.7 &   1.1 &   4.3 & 7.1 &  8.1E+00  &  5.5E-02  &  3.4E-03 \\
 310 &  83.76282 &  -5.46122 &   0.17 &  0.020 & 0.58 & N &  1.007 &  0.032 &  0.087 &  29.3 &   1.5 &   7.1 & 7.0 &  3.8E-01  &  6.4E-03  &  8.5E-03 \\
 312 &  83.71450 &  -5.64783 &   0.80 &  0.036 & 0.49 & N &  0.455 &  0.008 &  0.110 &  17.8 &   0.5 &   1.4 & 7.0 &  2.0E+00  &  6.8E-02  &  1.8E-02 \\
 317 &  83.75261 &  -5.32364 &   0.59 &  0.022 & 0.46 & N &  0.432 &  0.017 &  0.109 &  15.9 &   1.1 &   3.0 & 7.2 &  1.1E+00  &  9.1E-02  &  4.5E-02 \\
 319 &  83.70721 &  -4.91835 &   0.28 &  0.026 & 0.62 & N &  0.207 &  0.009 &  0.005 &  21.4 &   0.8 &   0.4 & 6.7 &  2.6E+00  &  8.2E-02  &  1.6E-02 \\
 323 &  83.81324 &  -5.28590 &   0.21 &  0.024 & 0.63 & N &  0.721 &  0.017 &  0.083 &  25.0 &   0.7 &   1.7 & 7.3 &  1.9E+00  &  1.2E-02  &  3.1E-03 \\
 327 &  83.87514 &  -5.08087 &   0.41 &  0.026 & 0.63 & N &  0.204 &  0.001 &  0.019 &  16.4 &   0.1 &   0.8 & 7.2 &  7.1E+00  &  1.4E-01  &  1.0E-02 \\
 330 &  83.81624 &  -5.25487 &   1.71 &  0.058 & 0.53 & N &  0.311 &  0.002 &  0.046 &  20.3 &   0.1 &   1.2 & 7.2 &  8.8E+00  &  1.5E-01  &  8.7E-03 \\
 331 &  83.82015 &  -5.24576 &   0.25 &  0.017 & 0.53 & N &  0.224 &  0.002 &  0.074 &  21.0 &   0.2 &   1.4 & 7.1 &  2.2E+00  &  1.0E-01  &  2.4E-02 \\
 342 &  83.79221 &  -5.63126 &   1.75 &  0.052 & 0.48 & N &  0.567 &  0.006 &  0.124 &  16.9 &   0.3 &   2.0 & 7.0 &  2.0E+00  &  7.3E-02  &  1.9E-02 \\
 353 &  83.73182 &  -5.33475 &   0.10 &  0.026 & 0.74 & N &  0.566 &  0.030 &  0.148 &  22.3 &   2.3 &   6.0 & 7.0 &  4.3E+00  &  7.9E-03  &  9.1E-04 \\
 355 &  83.91499 &  -5.10713 &   1.54 &  0.079 & 0.57 & N &  0.334 &  0.005 &  0.111 &  20.7 &   0.3 &   3.2 & 7.0 &  1.7E+01  &  9.1E-02  &  2.7E-03 \\
 357 &  84.08591 &  -6.22051 &   0.32 &  0.020 & 0.65 & N &  0.321 &  0.003 &  0.111 &  13.9 &   0.1 &   0.6 & 6.9 &  1.3E+00  &  8.9E-02  &  3.5E-02 \\
 361 &  83.68854 &  -5.73855 &   0.73 &  0.025 & 0.52 & N &  0.154 &  0.001 &  0.052 &  12.7 &   0.1 &   1.4 & 7.1 &  4.0E+00  &  3.8E-01  &  5.0E-02 \\
 362 &  83.75024 &  -5.64903 &   0.72 &  0.035 & 0.60 & N &  0.207 &  0.002 &  0.069 &  15.5 &   0.1 &   1.3 & 7.1 &  6.9E+00  &  1.9E-01  &  1.4E-02 \\
 363 &  83.72537 &  -5.62012 &   0.32 &  0.033 & 0.59 & N &  0.402 &  0.006 &  0.122 &  22.0 &   2.0 &   6.0 & 7.0 &  4.5E+00  &  3.4E-02  &  3.8E-03 \\
 367 &  83.82254 &  -5.18514 &   0.14 &  0.025 & 0.71 & N &  0.167 &  0.004 &  0.070 &  20.0 &   0.5 &   3.5 & 7.2 &  1.8E+01  &  5.1E-02  &  1.4E-03 \\
 371 &  83.73876 &  -5.69251 &   1.48 &  0.044 & 0.44 & N &  0.227 &  0.002 &  0.175 &  15.0 &   0.1 &   1.8 & 7.1 &  6.2E+00  &  2.8E-01  &  2.4E-02 \\
 387 &  83.70517 &  -5.69387 &   0.19 &  0.020 & 0.59 & Y &  0.842 &  0.016 &  0.148 &  17.0 &   0.7 &   1.8 & 7.1 &  5.7E-01  &  1.0E-02  &  8.8E-03 \\
 390 &  83.89300 &  -5.08923 &   0.12 &  0.021 & 0.68 & N &  0.466 &  0.012 &  0.149 &  22.2 &   0.7 &   2.3 & 7.2 &  3.5E+00  &  1.6E-02  &  2.4E-03 \\
 391 &  83.79428 &  -4.97123 &   0.11 &  0.021 & 0.65 & N &  0.446 &  0.005 &  0.082 &  21.6 &   0.2 &   3.3 & 7.1 &  3.5E+00  &  1.6E-02  &  2.2E-03 \\
 401 &  83.76262 &  -4.98428 &   0.34 &  0.026 & 0.60 & N &  0.274 &  0.004 &  0.092 &  16.6 &   0.3 &   2.1 & 7.0 &  3.3E+00  &  8.4E-02  &  1.3E-02 \\
 402 &  83.71703 &  -4.94449 &   0.47 &  0.034 & 0.55 & N &  0.235 &  0.004 &  0.056 &  16.7 &   0.3 &   1.3 & 6.8 &  3.9E+00  &  1.1E-01  &  1.4E-02 \\
 405 &  83.74210 &  -5.70125 &   1.13 &  0.051 & 0.55 & N &  0.201 &  0.002 &  0.123 &  15.1 &   0.1 &   1.7 & 7.1 &  1.5E+01  &  2.1E-01  &  7.1E-03 \\
 407 &  83.75100 &  -5.41395 &   0.85 &  0.070 & 0.61 & N &  0.662 &  0.011 &  0.162 &  23.7 &   0.6 &   4.3 & 7.2 &  1.0E+01  &  1.9E-02  &  9.4E-04 \\
 413 &  83.73518 &  -5.56496 &   0.76 &  0.022 & 0.43 & N &  0.160 &  0.002 &  0.043 &  13.3 &   0.3 &   1.5 & 6.9 &  1.8E+00  &  4.2E-01  &  1.3E-01 \\
 420 &  83.74128 &  -5.72616 &   0.17 &  0.026 & 0.67 & N &  0.225 &  0.002 &  0.104 &  15.5 &   0.1 &   1.2 & 7.1 &  1.2E+01  &  5.3E-02  &  2.2E-03 \\
 422 &  83.89687 &  -5.17238 &   0.29 &  0.038 & 0.57 & N &  0.204 &  0.004 &  0.062 &  20.9 &   0.3 &   2.2 & 7.1 &  2.0E+01  &  6.0E-02  &  1.5E-03 \\
 431 &  83.75942 &  -5.63871 &   0.49 &  0.030 & 0.63 & N &  0.216 &  0.001 &  0.049 &  14.8 &   0.1 &   0.8 & 7.1 &  6.5E+00  &  1.4E-01  &  1.1E-02 \\
 432 &  83.79646 &  -5.62771 &   0.45 &  0.020 & 0.48 & N &  0.522 &  0.007 &  0.100 &  16.0 &   0.4 &   1.8 & 7.0 &  5.1E-01  &  5.6E-02  &  5.8E-02 \\
 434 &  83.73855 &  -4.95586 &   0.18 &  0.021 & 0.58 & N &  0.290 &  0.009 &  0.094 &  17.7 &   0.7 &   1.9 & 6.9 &  2.3E+00  &  5.1E-02  &  1.1E-02 \\
 440 &  83.78488 &  -5.96492 &   0.86 &  0.042 & 0.61 & N &  0.183 &  0.001 &  0.044 &  13.9 &   0.1 &   0.7 & 7.0 &  1.1E+01  &  2.2E-01  &  1.0E-02 \\
 441 &  83.71771 &  -5.73641 &   0.81 &  0.045 & 0.54 & N &  0.931 &  0.007 &  0.094 &  15.0 &   0.2 &   2.5 & 7.1 &  1.5E+00  &  1.6E-02  &  5.3E-03 \\
 461 &  83.80562 &  -5.16646 &   0.22 &  0.018 & 0.47 & N &  0.470 &  0.017 &  0.139 &  21.1 &   1.1 &   4.3 & 7.1 &  9.7E-01  &  3.5E-02  &  1.8E-02 \\
 462 &  83.90534 &  -5.10844 &   0.44 &  0.053 & 0.64 & N &  0.372 &  0.006 &  0.162 &  20.6 &   0.3 &   3.5 & 7.1 &  1.9E+01  &  3.3E-02  &  9.0E-04 \\
 463 &  83.87745 &  -5.04815 &   0.28 &  0.036 & 0.56 & N &  0.483 &  0.016 &  0.152 &  20.0 &   0.9 &   3.4 & 7.2 &  8.3E+00  &  2.1E-02  &  1.3E-03 \\
 464 &  83.77074 &  -4.94613 &   0.40 &  0.031 & 0.56 & N &  0.219 &  0.001 &  0.044 &  17.5 &   0.1 &   0.8 & 7.0 &  5.8E+00  &  1.0E-01  &  9.1E-03 \\
 469 &  83.80346 &  -5.92901 &   0.33 &  0.024 & 0.59 & N &  0.238 &  0.001 &  0.037 &  14.4 &   0.1 &   0.7 & 7.0 &  4.1E+00  &  1.1E-01  &  1.4E-02 \\
 470 &  83.74294 &  -5.68198 &   0.68 &  0.051 & 0.54 & N &  0.825 &  0.008 &  0.124 &  17.0 &   0.2 &   2.8 & 7.1 &  2.7E+00  &  1.5E-02  &  2.7E-03 \\
 471 &  83.81435 &  -5.55115 &   0.88 &  0.058 & 0.55 & N &  0.785 &  0.024 &  0.095 &  19.3 &   1.0 &   3.7 & 6.9 &  2.1E+00  &  1.8E-02  &  4.4E-03 \\
 478 &  83.82794 &  -5.93413 &   0.60 &  0.035 & 0.60 & N &  0.170 &  0.002 &  0.050 &  12.7 &   0.9 &   2.0 & 7.0 &  9.2E+00  &  2.1E-01  &  1.1E-02 \\
 486 &  83.92940 &  -5.09902 &   0.16 &  0.037 & 0.64 & N &  0.597 &  0.028 &  0.104 &  28.2 &   1.9 &   4.6 & 6.9 &  4.8E+00  &  8.0E-03  &  8.3E-04 \\
 493 &  84.15930 &  -6.26998 &   0.14 &  0.027 & 0.63 & N &  0.443 &  0.057 &  0.037 &  20.2 &   4.4 &   0.0 & 6.9 &  3.6E+00  &  1.7E-02  &  2.3E-03 \\
 498 &  83.77784 &  -5.26313 &   0.52 &  0.059 & 0.65 & N &  0.355 &  0.018 &  0.214 &  18.4 &   1.4 &   5.6 & 7.0 &  1.9E+01  &  4.0E-02  &  1.0E-03 \\
 500 &  83.90644 &  -5.09942 &   0.36 &  0.030 & 0.49 & N &  0.344 &  0.006 &  0.103 &  21.0 &   0.3 &   2.6 & 7.1 &  4.3E+00  &  5.4E-02  &  6.4E-03 \\
 508 &  83.88870 &  -6.10200 &   0.10 &  0.017 & 0.65 & Y &  0.774 &  0.019 &  0.145 &  17.6 &   1.9 &   1.9 & 6.9 &  5.6E-01  &  7.2E-03  &  6.5E-03 \\
 518 &  83.89446 &  -5.18567 &   0.09 &  0.027 & 0.67 & N &  0.275 &  0.016 &  0.081 &  23.2 &   1.4 &   1.9 & 7.1 &  1.8E+01  &  1.8E-02  &  5.2E-04 \\
 524 &  83.75788 &  -5.88522 &   0.26 &  0.034 & 0.62 & N &  0.767 &  0.017 &  0.219 &  15.7 &   0.9 &   3.7 & 7.0 &  1.9E+00  &  9.7E-03  &  2.6E-03 \\
 535 &  83.92459 &  -6.12882 &   0.56 &  0.048 & 0.58 & N &  0.248 &  0.003 &  0.198 &  13.6 &   0.3 &   1.2 & 6.9 &  1.3E+01  &  9.3E-02  &  3.5E-03 \\
 537 &  83.78815 &  -5.91879 &   0.35 &  0.033 & 0.68 & N &  0.254 &  0.001 &  0.039 &  14.1 &   0.1 &   0.6 & 7.0 &  8.6E+00  &  8.0E-02  &  4.7E-03 \\
 538 &  83.71536 &  -5.70617 &   0.29 &  0.031 & 0.61 & N &  0.734 &  0.007 &  0.199 &  16.9 &   0.2 &   2.2 & 7.1 &  2.0E+00  &  1.3E-02  &  3.2E-03 \\
 539 &  83.95736 &  -5.51178 &   0.28 &  0.034 & 0.65 & N &  0.854 &  0.064 &  0.127 &  20.5 &   2.3 &   6.1 & 6.5 &  4.6E-01  &  8.5E-03  &  9.4E-03 \\
 540 &  83.72819 &  -5.40841 &   0.17 &  0.030 & 0.63 & N &  0.798 &  0.016 &  0.197 &  22.5 &   0.7 &   2.4 & 7.0 &  1.8E+00  &  6.5E-03  &  1.8E-03 \\
 543 &  83.87811 &  -5.12074 &   0.29 &  0.035 & 0.50 & N &  0.212 &  0.002 &  0.046 &  18.4 &   0.2 &   1.9 & 7.3 &  2.2E+01  &  6.7E-02  &  1.5E-03 \\
 545 &  83.76186 &  -4.93142 &   0.18 &  0.026 & 0.61 & N &  0.262 &  0.002 &  0.074 &  18.0 &   0.1 &   1.1 & 6.9 &  6.0E+00  &  4.5E-02  &  3.7E-03 \\
 546 &  83.73587 &  -4.90142 &   0.10 &  0.022 & 0.69 & N &  0.162 &  0.004 &  0.019 &  18.4 &   0.4 &   1.8 & 6.8 &  6.2E+00  &  4.6E-02  &  3.7E-03 \\
 551 &  83.88134 &  -5.89270 &   1.13 &  0.050 & 0.56 & N &  0.168 &  0.001 &  0.066 &  11.3 &   0.1 &   0.9 & 6.8 &  1.1E+01  &  2.9E-01  &  1.4E-02 \\
 554 &  83.75212 &  -5.30263 &   0.68 &  0.046 & 0.54 & N &  0.281 &  0.007 &  0.108 &  16.0 &   0.4 &   1.5 & 7.0 &  1.1E+01  &  9.4E-02  &  4.4E-03 \\
 570 &  84.16045 &  -6.26304 &   0.45 &  0.033 & 0.55 & N &  0.304 &  0.046 &  0.032 &  12.6 &   3.1 &   1.6 & 6.9 &  4.0E+00  &  8.6E-02  &  1.1E-02 \\
 571 &  84.01832 &  -6.16034 &   0.22 &  0.036 & 0.56 & N &  0.473 &  0.007 &  0.114 &  17.8 &   0.4 &   1.5 & 6.8 &  4.3E+00  &  1.8E-02  &  2.1E-03 \\
 572 &  83.72882 &  -5.72719 &   1.79 &  0.089 & 0.60 & N &  0.299 &  0.002 &  0.199 &  15.4 &   0.1 &   1.4 & 7.1 &  3.3E+01  &  1.2E-01  &  1.8E-03 \\
 584 &  83.85491 &  -5.98366 &   0.31 &  0.020 & 0.45 & N &  0.186 &  0.004 &  0.115 &  15.5 &   0.5 &   1.4 & 7.0 &  2.7E+00  &  1.5E-01  &  2.9E-02 \\
 586 &  83.67780 &  -5.47204 &   0.58 &  0.041 & 0.47 & N &  0.294 &  0.008 &  0.139 &  17.0 &   0.9 &   3.6 & 6.7 &  3.7E+00  &  8.2E-02  &  1.1E-02 \\
 588 &  83.92395 &  -5.08629 &   0.10 &  0.028 & 0.60 & N &  0.185 &  0.037 &  0.111 &  24.6 &   5.5 &   0.0 & 6.9 &  1.5E+01  &  2.7E-02  &  8.9E-04 \\
 592 &  83.86800 &  -5.99738 &   0.55 &  0.043 & 0.57 & N &  0.237 &  0.002 &  0.051 &  14.0 &   0.1 &   1.1 & 6.9 &  1.1E+01  &  1.0E-01  &  4.7E-03 \\
 594 &  83.70678 &  -5.74392 &   0.37 &  0.024 & 0.42 & N &  0.549 &  0.006 &  0.169 &  17.9 &   0.2 &   1.1 & 7.1 &  1.2E+00  &  3.5E-02  &  1.5E-02 \\
 595 &  83.69590 &  -5.70022 &   0.39 &  0.040 & 0.55 & Y &  0.718 &  0.010 &  0.154 &  16.6 &   0.4 &   1.7 & 7.1 &  3.0E+00  &  1.4E-02  &  2.3E-03 \\
 601 &  83.64971 &  -5.68440 &   0.25 &  0.023 & 0.60 & N &  0.270 &  0.023 &  0.266 &  13.5 &   4.1 &   0.0 & 6.9 &  2.7E+00  &  7.8E-02  &  1.5E-02 \\
 603 &  83.87687 &  -5.06588 &   0.49 &  0.037 & 0.60 & N &  0.216 &  0.003 &  0.069 &  16.9 &   0.3 &   2.3 & 7.2 &  1.6E+01  &  1.1E-01  &  3.4E-03 \\
 610 &  84.12174 &  -6.25239 &   0.81 &  0.041 & 0.49 & N &  0.357 &  0.003 &  0.042 &  13.9 &   0.2 &   1.8 & 6.9 &  3.6E+00  &  9.5E-02  &  1.3E-02 \\
 611 &  83.82141 &  -5.99614 &   0.43 &  0.049 & 0.62 & N &  0.155 &  0.002 &  0.109 &  14.0 &   0.2 &   1.2 & 7.0 &  3.9E+01  &  1.1E-01  &  1.4E-03 \\
 612 &  83.68629 &  -5.78575 &   1.56 &  0.130 & 0.64 & N &  0.606 &  0.003 &  0.102 &  15.9 &   0.1 &   1.6 & 7.0 &  3.3E+01  &  2.3E-02  &  3.6E-04 \\
 616 &  83.87311 &  -5.05444 &   0.26 &  0.031 & 0.56 & N &  0.255 &  0.003 &  0.066 &  17.2 &   0.2 &   1.4 & 7.2 &  1.5E+01  &  5.8E-02  &  2.0E-03 \\
 622 &  83.76991 &  -5.92961 &   0.40 &  0.026 & 0.52 & N &  0.217 &  0.002 &  0.069 &  13.5 &   0.1 &   1.0 & 7.0 &  4.8E+00  &  1.4E-01  &  1.5E-02 \\
 623 &  83.71152 &  -5.71424 &   0.86 &  0.061 & 0.52 & N &  0.659 &  0.005 &  0.146 &  16.5 &   0.2 &   1.8 & 7.1 &  6.2E+00  &  2.3E-02  &  1.9E-03 \\
 630 &  83.81553 &  -5.05844 &   0.13 &  0.034 & 0.66 & N &  0.525 &  0.005 &  0.056 &  23.0 &   0.8 &   2.1 & 7.2 &  1.1E+01  &  9.1E-03  &  4.1E-04 \\
 632 &  83.93491 &  -6.13118 &   0.58 &  0.045 & 0.59 & N &  0.252 &  0.003 &  0.202 &  14.0 &   0.2 &   1.5 & 6.9 &  1.0E+01  &  9.9E-02  &  4.8E-03 \\
 637 &  83.80293 &  -5.21256 &   0.12 &  0.034 & 0.64 & N &  0.725 &  0.052 &  0.128 &  24.9 &   2.3 &   4.3 & 7.1 &  5.5E+00  &  4.7E-03  &  4.2E-04 \\
 648 &  83.80923 &  -5.17569 &   0.19 &  0.027 & 0.53 & N &  0.446 &  0.019 &  0.076 &  22.0 &   1.3 &   3.6 & 7.1 &  4.1E+00  &  2.2E-02  &  2.6E-03 \\
 686 &  83.73658 &  -5.80658 &   0.22 &  0.036 & 0.57 & N &  0.782 &  0.011 &  0.149 &  16.3 &   0.7 &   2.3 & 7.0 &  2.7E+00  &  7.6E-03  &  1.4E-03 \\
 691 &  83.75529 &  -5.46870 &   0.26 &  0.033 & 0.63 & N &  1.111 &  0.073 &  0.079 &  15.7 &   4.0 &   0.0 & 6.9 &  8.2E-01  &  4.9E-03  &  3.0E-03 \\
 699 &  83.75322 &  -4.89717 &   0.48 &  0.067 & 0.59 & N &  0.146 &  0.005 &  0.035 &  17.6 &   0.8 &   3.0 & 6.8 &  4.2E+01  &  7.8E-02  &  9.3E-04 \\
 701 &  83.75135 &  -5.89591 &   0.58 &  0.065 & 0.59 & N &  0.738 &  0.010 &  0.112 &  15.9 &   0.5 &   3.5 & 7.0 &  6.2E+00  &  1.2E-02  &  9.7E-04 \\
 702 &  83.72680 &  -4.96436 &   0.21 &  0.030 & 0.58 & N &  0.302 &  0.008 &  0.118 &  17.2 &   0.5 &   1.5 & 6.8 &  4.4E+00  &  4.0E-02  &  4.6E-03 \\
 707 &  83.75019 &  -5.81241 &   0.47 &  0.064 & 0.61 & N &  0.619 &  0.007 &  0.115 &  16.9 &   0.4 &   2.7 & 7.0 &  1.1E+01  &  1.4E-02  &  6.3E-04 \\
 721 &  83.67246 &  -5.69374 &   0.32 &  0.042 & 0.66 & N &  0.248 &  0.006 &  0.126 &  13.3 &   0.7 &   2.2 & 7.0 &  1.8E+01  &  6.1E-02  &  1.7E-03 \\
 734 &  83.74316 &  -5.83421 &   0.55 &  0.072 & 0.62 & N &  0.688 &  0.010 &  0.148 &  17.1 &   0.6 &   2.0 & 7.0 &  9.8E+00  &  1.2E-02  &  6.1E-04 \\
 750 &  83.82236 &  -5.58375 &   0.21 &  0.047 & 0.63 & N &  0.793 &  0.065 &  0.071 &  23.0 &   2.5 &   6.7 & 6.9 &  4.7E+00  &  5.2E-03  &  5.5E-04 \\
 755 &  83.73468 &  -5.90930 &   0.37 &  0.069 & 0.67 & N &  0.649 &  0.009 &  0.092 &  16.1 &   0.4 &   3.0 & 6.9 &  1.3E+01  &  9.2E-03  &  3.6E-04 \\
 768 &  83.78508 &  -6.00744 &   0.56 &  0.063 & 0.66 & N &  0.248 &  0.006 &  0.118 &  14.2 &   0.7 &   1.9 & 7.0 &  3.5E+01  &  6.9E-02  &  9.8E-04 \\
 784 &  84.01330 &  -6.23102 &   0.13 &  0.034 & 0.64 & N &  0.562 &  0.007 &  0.050 &  17.4 &   0.4 &   2.1 & 6.8 &  5.0E+00  &  8.3E-03  &  8.4E-04 \\
 790 &  83.80356 &  -4.93682 &   0.43 &  0.108 & 0.59 & N &  0.464 &  0.011 &  0.128 &  27.4 &   0.8 &   5.2 & 7.0 &  7.5E+01  &  1.1E-02  &  7.3E-05 \\
 816 &  83.73568 &  -5.36760 &   0.10 &  0.030 & 0.72 & N &  0.624 &  0.011 &  0.111 &  22.0 &   0.5 &   1.9 & 7.1 &  6.4E+00  &  6.0E-03  &  4.7E-04 \\
 835 &  83.65690 &  -5.69668 &   0.45 &  0.036 & 0.57 & N &  0.277 &  0.012 &  0.184 &  13.2 &   2.2 &   0.2 & 6.9 &  6.4E+00  &  8.7E-02  &  6.9E-03 \\
 869 &  83.80022 &  -5.78781 &   0.16 &  0.061 & 0.60 & N &  0.374 &  0.032 &  0.059 &  26.4 &   5.9 &   0.0 & 6.7 &  2.7E+01  &  9.6E-03  &  1.8E-04 \\
 871 &  83.72573 &  -5.80079 &   0.40 &  0.064 & 0.61 & N &  0.881 &  0.011 &  0.114 &  16.6 &   0.4 &   1.4 & 7.1 &  7.5E+00  &  6.2E-03  &  4.1E-04 \\
 873 &  83.86276 &  -4.95504 &   0.67 &  0.057 & 0.53 & N &  0.396 &  0.007 &  0.090 &  27.5 &   0.4 &   2.8 & 6.9 &  8.5E+00  &  4.0E-02  &  2.3E-03 \\
 879 &  83.80411 &  -6.02245 &   0.42 &  0.085 & 0.72 & N &  0.930 &  0.053 &  0.072 &  15.8 &   2.8 &   0.0 & 7.0 &  1.3E+01  &  4.4E-03  &  1.7E-04 \\
 907 &  83.84457 &  -6.05341 &   0.95 &  0.096 & 0.66 & N &  0.597 &  0.007 &  0.126 &  15.2 &   0.4 &   2.4 & 6.9 &  1.7E+01  &  2.0E-02  &  5.9E-04 \\
\enddata
\vspace*{0.1in}
\tablenotemark{a}{Core properties based on the {\it getsources} catalogue of \citet{Lane16}.  
	Only cores
	which have kinematic properties measured by GAS are listed.  Columns are the core ID
	in \citet{Lane16}, central position, mass (estimated using eqn 1 from the total flux),
	effective radius (geometric mean of the major and minor axis lengths), 
	concentration as derived using equation 11 and whether the core has an associated
	protostar.}
\tablenotemark{b}{Velocity dispersion ($\sigma_{obs}$) and kinetic temperature ($T_{kin}$) measured
	for the cores, averaging over all core pixels where sufficient signal was present.
	The value reported for each quantity is the mean weighted by the inverse square of
	the uncertainty.  The formal error in the weighted mean value is also reported
	(second column), as is the weighted standard deviation (third column).  This
	latter quantity is more reflective of the variation between fitted values 
	across the core.}
\tablenotemark{c}{Estimated pressure on the core boundary due to the weight of the overlying
	cloud material.  See Section~3.2 for more information.}
\tablenotemark{d}{Virial parameters estimated according to Section~3.2.  The virial ratio is
	given by $-(\Omega_G+\Omega_{P,c})/2\Omega_K$.  $-\Omega_G/\Omega_K$ reflects the 
	balance of gravitational binding versus thermal pressure, while $\Omega_G/\Omega_{P,c}$
	reflects whether gravity or external cloud pressure dominates the confinement of the
	cores.}

\end{deluxetable}
\end{longrotatetable}

\end{document}